\documentclass{mn2e} 
\usepackage{epsf,times}
\usepackage{amsmath}
\usepackage{graphicx}
\usepackage{color}
\usepackage{natbib}
\usepackage{times}
\newcommand{\etal}{{et al}\/.}
\newcommand{\simnot}{\mathord{\sim}}

\begin{document}

\title[Spectral ageing in the lobes of FR-II radio galaxies]{Spectral ageing in the lobes of FR-II radio galaxies: New methods of analysis for broadband radio data}

\author[J.J.~Harwood \etal]{Jeremy J.\ Harwood$^{1}$\thanks{E-mail: jeremy.harwood@physics.org}, Martin J.\ Hardcastle$^{1}$, Judith H.\ Croston$^{2}$ and Joanna L.\ Goodger$^{1}$
\\$^{1}$School of Physics, Astronomy and Mathematics, University of Hertfordshire, College Lane, Hatfield, Hertfordshire AL10 9AB, UK
\\$^{2}$School of Physics and Astronomy, University of Southampton, Southampton SO17 1BJ, UK}

\maketitle

\begin{abstract}

The broad-bandwidth capabilities of next generation telescopes such as the JVLA mean that the spectrum of any given source varies significantly within the bandwidth of any given observation. Detailed spectral analysis taking this variation into account is set to become standard practice when dealing with any new broadband radio observations; it is therefore vital that methods are developed to handle this new type of data. In this paper, we present the Broadband Radio Astronomy ToolS ({\sc brats}) software package and, use it to carry out detailed analysis of JVLA observations of three powerful radio galaxies. We compare two of the most widely used models of spectral ageing, the Kardashev-Pacholczyk and Jaffe-Perola models and also results of the more complex, but potentially more realistic, Tribble model. We find that the Tribble model provides both a good fit to observations as well as providing a physically realistic description of the source. We present the first high-resolution spectral maps of our sources and find that the best-fitting injection indices across all models take higher values than has previously been assumed. We present characteristic hot spot advance speeds and make comparison to those derived from dynamical ages, confirming the previously known discrepancy in speed remains present when determined at high spectral resolutions. We show that some previously common assumptions made in determining spectral ages with narrow-band radio telescopes may not always hold and strongly suggest these are accounted for in future investigations.

\end{abstract}


\section{Introduction}
\label{intro}

\subsection{Radio galaxies in the modern era}
\label{modernera}

\begin{figure*}
\centering
\includegraphics[angle=0,width=8.8cm]{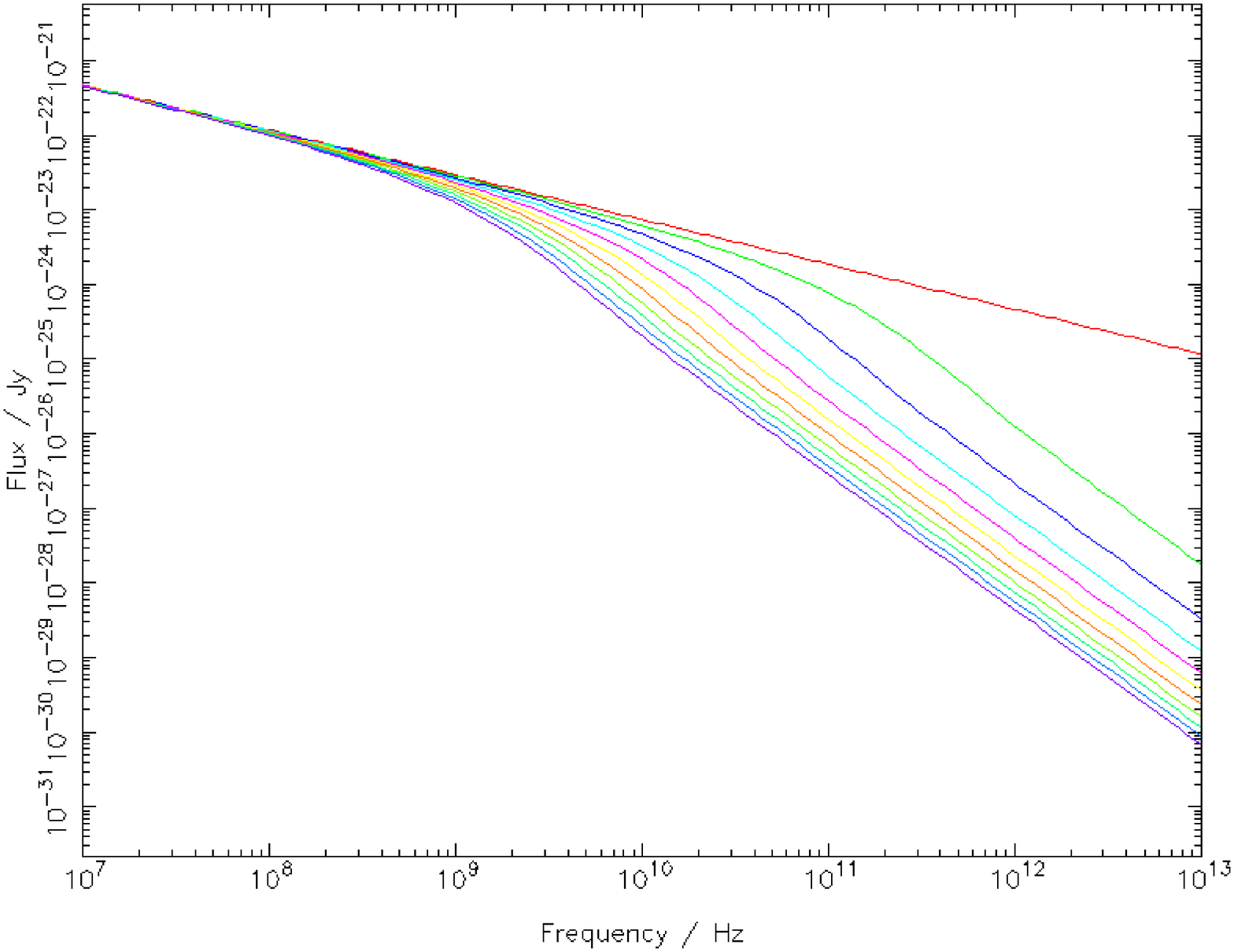}
\includegraphics[angle=0,width=8.8cm]{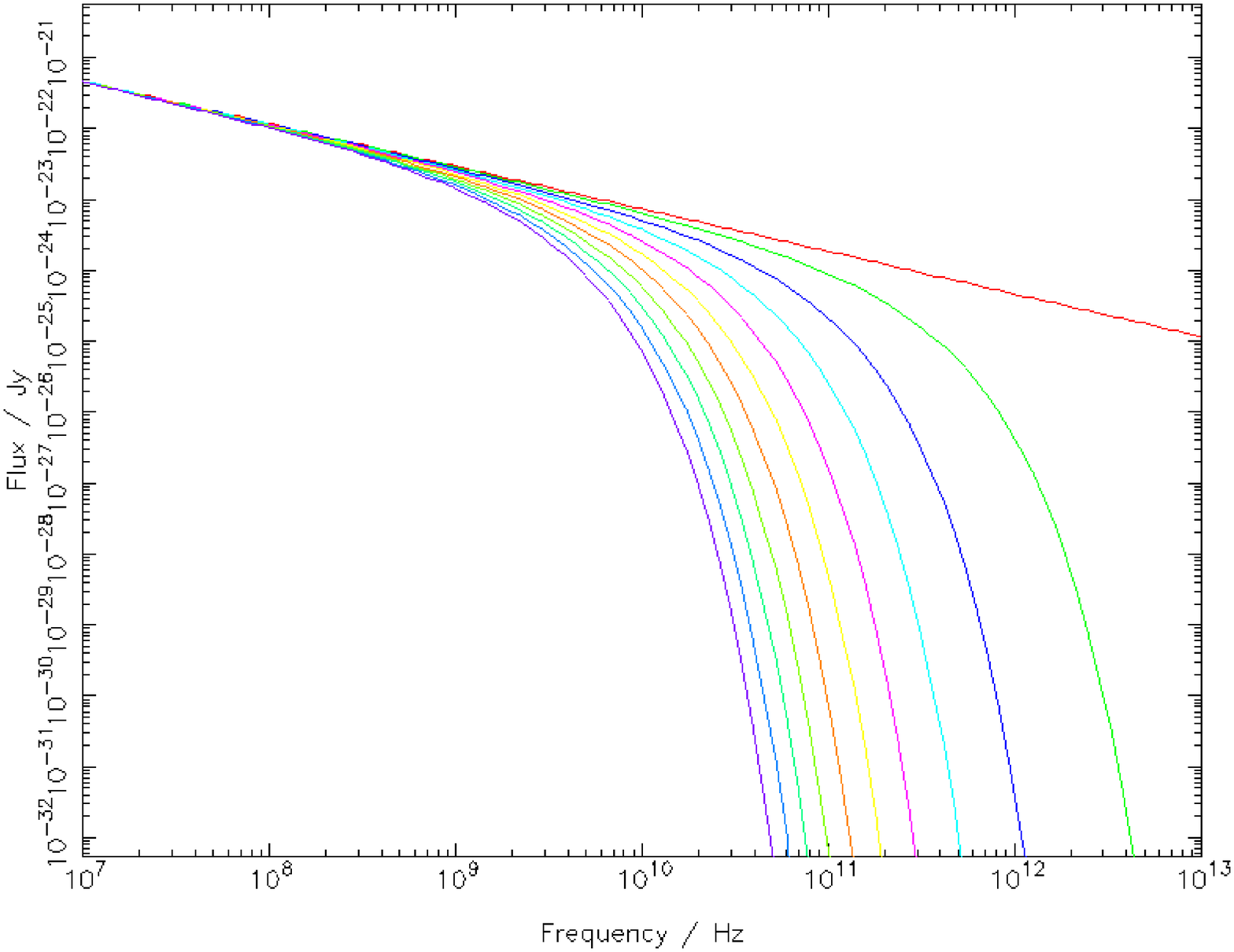}\\
\includegraphics[angle=0,width=8.8cm]{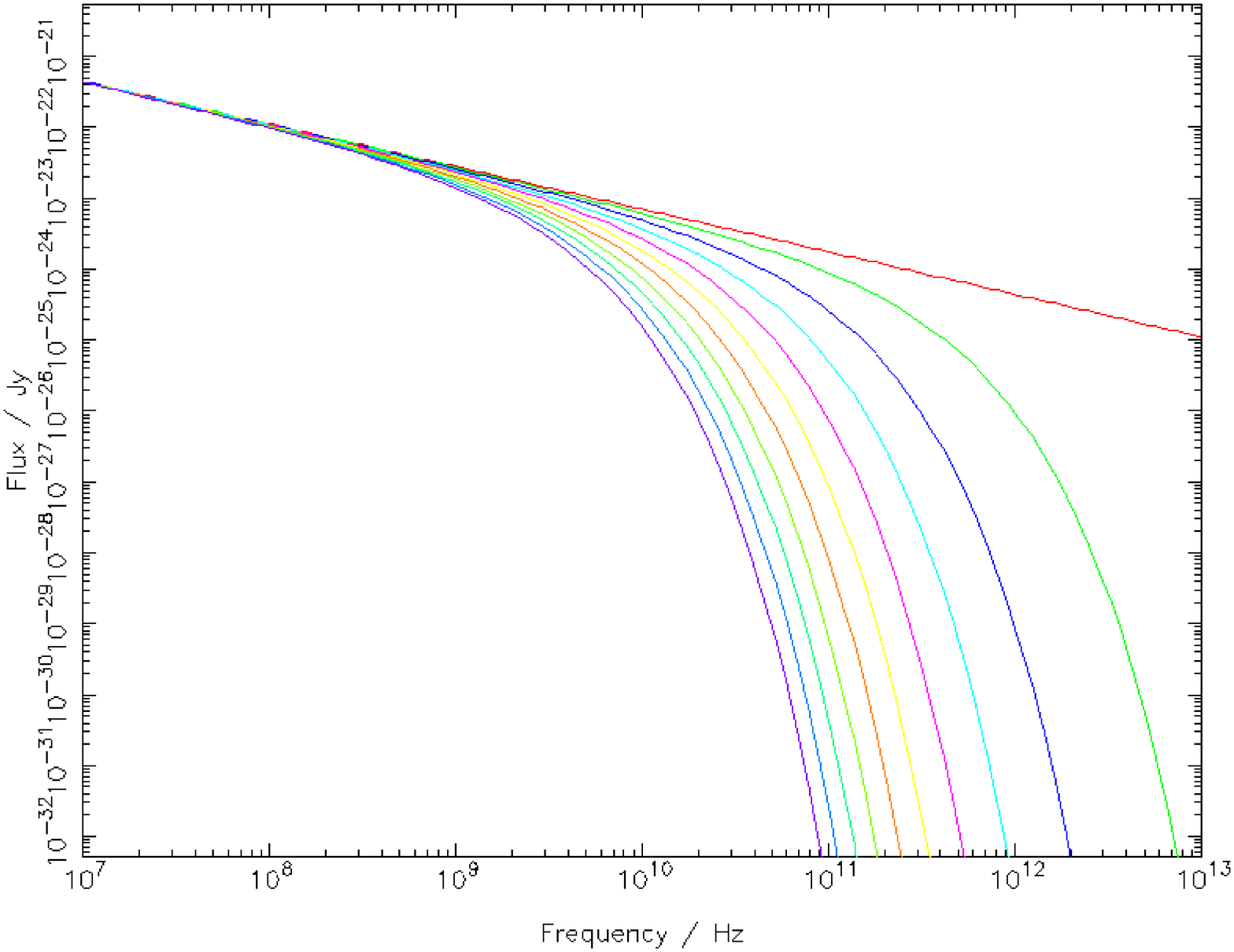}\\
\caption{Example KP (top left), JP (top right) and Tribble (bottom middle) models between 10 MHz and 10 THz for an arbitrary normalisation with 0.6 injection index. Each plot shows model ages between 0 (red) and 10 (purple) megayears.}
\label{examplemodels}
\end{figure*}

Found predominantly in massive ellipticals, radio-loud active galaxies are observed to have structure which can extend to scales of hundreds of kiloparsecs \citep{birkinshaw81, alexander87a, konar06, machalski09} to megaparsecs \citep{mullin06, marscher11}, meaning that they are not only able to interact with the interstellar medium, but also the environment in which the galaxy itself resides. As these galaxies have not only the potential to have such a large influence on their environment but can also be the primary channel for energy loss from supermassive black holes, understanding the dynamics, energetics and processes which occur within such systems is vital if we are to understand galaxy evolution.

The beam model first suggested by \citet{longair73} and refined by \citet{blandford74} and \citet{scheuer74} is commonly accepted as explaining the observed large scale structure of these sources. Powerful \citet{fanaroff74} class II (FR-II) radio galaxies (with which this paper is concerned) broadly consist of three large scale components; jets, lobes and hot spots. Collimated jets are the primary transport mechanism of material from the central AGN out to large distances in the form of a fluid flow. In FR-II galaxies, this material remains relativistic until terminating in a strong shock resulting in a region of high luminosity through synchrotron and inverse-Compton emission, known as the hot spot. As the hot spot advances through the external medium \citep{burch77, burch79, winter80, meisenheimer89}, the previously accelerated regions of plasma is then constrained to form the characteristic lobes associated with FR-II type radio galaxies (e.g. \citealp{scheuer74, begelman89, kaiser97, krause12}). Subsequent modelling of these galaxies has also shown good agreement with both the beam model and observations (see \citealt{ferrari98} for a review) and continues to this day \citep{tregillis01, heinz06, krause12, mendygral12, hardcastle13, morsony13}, although many outstanding questions remain unanswered.

Determining the shape of the energy spectrum for an electron population can often give important insights in to the underlying physics of the radio source. A region producing synchrotron radiation in a fixed magnetic field will have energy losses which scale as $\frac{dE}{dt} \propto \nu^{2}$; hence, in the absence of particle acceleration (such as may be the case in the lobes of FR-IIs) we expect preferential cooling of higher energy electrons leading to a steeper, more strongly curved spectrum in older regions of plasma (e.g. Figure \ref{examplemodels}). Models of this `spectral ageing' have therefore become a commonly used tool when describing the processes involved in these galaxies. Much of the early work on spectral ageing (e.g. \citealp{alexander87}) suggested that there was a physical reality to these models, so giving a measure of the age, and hence dynamics, of these powerful radio galaxies. Work on these models continues to this day \citep*{machalski09}; however, they are not the only possible solution and the problems arising from these models have been widely discussed (e.g. \citealp{eilek96a, blundell01}). Most notably, \citet*{katz93} showed that for Cygnus A a single electron energy spectrum in a varying magnetic field can produce the same results as those described by a spectral ageing model. Recent studies have shown \citep{hardcastle05, goodger08} that this single electron interpretation also has its problems as it is unable to explain observations of the spectrum at X-ray wavelengths, but this type of investigation is only currently possible for the largest and brightest sources due to the need for resolved inverse-Compton observations, so no consensus has yet been reached.

\begin{table*}
\caption{List of target sources, galaxy properties and observation IDs}
\label{targets}
\begin{tabular}{llcccccc}
\hline
\hline

Name&IAU Name&Redshift&5 GHz Core Flux&Spectral Index&LAS&JVLA Project ID&Frequencies\\
&&&(mJy)&(178 to 750 MHz)&(arcsec)&(GHz)&\\
\hline
3C436&J2144$+$281&0.215&19&0.86&109.1&AH1004&1.47\\
&&&&&&10B-137&4.04, 4.47, 4.92, 5.41,\\
&&&&&&&5.86, 7.21, 7.95\\
&&&&&&AC0138&8.47\\
3C300&J1422$+$195&0.272&9&0.78&100.9&AH1004&1.49\\
&&&&&&10B-137&4.04, 4.47, 4.92\\
&&&&&&&5.86, 7.21\\
&&&&&&AS0179 \& AV0088&7.94\\
3C234&J1001+287&0.1848&90&0.86&112&10B-137&4.92, 5.86, 7.21\\
&&&&&&AH1004&8.42\\
\hline

\end{tabular}

\vskip 5pt
\begin{minipage}{17.5cm}
`Name' and `IAU Name' list the 3C and IAU names of the galaxies as used within this paper. `Spectral Index' lists the low frequency spectral index between 178 to 750 MHz and `LAS' the largest angular size of the source. The `Redshift', `5 GHz Core Flux', `Spectral Index' and `LAS' column values are taken directly from the 3CRR database \citep{laing83} (http://3crr.extragalactic.info/cgi/database). `JVLA Project ID' refers to the project identifier as used by the NRAO archive search facility (https://archive.nrao.edu/). Where two IDs are listed in a single row, the data sets have been combined in the \emph{uv} plane during reduction. `Frequencies' lists the final image frequencies. Note that due to correlator issues, 3C234 has an insufficient number of radio maps for a full spectral analysis to be performed.
\end{minipage}

\end{table*}

\subsection{Models of spectral ageing}
\label{spectralageing}

Two of the most widely used models of spectral ageing are those first proposed by \citet*{kardashev62} and \citet*{pacholczyk70} (the KP model) and by \citet*{jaffe73} (the JP model). Both assume a single injection electron energy distribution, thought to be generated at the termination point of the jet (the hot spot) (e.g. \citealp{myers85, meisenheimer89, carilli91}). At the point of acceleration, the electron population is initially assumed to take the form of a power law given by \begin{equation}\label{initialpowerlaw}N(E) = N_0 E^{-\delta}\end{equation} where $N_{0}$ is a normalization factor and $\delta$ the power law index of the initially injected electron energy distribution. The back-flow is then subject to synchrotron losses as it propagates along the lobes potentially allowing the age and hence power of these outflows to be derived. For the KP model, an electron energy distribution, $N$, subject to synchrotron and inverse-Compton losses is given by \citet{pacholczyk70} to be \begin{equation}\label{kpdistribution} N(E,\theta,t) = N_0 E^{-\delta} (1 - E_{T} E)^{-\delta - 2}\end{equation} and the intensity at a given frequency by \begin{multline}\label{kpintensity} I_{\nu}(t) = 4\pi C_{3} N_{0} sB \int^{\pi / 2}_{0}\,d\theta \sin^{2} \theta \int^{E_{T}^{-1}}_{0}\,dE F(x) E^{-\delta} \\ \times (1 - E_{T} E)^{\delta -2}\end{multline} where \begin{equation}\label{kpetdef} E_{T} \equiv C_{2} B^{2} (\sin^{2} \theta) t\end{equation} with energy $E$, pitch angle $\theta$ and time since initial acceleration $t$. $C_{2}$, $C_{3}$, $\nu_{c}$ and the function $F(x)$ are constants defined by \citet*{pacholczyk70} where $x \equiv \nu / \nu_{c}$. We see from Equation \ref{kpetdef} that for the KP model the pitch angle $\theta$ is assumed to be both isotropic and constant over the radiative lifetime of the electrons. The JP model proposed by \citet*{jaffe73} instead assumes that the pitch angle is only isotropic on short time scales relative to the radiative life time, so giving \begin{equation}\label{jpetdef} E_{T} \equiv C_{2} B^{2} \langle \sin^{2} \theta \rangle t\end{equation} where $\langle \sin^{2} \theta \rangle$ represents the time averaged pitch angle.

It has long been recognised (e.g. \citealp{eilek96}) that multi-frequency observations for which spectra can be determined over a large number of regions within the source can provide vital advances in determining which, if any, of the current models is correct. However, this has largely been limited by the fact that only three or four narrow-band frequencies have been available at GHz wavelengths where the curvature is expected to be most obvious. The historic inability of radio interferometers to produce high sensitivity, high bandwidth observations with good \emph{uv} coverage has meant that many ambiguities exist in our current understanding of radio-loud active galaxies. Indirect methods such as the colour-colour diagrams used by \citet{katz93} and \citet{hardcastle01} have therefore been required to determine the validity of these spectral ageing models. The new generation of radio telescopes currently becoming operational is set to change this situation dramatically. The broad-bandwidth capabilities of instruments such as the JVLA mean that the spectrum of any given source can be determined within the bandwidth of any given observation, producing a detailed spectral shape and highlighting any curvature present. In the case of the JVLA, current capabilities allow curvature across the entire C-band (4-8 GHz) to be observed. This type of detailed spectral analysis is set to become standard practice when dealing with any new broadband radio observations; hence it is vital that methods are developed to handle this new type of data.

The improved bandwidth capabilities of instruments such as the JVLA also provide the first opportunity of investigating more complex, but possibly more realistic, spectral ageing models. Both the KP and JP models assume a fixed magnetic field strength throughout the source, but this is unlikely to be the case. A model in which an electron population ages within in a variable magnetic field was therefore proposed by \citet{tribble93} (herein the Tribble model). This family of models attempts to account for a more realistic magnetic field structure by instead assuming a Gaussian random field, allowing electrons to diffuse across regions of varying field strength. In the weak field, high diffusion case in which electrons are free-streaming, the spectrum can be modelled by integrating the standard JP losses over a Maxwell-Boltzmann distribution \citep{hardcastle13a}. \citet{tribble93} noted that the pitch angle scattering of the JP model is a more likely scenario than that of the fixed pitch angles of the KP model and that in a varying magnetic field such as this, the spectrum more closely resembles the KP model than the classical JP case. They suggested that this might explain results where the KP model provides the best fit, such as that of \citet{carilli91} in their investigation of Cygnus A. It is clear from Figure \ref{examplemodels} that the Tribble model provides a spectral shape which differs from either of the classical models. Historically, the small number of narrow-bandwidth frequencies available have not been sufficient to discern between the classical cases and the more complex models proposed by Tribble, but, with broadband data now available we are in a position to investigate the viability of these potentially more realistic models.

\begin{figure*}
\centering
\includegraphics[angle=0,height=4.7cm]{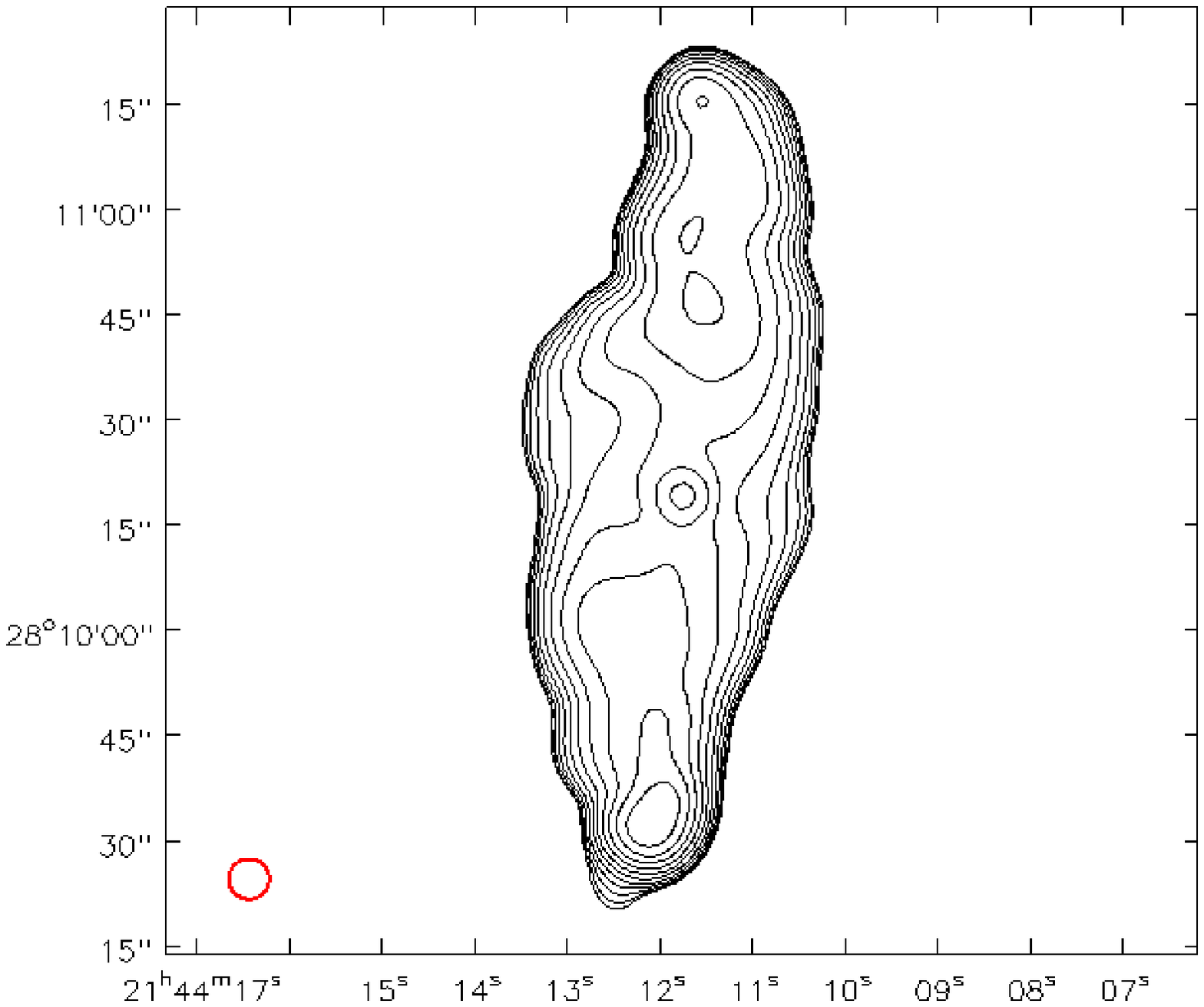}
\includegraphics[angle=0,height=4.7cm]{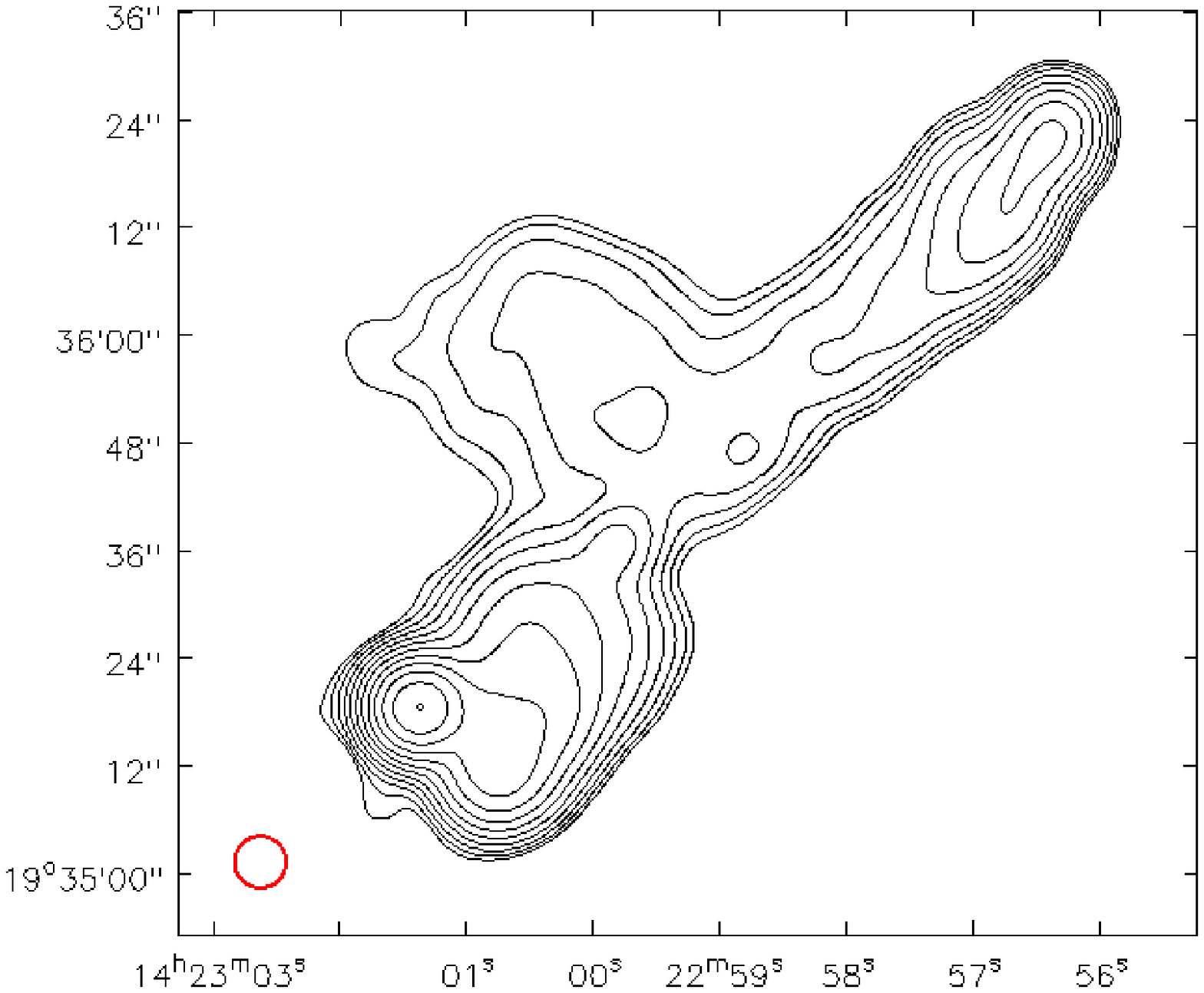}
\includegraphics[angle=0,height=4.7cm]{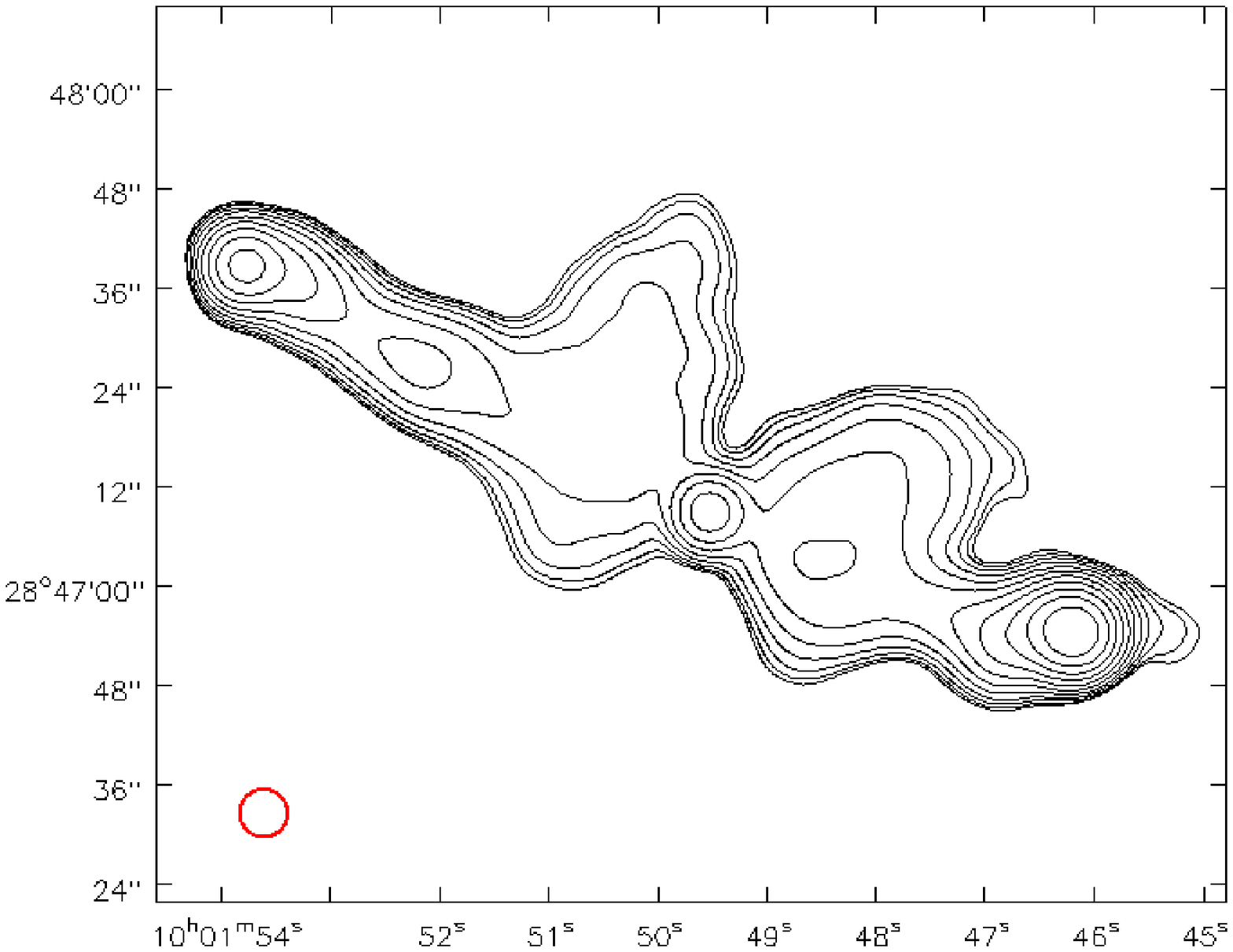}\\
\caption{Combined frequency radio maps of 3C436 (left), 3C300 (middle) and 3C234 (right) between 4 and 8 GHz scaled to 6.0 GHz. The off-source RMS of the combined maps is $27$, $33$ and $89$ $\mu$Jy beam$^{-1}$ for 3C436, 3C300 and 3C234 respectively. 12 contours are spaced logarithmically between $5\sigma$ based on the on-source RMS noise and the peak flux density. The restoring beam (red) is indicated in the bottom left corner of each image.}
\label{combinedmaps}
\end{figure*}

Within this paper we therefore address two primary aims; we detail new methods of analysis for broad bandwidth radio data and present the results of this analysis for the lobes of FR-II radio galaxies comparing JP, KP and Tribble models of spectral ageing. We go on to discuss the implications of our results in the context of spectral ageing and the dynamics of FR-II radio galaxies as a whole. Section \ref{method} gives details of target selection, data reduction and the spectral analysis methods used. In Section \ref{results} we present our results and Section \ref{discussion} discusses the implications for our current understanding of the dynamics of FR-II radio galaxies. Throughout this paper, a concordance model in which $H_0=71$ km s$^{-1}$ Mpc$^{-1}$, $\Omega _m =0.27$ and $\Omega _\Lambda =0.73$ is used.

\section{Data Reduction and Spectral Analysis}
\label{method}

\subsection{Target selection and observations}
We selected our targets from the 3CRR sample \citep*{laing83}, which contains many well-studied FR-II sources. Previous studies have shown (\citealp{alexander87}; \citealp{alexander87a}; \citealp{carilli91}; \citealp*{perley97}; \citealp{hardcastle01}) that spectral steepening becomes observable at frequencies above a few GHz. The resolution obtained by the JVLA at C-band frequencies (4.0-8.0 GHz) was therefore a natural choice for making the required observations. In choosing our targets it was important that we were confident of seeing all known structure and so the largest angular size (LAS) of our targets must be well sampled by the shortest spacing available at the top of C-band. This restricted us to sources with angular sizes $\leq 180$ arcsec. At the same time, we required a reasonably large angular size in order to be able to make well-resolved images with the resolution available at the bottom of the C-band. In order to extend our investigation beyond the C-band range, we restricted ourselves to $z < 0.3$, since the sources of appropriate angular sizes also have good X-band (8.0 - 8.8 GHz) VLA observations \citep{leahy97, hardcastle97} as well as archival VLA L-band (1.34 - 1.73 GHz) data available. Within this redshift range, there were 6 3CRR sources with 100 $<\,\theta_{LAS}\,<$ 180 arcsec. In order to obtain the best signal to noise ratio the brightest three sources were chosen for observation (Table \ref{targets}).

At the time of proposal the JVLA was still at an intermediate stage of development. In order to simulate the broadband nature of the fully upgraded instrument, observations were made using 128-MHz bandwidth spread logarithmically throughout the whole C-band. In order to ensure both compact and diffuse emission was sampled, observations were obtained in both C and D configurations. Details of the datasets obtained along with the archival VLA datasets used can be found in Table \ref{targets}. Observations at X-band frequencies used data reduced previously by \citet{hardcastle97}. Unfortunately, due to issues with the JVLA WIDAR correlator at the time of observation, 3C234 has an insufficient number of radio maps available for a full spectral analysis to be performed; however, we include details of these observations and present combined radio map images in Section \ref{datareduction} for those frequencies which were available, as these still provide the broadest frequency radio maps of the source to date.

\subsection{Data Reduction}
\label{datareduction}

The C- and D-band configuration data were downloaded from the NRAO archive and reduced using {\sc aips}\footnote{http://www.aips.nrao.edu/index.shtml} in the standard manner accounting for the special considerations for JVLA data calibration in Appendix E of the {\sc aips} cookbook\footnote{http://www.aips.nrao.edu/cook.html}. CLEAN components for self-calibration which was applied in phase only were obtained using the Cotton-Schwab algorithm \citep{schwab84}. The final image was then produced from the calibrated data set by CLEANing to convergence. Before the data were ready for analysis, the C and D configuration observations first had to be combined in the \emph{uv} plane and the final images created. The D configuration data were re-calibrated in phase only using the higher resolution C configuration map as a model. The data sets were then combined in the \emph{uv} plane using the DBCON task. As we require all of the maps to be well matched in terms of both image and beam size, the combined data were imaged to the resolution of the lowest frequency C-band data. A summary of parameters used for imaging is shown in Table \ref{imagrprms}.

The X-band data of \citet{hardcastle97} were already pre-calibrated to a high standard; however, regridding to J2000 co-ordinates was required along with re-imaging to match the parameters of Table \ref{imagrprms}. For reduction of the archival VLA L-band observations, standard methods were used as per the {\sc aips} cookbook. The parameters used in reduction of the C-band observations were applied where appropriate and the array configurations combined in the \emph{uv} plane. As this is a long standing and well established process we shall not go in to any further detail here, but we note that B and C configurations were used to allow comparable resolution to be obtained at these longer wavelengths.

Ensuring the radio maps were accurately aligned across all frequencies was of the utmost importance in allowing spectral features as a function of position to be determined. The core and hot spots were therefore used as reference and each frequency aligned during imaging to correct for small deviations in position between maps. On completion of the JVLA upgrade, fully broadband observations will partially resolve this issue as all frequencies within a given band will be taken in a single pointing; however, it will still be necessary to correct for these errors due to frequency-dependent phase shifts causing offsets and when combining data over more than one frequency band.

Integrated fluxes were measured (Table \ref{mapdetails}) using DS9\footnote{http://hea-www.harvard.edu/RD/ds9/site/Home.html} as a check for our assertion that all known structure is being observed on short baselines. Taking measurements from the single dish observations by \citet{becker91} and the updated Parkes catalogue (PKSCAT90\footnote{http://www.parkes.atnf.csiro.au/observing/databases/pkscat90.html}) of \citet{wright90} and scaling to 4.92 GHz, we found integrated flux densities of $1.01 \pm 0.15$ Jy and $1.09 \pm 0.05$ Jy for 3C436 and 3C300 respectively. Comparing these values to those of Table \ref{mapdetails} we found a a difference of less than 5 per cent in both cases. As this is well within the measurement errors, we can be confident that we are observing all known structure on short baselines.

In order to help distinguish between possible deconvolution and image fidelity effects and real structure, as well as to test the best form of reduction for spectral ageing studies, additional methods of imaging were also used. As good \emph{uv} coverage is important for spectral studies such as this, multi-scale CLEANing was an obvious choice for such a test. We therefore imaged both sources in {\sc aips} using multi-scale CLEAN on scales of 0, 1, 3 and 9 times the beam size. As our targets contain large areas of extended emission, especially for the oldest regions of plasma which are of particular interest, the data were also deconvolved using the maximum entropy method (MEM) which typically handles diffuse emission well. Images were made using the VTESS task as described in the {\sc aips} cookbook; however, our sources are known to have both extended (lobes) and compact structure (hotspots and core). As the CLEAN algorithm is known to work particularly well on compact structure, a hybrid MEM was also used in which a small number of CLEAN iterations were first run to deconvolve the small scale structure. The CLEAN components found were then subtracted from the \emph{uv} data and VTESS used to deconvolve the remaining visibilities. The CLEAN components were then restored to create the final images. This provided a total of 3 additional image sets for comparison to our standard CLEAN algorithm imaging.

\begin{table}
\caption{Summary of imaging parameters}
\label{imagrprms}
\begin{tabular}{llcl}
\hline
\hline

Parameter&AIPS Name&Value&Units\\

\hline
Polarization&STOKES&I&\\
Image Size&IMSIZE& 512 512&Pixels\\
Cell Size&CELLSIZE&0.8 0.8&Arcsec\\
Weighting&ROBUST&3&\\
Beam Major Axis&BMAJ&5.75&Arcsec\\
Beam Minor Axis&BMIN&5.75&Arcsec\\
\hline

\end{tabular}

\vskip 5pt
\begin{minipage}{8.5cm}
`Parameter' refers to the imaging parameter used in making of radio maps within this paper. `{\sc aips} Name' refers to the {\sc aips} parameter name; this parameter takes the value stated in the `Values' column.
\end{minipage}

\end{table}

\subsection{Spectral Analysis}

With the new generation of radio telescopes such as the JVLA coming online, the analysis of spectral energy distributions over broad bandwidths is set to become common practice; however, the tools to effectively deal with such data are currently not yet available. We therefore developed the Broadband Radio Analysis ToolS\footnote{Contact lead author for details on using this software or visit http://www.askanastronomer.co.uk/brats} ({\sc brats}) software package to address this issue. The {\sc brats} software provides a wide range of model fitting, visualisation and statistical tools for analysing broadband radio maps, key of which to this paper are:

\begin{enumerate}

\item Automatic selection of radio map regions for a given set of parameters.
\item Fitting and statistical testing of spectral ageing models.
\item Determination of source properties as a function of position (e.g. spectral index, spectral age, goodness-of-fit).
\item Determination of model parameters (e.g. injection index).
\item Source reconstruction and subtraction for a given spectral ageing model.

\end{enumerate}

The {\sc brats} software is written in the C programming language and for visualisation the software makes use of the {\sc pgplot}\footnote{http://www.astro.caltech.edu/\textasciitilde tjp/pgplot/} graphics subroutine library, with reading and manipulation of FITS files making use of the Funtools\footnote{https://www.cfa.harvard.edu/\textasciitilde john/funtools/} library. A detailed account of the full range of features available within the software package is beyond the scope of this paper, but instead we detail the functions and underlying methods used in the context of spectral ageing.

The radio maps at each frequency were first loaded in to {\sc brats} and, to save on processing time at later stages of analysis, a region fully encompassing the source (but excluding the core) was defined using DS9 along with a background region. During the load process the maps were automatically checked to ensure that they were matched in terms of dimensions, target source, beam size as well as ensuring no two maps were of the same frequency. The defined background region was then used to calculate a value for the RMS of each map which we can take to represent the off-source thermal noise. Initial source detection was then performed, for which a $5 \sigma$ cut-off was used based on this RMS value.

The radio maps at each C-band frequency were also combined in order to give the widest bandwidth images of these sources available to date. The integrated flux density of the source at each frequency was first plotted using the {\sc brats} \emph{`totalflux'} command and a suitable scaling factor between frequencies determined based on a power law model. The flux density was taken from the {\sc BRATS} data tables (as loaded from the FITS images) and this scaling factor applied on a pixel by pixel basis to an intermediate frequency of 6 GHz. The flux density was then averaged across the number of maps merged and a FITS image output using the {\sc funtools} source library. The details of the individual frequency maps are given in Table \ref{mapdetails} with resultant combined radio maps shown in Figure \ref{combinedmaps}.

\subsubsection{{\sc brats}: Adaptive regions}
\label{adaptiveregions}

Traditionally, the regions over which integrated flux density values are obtained in the analysis of spectral ageing are large and often encompass the entire width of the lobe. With the improved performance of the JVLA and subsequent improvement in image fidelity, it is possible to consider regions much smaller regions than has previously been viable. In the case of bright sources with good \emph {uv} coverage, it is often possible to consider the source spectra on a pixel by pixel basis but in other cases we may need to consider larger regions to get the required signal to noise. A function was therefore written to group pixels in to regions based on a specified set of parameters. These adaptive regions were constrained in two primary ways; signal to noise ratio and maximum search area.

Although we have so far assumed the off-source RMS to be a good approximation of background noise, it is likely to be the case that the noise on-source is considerably higher due to the increased uncertainty in the modelling of the extended emission. This has implications for both region selection and for statistical testing of models (Section \ref{modelfitting}). For all but the lowest flux regions this effect is likely to be negligible with the off-source noise being at least an order of magnitude smaller than the standard 2 per cent flux calibration error for C-band observations with the JVLA. However, the most diffuse, low flux regions should also contain the oldest electron population. To correct for this uncertainty, the flux density and spectral index maps were carefully inspected to locate areas of diffuse emission within the source for which we could be confident that flux density was approximately constant and free from any significant structure. Multiple thin regions were then selected in DS9 and an RMS of the deviation from the mean flux density obtained. Dividing each region through by the off-source RMS we found a tight clustering of values $\simnot\,3$ times the RMS for both sources. This multiplier was then applied during both the region selection and the the model fitting process through the {\sc brats} `\emph{onsource}' command.

For each pixel for which a region has not yet been defined, and has not been excluded by one of the bad pixel detection techniques (see below), its flux density  $S_{reg}$ was compared to a minimum flux density given by \begin{equation}\label{adaptselect}S_{reg} \geq R_{SN} \left\{(J \times S_{RMS}) \sqrt{n_{reg} / a_{beam}}\right\}\end{equation} where $R_{SN}$ is a user defined signal to noise ratio, $J$ is the on-source noise multiplier, $S_{RMS}$ is the thermal noise, $n_{reg}$ is the number of pixels currently in a given region and $a_{beam}$ is the primary beam area. If the pixel flux density is below that of the minimum value, an adjacent pixel is added to the region and the test repeated with an increased value of $n_{reg}$. This process was then repeated adding the closest unused pixel until either the inequality of Equation \ref{adaptselect} is satisfied, or, $n_{reg}$ is greater than a defined maximum search area. In the case that a pixel failed to satisfy these criteria it was marked as bad and was no longer considered in any subsequent analysis.

\begin{table}
\caption{Summary of maps by frequency}
\label{mapdetails}
\begin{tabular}{lcccc}
\hline
\hline

Source&Frequency&Off-source RMS&On-source RMS&$S_{int}$\\
&(GHz)&($\mu$Jy beam$^{-1}$)&($\mu$Jy beam$^{-1}$)&(Jy)\\

\hline
3C436&1.47&$247$&$741$&3.44\\
&4.04&$141$&$423$&1.15\\
&4.47&$78.5$&$236$&1.08\\
&4.92&$111$&$334$&0.96\\
&5.41&$98.5$&$295$&0.87\\
&5.86&$68.1$&$204$&0.83\\
&7.21&$63.1$&$189$&0.66\\
&7.95&$64.3$&$193$&0.60\\
&8.47&$49.9$&$150$&0.56\\
3C300&1.49&$266$&$799$&3.55\\
&4.04&$107$&$321$&1.31\\
&4.47&$67.6$&$203$&1.23\\
&4.92&$67.7$&$203$&1.11\\
&5.86&$69.7$&$209$&0.92\\
&7.21&$79.2$&$237$&0.73\\
&7.94&$90.4$&$271$&0.66\\
3C234&4.92&$219$&$657$&1.58\\
&5.86&$125$&$375$&1.28\\
&7.21&$103$&$309$&1.03\\
&8.42&$114$&$342$&0.91\\
\hline

\end{tabular}

\vskip 5pt
\begin{minipage}{8.5cm}
`Frequency' refers to the frequency of the map listed in the `Source' column. `Off-source RMS' refers to RMS noise measured over a large region well away from the source and `On-source RMS' the noise used for region selection and statistics as per Section \ref{adaptiveregions}. `$S_{int}$' refers to the background subtracted total flux density of the source as measured in DS9. The integrated flux density errors are taken to be the standard JVLA flux calibration uncertainty of 2 per cent.
\end{minipage}

\end{table}

We also note a word of caution when selecting the parameters for this method of region selection, which is of particular importance when one wishes to consider each as a potential fit to an spectral ageing model. As models of spectral ageing have multiple free parameters (see Section \ref{modelfitting}), the superposition of two JP or KP synchrotron spectra will not always result in the expected spectrum for a simple variation in age. If we consider the extreme case in which a region is defined such that it covers an area over which there is a sharp gradient between a very old spectrum with a high normalization and a young region with a low normalization, the superposition of these two ages will only provide a JP or KP `like' spectrum rather than the true spectrum which was intended to be observed. This will depend heavily on the frequencies of the observations that have been used in the fitting process and can lead to unreliable ages and inflated statistical values for the model fits. In reality, we are nearly always observing the sum of different synchrotron spectra in this way as we are integrating along the line of sight through the source, but we assume that these variations will be small enough to be negligible. For the most part, this is also likely to be true of the point to point variations across the length and width of the lobes for most sources; however, it is in general best practice to use the smallest possible regions for the given quality of data available to minimise these effects. It should be noted that although we cannot consider structure on scales smaller than the beam size, each pixel will still be weighted towards the true spectrum, hence by using the smallest regions possible for a given set of observations, the overall reliability of the results in terms of model fitting will be increased. As our observations were planned specifically with these considerations in mind (e.g. good \emph{uv} coverage for high image fidelity) in the analysis that follows we considered each source on a pixel by pixel basis to minimise the risk of the superposition of spectra affecting the results in this way. A summary of the values chosen for this region selection is shown in Table \ref{regprms}.

The adaptive regions function also provides additional bad pixel detection techniques above those of the signal to noise ratio and sigma level detections. During our analysis, we apply the {\sc brats} `hot' and `cold' pixel detection function. When a pixel was tested for inclusion within a region its flux density value was compared with the surrounding pixels. If the tested flux density differs from those surrounding it by a specified multiple more then 50 per cent of the time, or, too many of the surrounding pixels were already marked as bad so the test cannot be performed reliably, the tested pixel was also excluded. It was important that the parameters for this detection method (see Table \ref{regprms}) were set so as to only exclude large variations in flux density which are unphysical and highly likely to be a result of spurious data.

\begin{table}
\caption{Summary of adaptive region parameters}
\label{regprms}
\begin{tabular}{lcll}
\hline
\hline

Parameter&Value&Units&Description\\

\hline
Signal to noise&1&&SNR (pixel to pixel)\\
Search area&1&Pixels$^{2}$&Max. search area\\
On-source noise&3&&On-source noise multiplier\\
Hot pixel limit&20&Per cent&Max. pixel variation\\
Map variations&-1&&Maximum map variation (off)\\
\hline

\end{tabular}

\vskip 5pt
\begin{minipage}{8.5cm}
`Value' refers to the values applied within {\sc brats} for the corresponding `Parameter'. The `Description' column provides further details of the value meaning. Note that a signal to noise of 1 is still subject to the 5$\sigma$ cutoff and the on-source noise multiplier.
\end{minipage}

\end{table}

\subsubsection{{\sc brats}: Model fitting and parameter determination}
\label{modelfitting}

The {\sc brats} software provides a number of tools designed specifically for the analysis of spectra and fitting of spectral ageing models to broadband radio data. These range from the fitting of a simple power law between the top and bottom frequencies within a given data set, through to the much more complex spectral ageing models of Figure \ref{examplemodels}. Here, we tested two of the most commonly used JP and KP models, along with the more complex Tribble model with free-streaming JP losses as detailed in Section \ref{spectralageing}.

In determining the model fluxes for comparison to observations, we used the standard synchrotron equations outlined by \citet{longair11} where the emissivity of a single electron as a function of frequency is given by \begin{equation}\label{jv}J(\nu)  =  \frac{\sqrt{3} B e^{3} \sin^2\alpha}{8 \pi \epsilon_{0} c m_{e}} F(x)\end{equation} To reduce computational time, a lookup table of the single-electron synchrotron radiation spectrum values, $F(x)$ (see Equation \ref{kpintensity}), was made between $x = 1\times10^{-4}$ and $22$ at 100 logarithmic intervals. We took $F(x)$ as defined by \citet{rybicki79} to be \begin{equation}\label{fx} F(x) = x \int^{\infty}_{x}\,K_{5/3}(z)dz\end{equation} where $K$ is the Bessel function of order $5/3$, $x \equiv \nu / \nu_{c}$ and $\nu_{c} = {3 \gamma^{2} e B_{\perp}} / {4 \pi m_e c}$ is the critical frequency. These points were then used to find any value of $x$ using a log-linear two-point interpolation. For cases where $x$ falls outside of the minimum tabulated range, we use the asymptote of \citet{pacholczyk70} \begin{equation}\label{asymptote}F(x) = 2.15 x^{1/3}\end{equation} and for large values of $x$ we assumed that $F(x) = 0$. The pitch angles were taken to be isotropic, hence integrating over the probability distribution $(1/2) \sin{\alpha}$ and between the minimum ($E_{min}$) and maximum ($E_{max}$) electron energies we find the total model emission to be \begin{equation}\label{emiss}S_{mod}(\nu) = \frac{\sqrt{3} e^{3} B} {8 \pi \epsilon_{0} c m_{e}} \int^{\pi}_{0} \int^{E_{max}}_{E_{min}}\, F(x) \sin^2\alpha \,n_{e}(E)\, de\, d\alpha \end{equation} where $n_e$ is the electron energy distribution subject to losses given by \begin{equation}\label{eloss}n_e(E) = N_{0} E^{-2\alpha+1} (1 - \beta)^{(2\alpha+1)-2}\end{equation} and $\beta$ are the model dependent losses. For the KP model \begin{equation}\label{kploss} \beta_{KP} = B^{2} E t \left(\frac{4 \sigma_{t} \sin \alpha}{6 m^{2}_{e} \nu^3_{c} \mu_0}\right) \end{equation} is analogous to Equation \ref{kpetdef} and for the JP model \begin{equation}\label{jploss} \beta_{JP} = B^{2} E t \left(\frac{4 \sigma_{t}}{6 m^{2}_{e} \nu^3_{c} \mu_0}\right) \end{equation} is analogous to Equation \ref{jpetdef}. 

In the case of the free-streaming Tribble model, the JP losses are integrated over a Maxwell-Boltzmann distribution \citep{hardcastle13a}, hence where $B_{0}$ is the mean magnetic field strength of the source Equation \ref{emiss} becomes \begin{multline}\label{emissmb}S_{mod}(\nu) = \sqrt{\frac{2}{\pi}} \frac{\sqrt{3} e^{3}} {8 \pi \epsilon_{0} c m_{e}} \int^{\infty}_{0} \int^{\pi}_{0} \int^{E_{max}}_{E_{min}} F(x) \sin^2\alpha \,n_{e}(E) \\\times \frac{B^{3} \exp{\left(-B^{2} / 2B_{0}^{2}\right)}}{B^{3}_{0}} \, dE\, d\alpha \, dB\end{multline}

 Using the GNU Scientific Library\footnote{http://www.gnu.org/software/gsl/} ({\sc gsl}) for the required numerical integration, unnormalised flux density values were determined between a minimum and maximum age range. These values were then normalised (see below) and a $\chi^{2}$ statistical test performed using the standard equation \begin{equation}\label{chisquared}\chi^{2} = \sum\limits_{\nu=1}^{N} \left(\frac{S_{i, \nu} - S_{model, \nu}} {\Delta S_{i, \nu}}\right)^{2}\end{equation} where $N$ is the total number of observed frequencies, $S_{i, \nu}$ is the observed flux density of region $i$ at frequency $\nu$, $S_{model, \nu}$ is the model flux and $\Delta S_{i, \nu}$ is the total uncertainty of the observed region given by \begin{equation}\label{totalregerror}\Delta S_{i, \nu} = \left\{(J \times S_{RMS}) \sqrt{n_{reg} / a_{beam}}\right\} + (S_{i, \nu} \times \Delta S_{E})\end{equation} where $\Delta S_{E} = 0.02$ is the standard 2 per cent flux calibration error for the JVLA at C- and L-bands and $J$ is the on-source noise multiplier as defined in Section \ref{adaptiveregions}.

As we could not be certain that in all cases the spectral age would not contain local minima, a grid search was used to find the best fitting spectral age for each model. In order to obtain a high accuracy, but at a reasonable computing cost, a broad search was first run using large age intervals (1 Myr). This process was then automatically repeated over a series of decreasing age intervals bounded by the best fitting model of the previous cycle until the required accuracy was obtained. For each spectral age tested the normalization was determined by a golden ratio search based on the methods of \citet{press07}. Using this approach over a 2-dimensional grid search vastly improves the speed of the model fitting and allowed accuracy to be achieved to the floating point limit.

\begin{figure*}
\centering
\includegraphics[angle=-90,width=8.75cm]{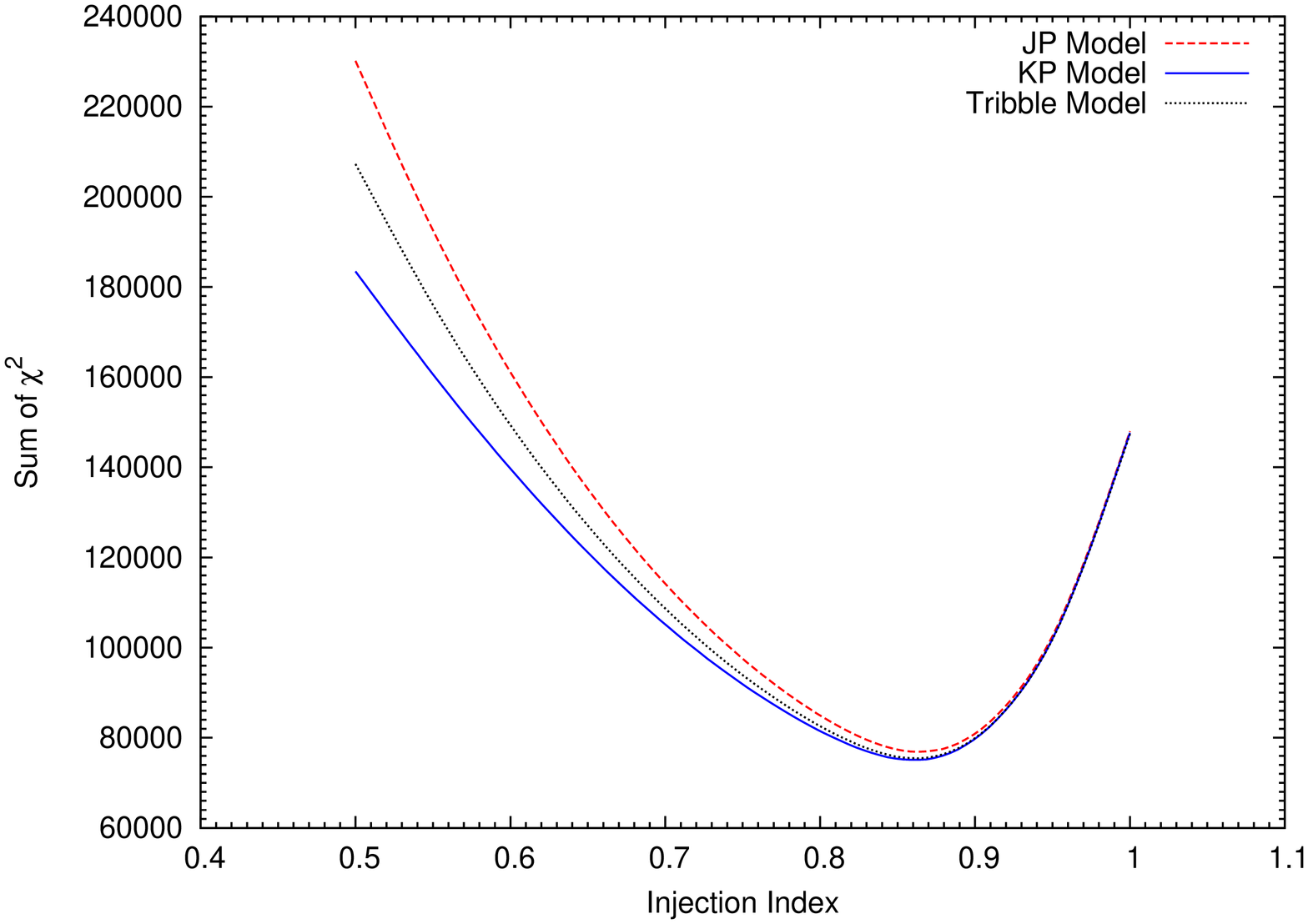}
\includegraphics[angle=-90,width=8.75cm]{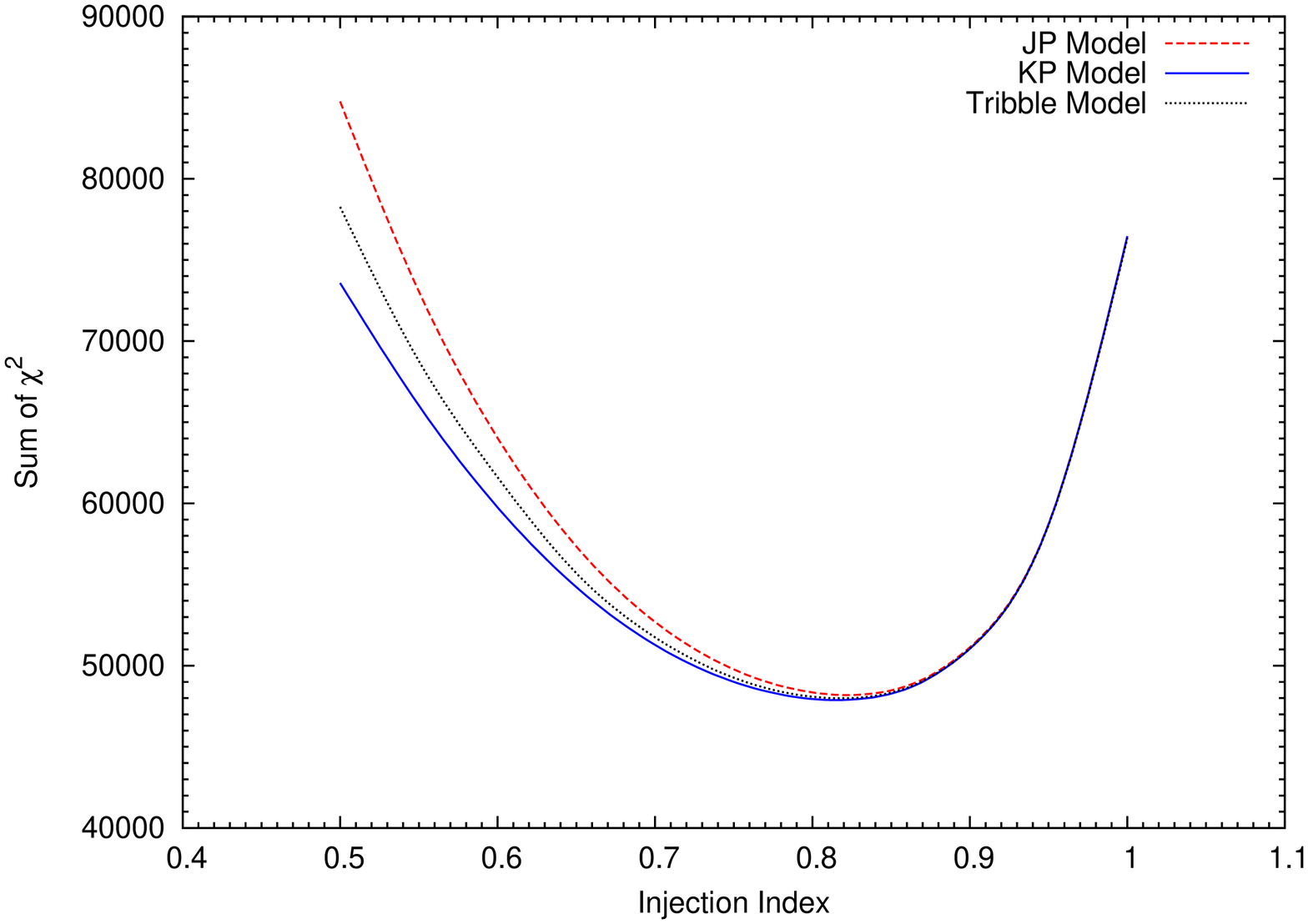}\\
\caption{$\chi^{2}$ values for 3C436 (left) and 3C300 (right) for varying injection index values fitted with a natural cubic spline. The solid blue line represents the KP model, the red dashed line the JP model and the black dotted line the Tribble model. Data points are taken at intervals of 0.1 between 0.5 and 1.0 inclusive and intervals of 0.01 between 0.8 and 0.92 for 3C436 and between 0.8 and 0.87 for 3C300. As all data points lie on the fitted spline, they are excluded for clarity. Note that minima occur at 0.86 for 3C436 and at 0.82 for 3C300 for all models.}
\label{injectmin}
\end{figure*}

For both the JP and KP models, we assumed a fixed magnetic field which is in equipartition across the lobes (although see Section \ref{spectralages} for discussion on the validity of this assumption). We made use of the {\sc synch} code of \citet*{hardcastle98} and the parameters of Table \ref{synchparams} to determine the magnetic field strength for each source. As both the JP and KP models assume an initial electron energy distribution described by a power law, previous attempts to fit such models to observations have assumed an injection index of between $\simnot\,0.5$ and $0.7$ \citep{jaffe73, carilli91, katz93, orru10}. However, whilst this has historically been a reasonable assumption, the improved radio data provided by the JVLA both in terms of both quality and quantity gave us an ideal opportunity to investigate this assumption. {\sc brats} therefore provides the `\emph{findinject}' command which fits models over a range of injection indices outputting the sum of the $\chi^{2}$ over all regions. This in turn can be used to determine the best fitting injection index for a given model and source. A lower limit for the injection index of 0.5 is given by synchrotron theory, therefore both the KP and JP models were initially fitted at intervals of 0.1 between this value and a reasonable upper limit of 1.0. Once the best injection index had been determined at this level of accuracy, the \emph{findinject} command was run for a second time in a range around the minimum value at intervals of 0.01.  We took the minimum and maximum electron Lorentz factor to be $\gamma = 10$ and $\gamma = 1 \times 10^{6}$ respectively. The resulting values were then plotted and the minimum injection index found.

\begin{table}
\caption{Summary of SYNCH parameters}
\label{synchparams}
\begin{tabular}{llcll}
\hline
\hline

Source&Parameter&Value&Units&Comments\\

\hline
3C436&Length, Radius&110, 10.5&Arcsec&\\
3C300&Length, Radius&115, 12.0&Arcsec&\\
3C436&$\delta$&2.72&&$\delta = 2\alpha_{inj} + 1$\\
3C300&$\delta$&2.64&&$\delta = 2\alpha_{inj} + 1$\\
All&$E_{min}$&$5 \times 10^{7}$&eV&$\gamma \approx 10$\\
All&$E_{max}$&$1 \times 10^{11}$&eV&$\gamma \approx 1\times10^{6}$\\
\hline
\end{tabular}

\vskip 5pt
\begin{minipage}{8.5cm}
`Source' lists the target name for which the listed parameters were applied. `Parameter' refers to the parameter name where; $E_{min}$ is the minimum energy of the electron distribution; $E_{max}$ is the maximum energy of the electron distribution; `Length, Radius' are the source dimensions; and $\delta$ is the electron energy power law index given by $\delta = (\alpha_{inj} / 2) - 1$ where $\alpha_{inj}$ is the injection index of the source determined in Section \ref{modelprms}.
\end{minipage}

\end{table}

Once the model parameters had been determined, the final model fitting of the sources was performed. The combination of normalization and spectral age which provided the lowest $\chi^{2}$ value for each region were recorded. Plots of model against observed flux density along with mapping of spectral ages and corresponding $\chi^{2}$ values as function of position were produced using the {\sc brats} software package.

Although the plotting of the $\chi^{2}$ value as a function of position was used as the measure of the goodness-of-fit for regions within a source, additional statistical tests were also available for determining the fit of a given model over the source as a whole. The mean of the $\chi^{2}$ for an entire source can provide an initial indication of the overall goodness-of-fit; however, regions where there are high $\chi^{2}$ values due to known, non-physical reasons (such as dynamic range effects) can cause the average to be biased toward the rejection of a model. Two additional statistical checks were therefore made to account for this bias. The $\chi^{2}$ cumulative distribution function was first calculated at 68, 90, 95 and 99 per cent confidence levels using the computational techniques of \citet{press07} and the GNU Scientific Library functions. The $\chi^{2}$ value of each region was then binned to a given confidence level with a cut-off for the rejection of the region set at $\geq 95$ per cent confidence. If more than half of regions fell above this rejection cut-off, the model over the source is classed as a poor fit and rejected. The median of the individual rejected or non-rejected interval bins was then taken to give a level at which the (non-) rejection is made. For example, if the median of a non-rejection falls within the 68 to 90 per cent interval bin, we can say that the model for the source as a whole cannot be rejected at greater than a 90 per cent confidence level. This method of analysis allows the goodness-of-fit to be weighted towards the area over which a model is well or poorly fitted rather than a bias towards high $\chi^{2}$, but potentially small, regions of poor fit.

\section{Results}
\label{results}
\subsection{Model Parameters}
\label{modelprms}

\begin{table*}
\caption{Model Fitting Results}
\label{restab}
\begin{tabular}{lccccccccccc}
\hline
\hline

Source&$\alpha_{inj}$&Model&Mean $\chi^{2}$&Mean $\chi^{2}_{red}$&&&Confidence Bins&&&Rejected&Median Confidence\\
&&&&&$<$ 68&68 - 90&90 - 95&95 - 99&$\geq$ 99&&\\

\hline
3C436&0.86&KP&15.64&2.23&1300&1057&437&663&1337&No&$<$ 68\\
&&JP&16.03&2.29&1238&1042&427&674&1413&No&$<$ 68\\
&&Tribble&15.74&2.25&1286&1055&429&670&1354&No&$<$ 68\\
&0.60&KP&29.21&4.17&472&372&227&567&3156&Yes&$>$ 99\\
&&JP&33.61&4.80&360&313&179&491&3451&Yes&$>$ 99\\
&&Tribble&31.16&4.45&395&347&194&542&3316&Yes&$>$ 99\\
3C300&0.82&KP&11.65&2.33&1705&724&276&476&931&No&$<$ 68\\
&&JP&11.72&2.34&1705&709&280&462&956&No&$<$ 68\\
&&Tribble&11.67&2.33&1706&719&273&476&938&No&$<$ 68\\
&0.60&KP&14.55&2.91&1593&728&254&388&1149&No&$<$ 68\\
&&JP&15.58&3.12&1544&748&246&374&1200&No&$<$ 68\\
&&Tribble&14.91&2.98&1574&743&239&375&1181&No&$<$ 68\\
\hline
\end{tabular}

\vskip 5pt
\begin{minipage}{17.5cm}
`Source' lists the target name of the listed results. `$\alpha_{inj}$' refers to the injection index used for fitting of the `Model' column. Mean $\chi^{2}$ lists the average $\chi^{2}$ over the entire source with an equivalent reduced value shown in the `Mean $\chi^{2}_{red}$' column. `Confidence Bins' lists the number of regions for which their $\chi^{2}$ values falls with the stated confidence range. `Rejected' lists whether the goodness-of-fit to the source as a whole can be rejected as per Section \ref{modelfitting} and `Median Confidence' the confidence level at which the model can or cannot be rejected.
\end{minipage}

\end{table*}

The plot of $\chi^{2}$ values for varying injection indices are shown in Figure \ref{injectmin} where we see that clear minima occur for all models and sources. In the case of 3C436 the JP, KP and Tribble models give an injection index of 0.86 with 3C300 having minima at 0.82 in all cases. It is immediately clear that these represent a much steeper injection index than has previously been assumed; however, given that there is no \emph{a priori} knowledge of the injection indices of FR-IIs at this level of spectral resolution, particularly at the low frequencies at which the power law distribution is assumed to occur, these best-fit values were used for model fitting and the determination of $\delta$ for the magnetic field strength as listed in Table \ref{synchparams} (It is interesting to note that in the recent LOFAR study of the nearby FR-I source M87 by \citet{degasperin12}, a similar high injection index was observed). Fits were also performed using the commonly used injection index of 0.60 to allow comparison. This deviation from the standard assumption of injection index and its implications are discussed further in Section \ref{injectionindex}.

Where $\delta = 2\alpha_{inj} + 1$ relates the injection index $\alpha$ to the initial electron energy power law distribution $\delta$, for the best-fitting injection indices we derive values of $\delta_{3C436} = 2.72$ and $\delta_{3C300} = 2.64$ respectively. Assuming equipartition and the parameters of Table \ref{synchparams}, we find magnetic field strengths of $1.04$ nT for 3C436 and $1.10$ nT for 3C300. For the commonly assumed $\alpha_{inj} = 0.6$ where $\delta = 2.2$, we derive magnetic fields strengths of $0.648$ nT for 3C436 and $0.755$ nT for 3C300. These values agree well with commonly assumed values of around $10^{-9}$ T for lobes of FR-II radio galaxies (e.g. \citealp{jaffe73, alexander87, alexander87a, carilli91, croston05}).

\begin{figure*}
\centering
\includegraphics[angle=0,width=8.80cm]{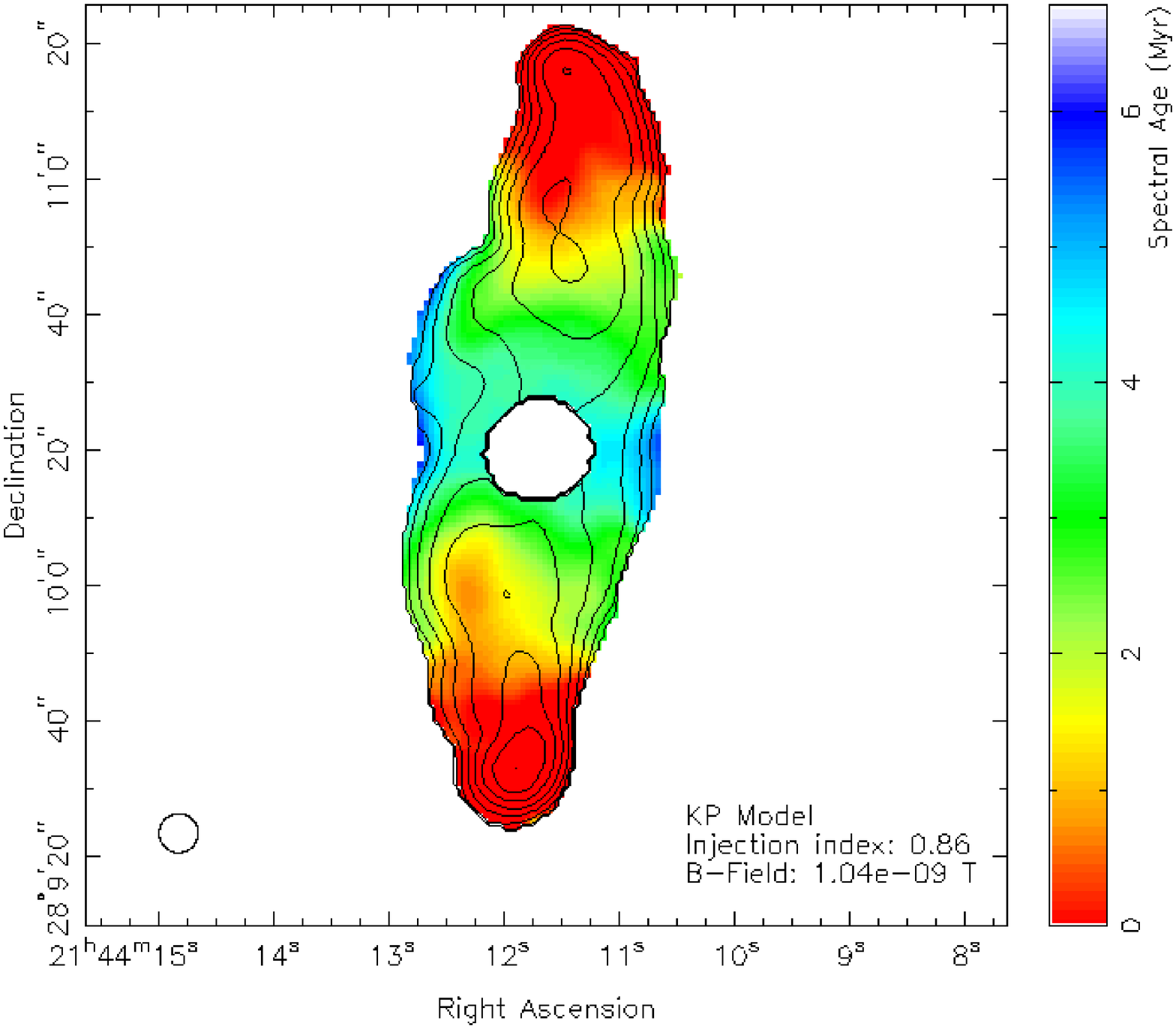}
\includegraphics[angle=0,width=8.80cm]{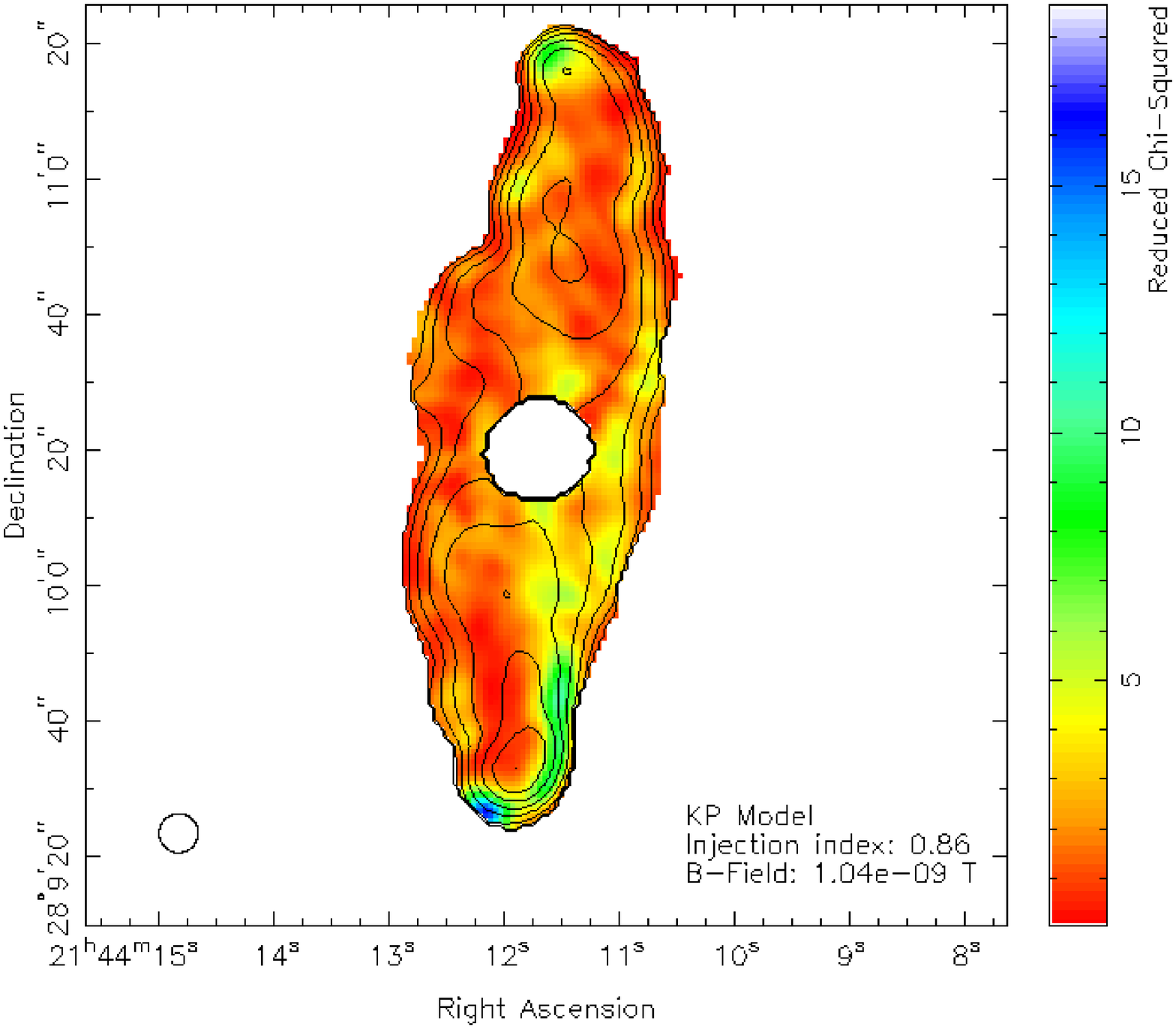}\\
\includegraphics[angle=0,width=8.80cm]{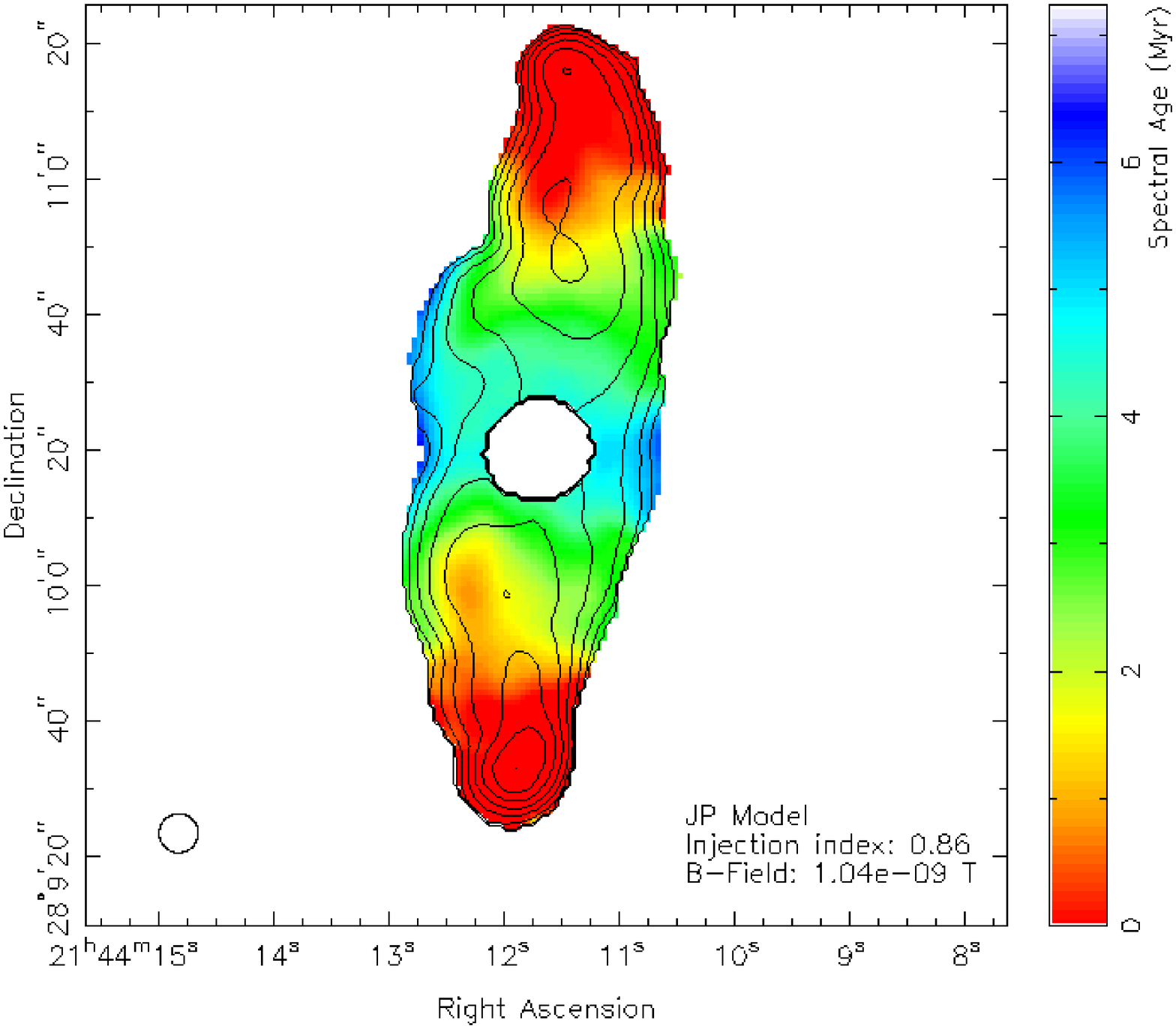}
\includegraphics[angle=0,width=8.80cm]{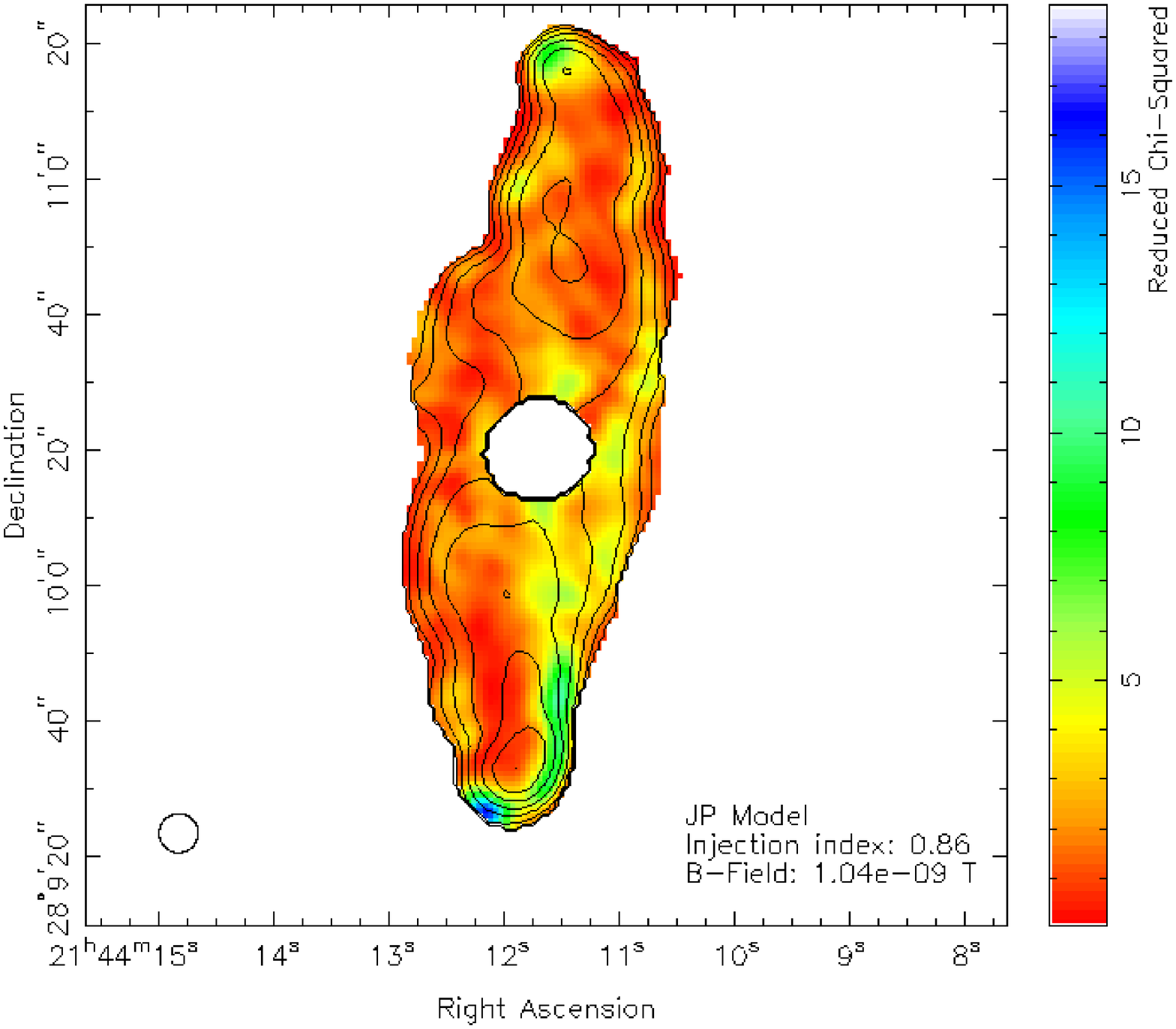}\\
\includegraphics[angle=0,width=8.80cm]{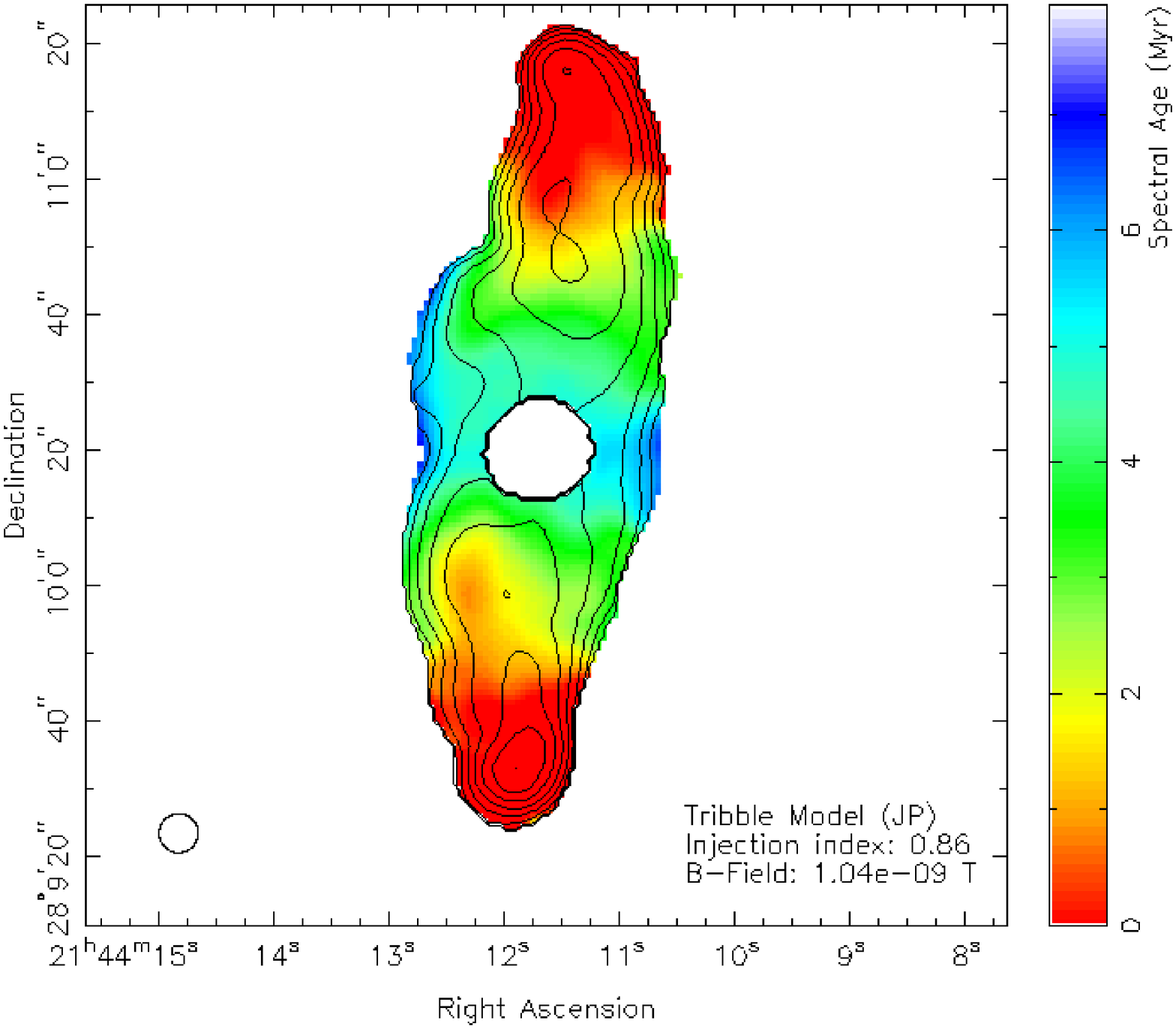}
\includegraphics[angle=0,width=8.80cm]{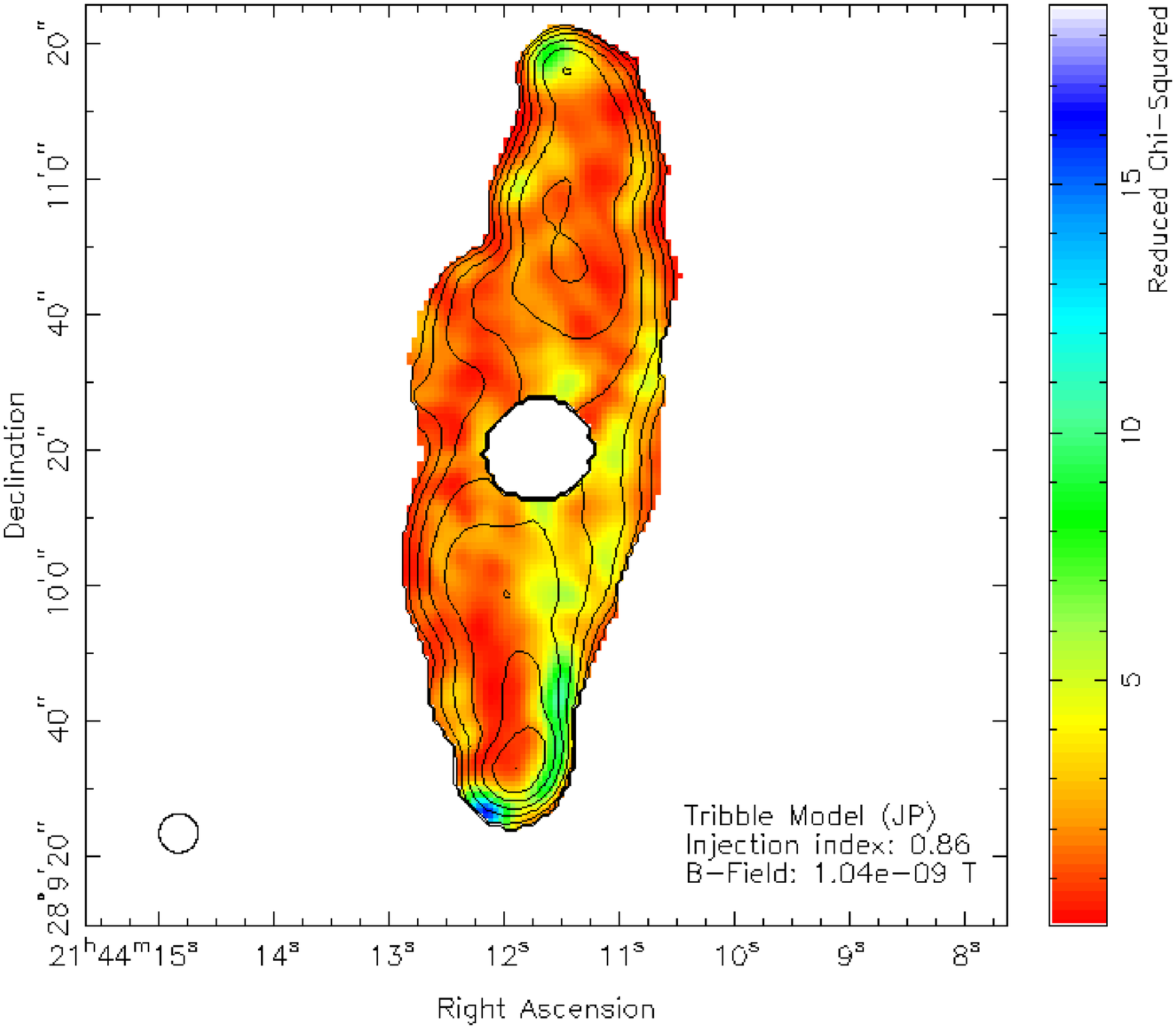}\\
\caption{Spectral ageing maps (left) and corresponding $\chi^{2}$ maps (right) of 3C436 with 7.2 GHz flux contours. Three model fits are shown; KP model (top), JP model (middle) and Tribble model (bottom) with injection index 0.86.}
\label{3C436specagemap}
\end{figure*}

\begin{figure*}
\centering
\includegraphics[angle=0,width=8.80cm]{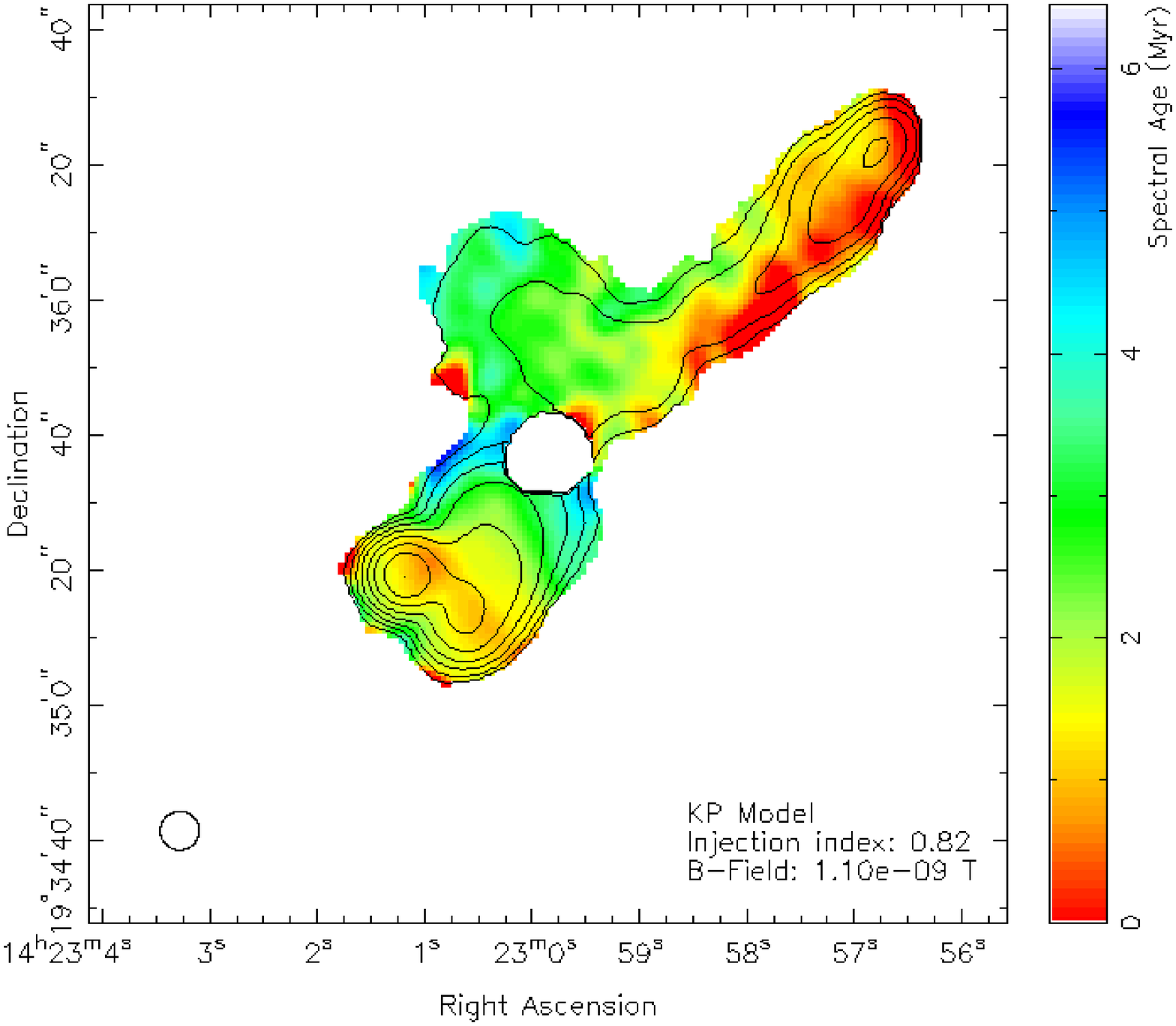}
\includegraphics[angle=0,width=8.80cm]{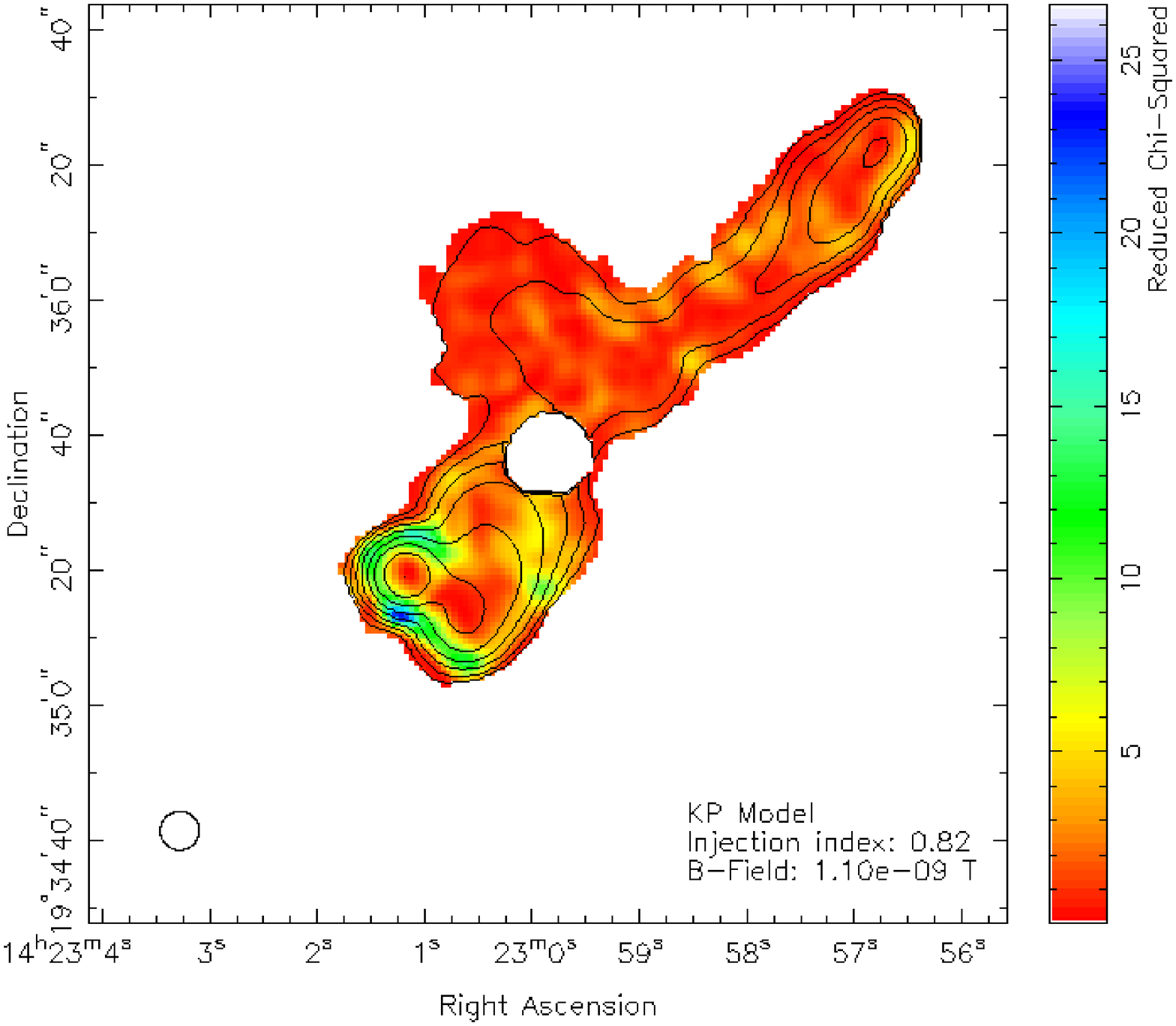}\\
\includegraphics[angle=0,width=8.80cm]{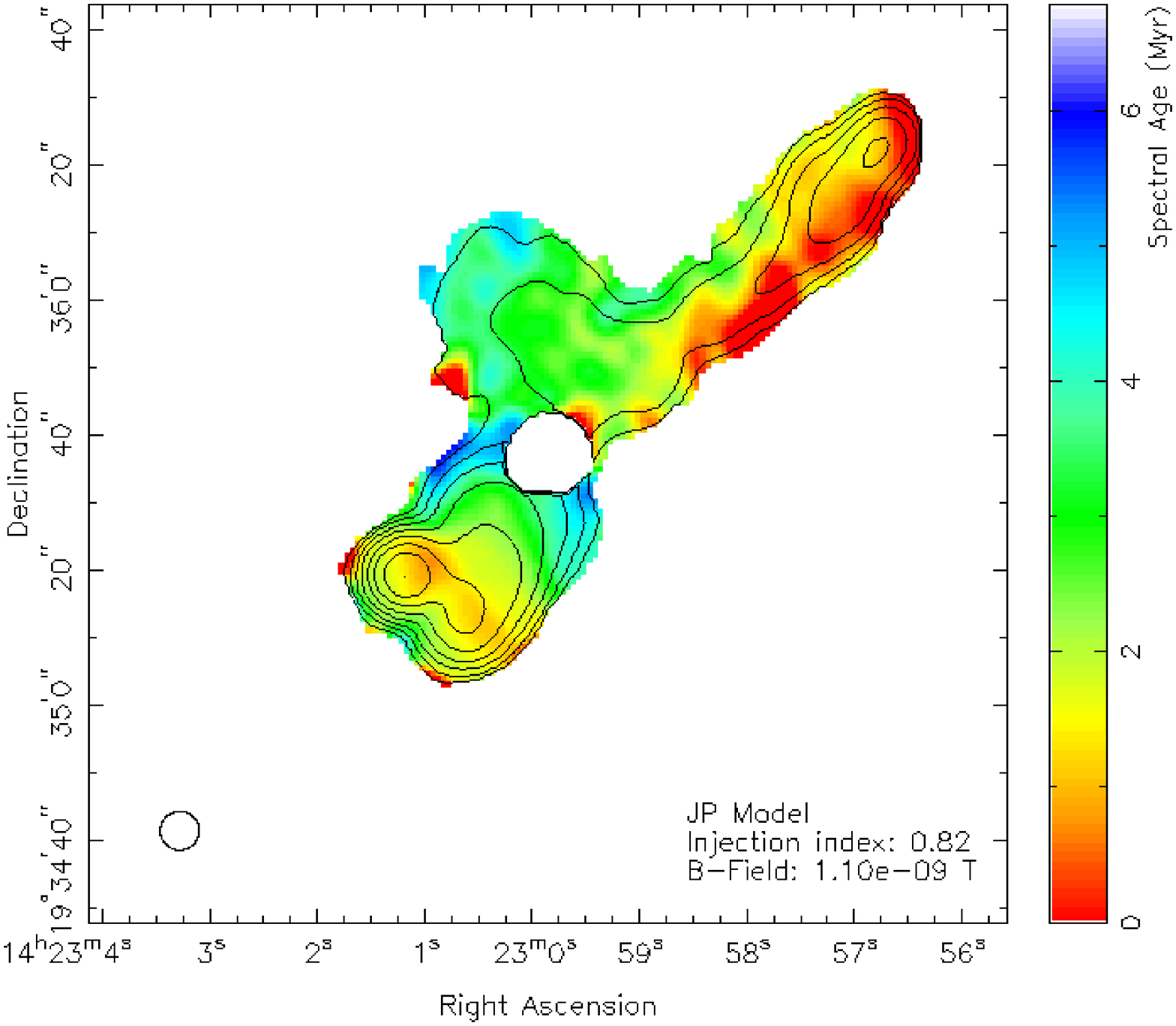}
\includegraphics[angle=0,width=8.80cm]{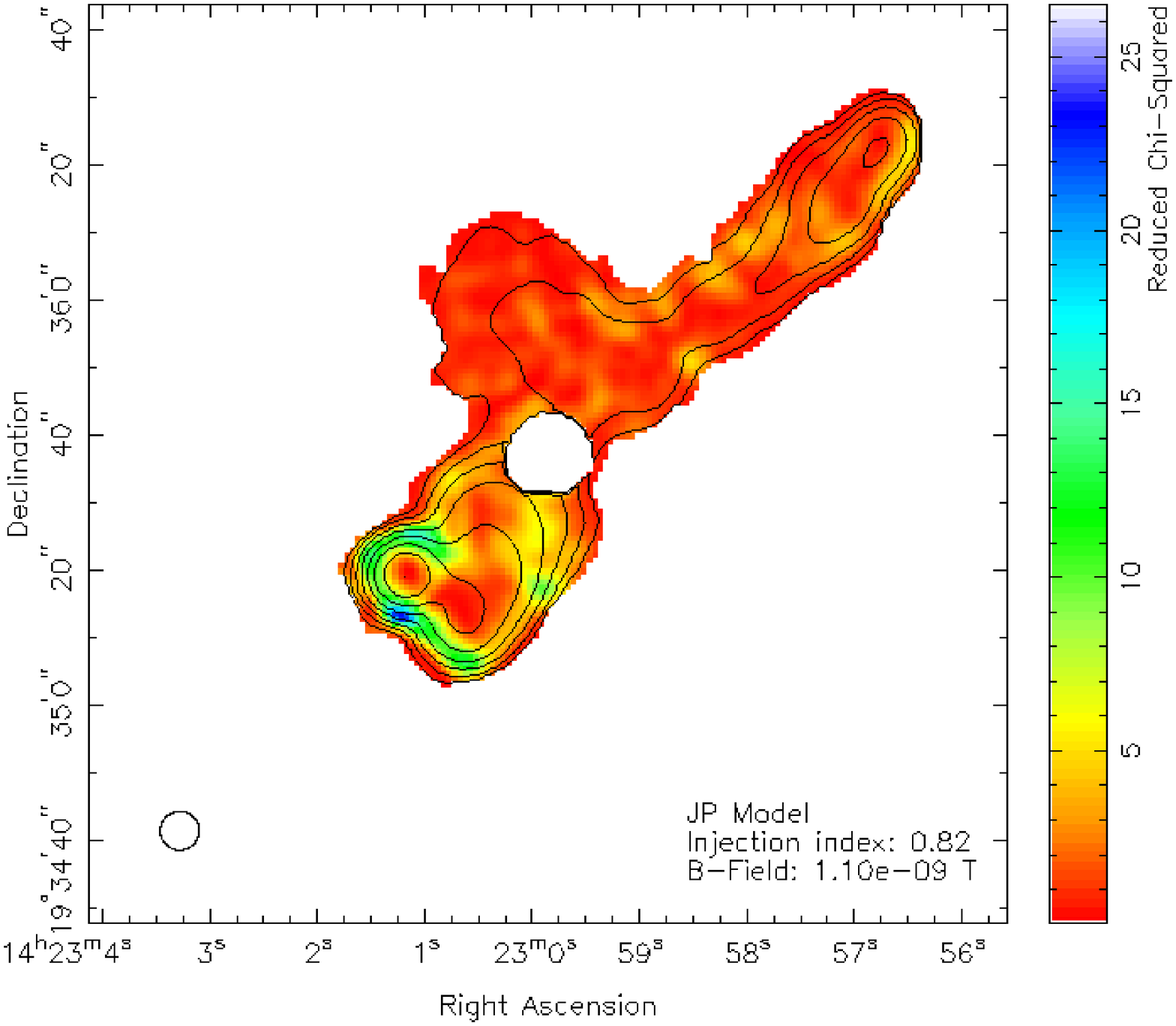}\\
\includegraphics[angle=0,width=8.80cm]{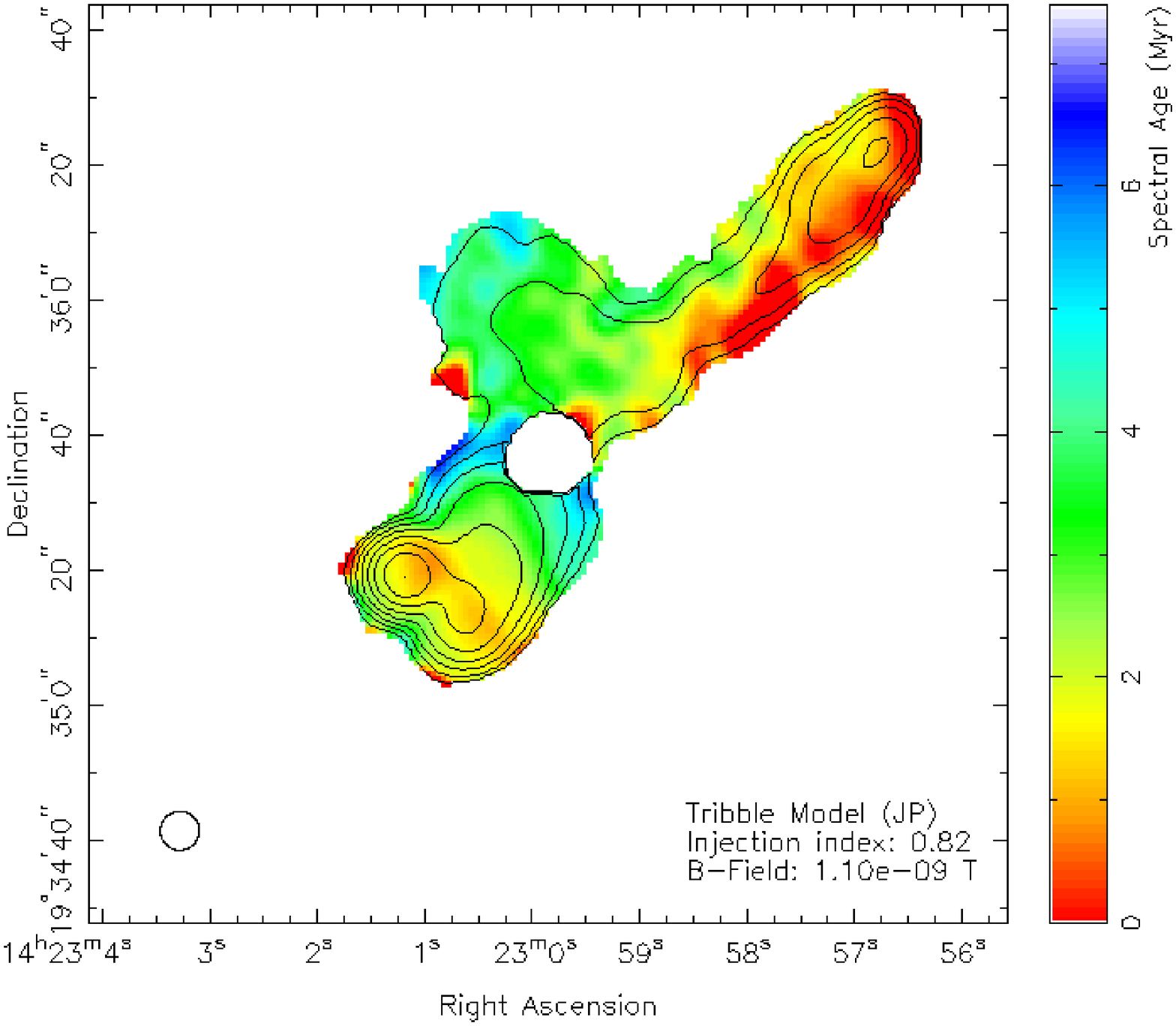}
\includegraphics[angle=0,width=8.80cm]{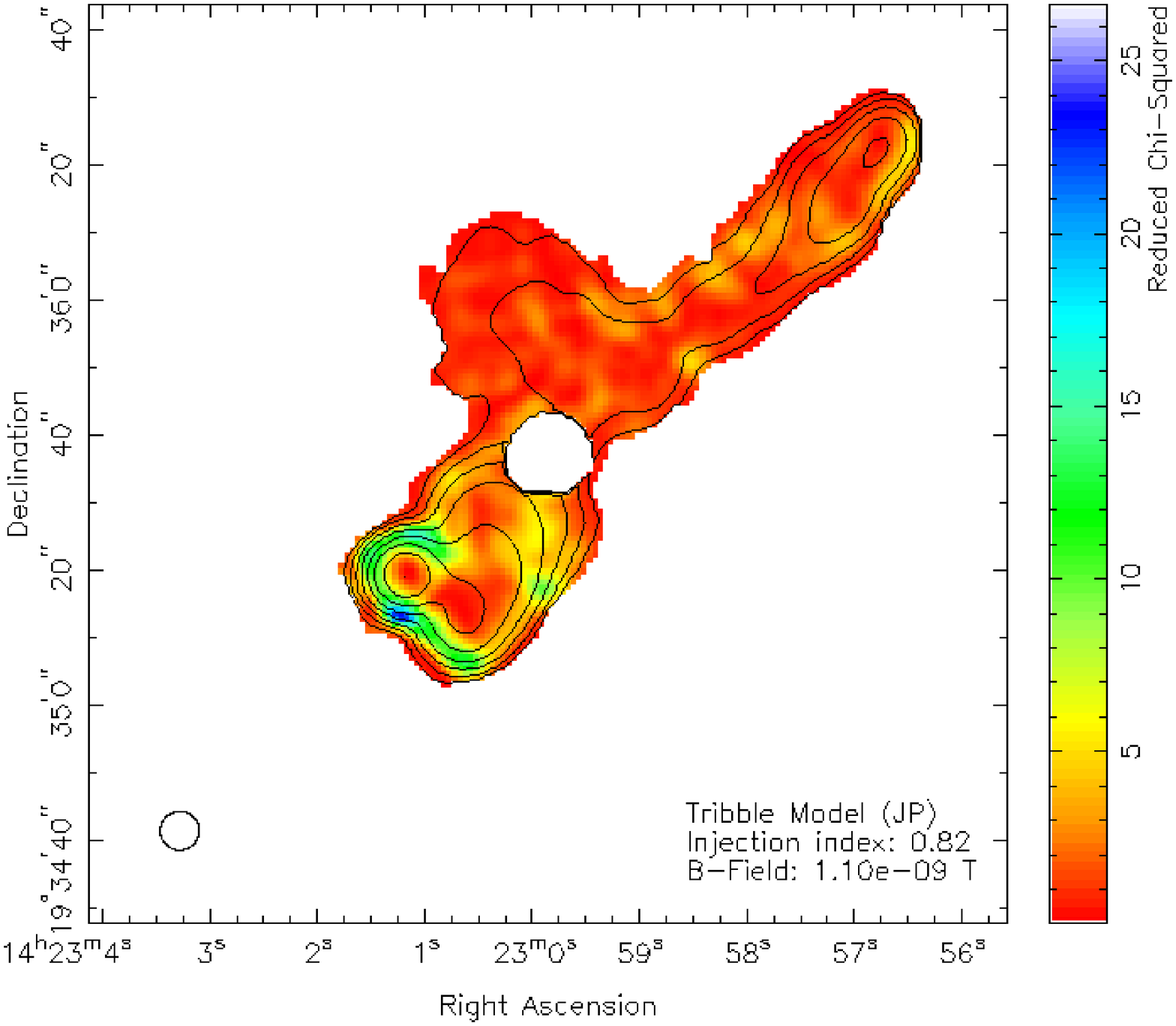}\\
\caption{Spectral ageing maps (left) and corresponding $\chi^{2}$ maps (right) of 3C300 with 7.2 GHz flux contours. Three model fits are shown; KP model (top), JP model (middle) and Tribble model (bottom) with injection index 0.82.}
\label{3C300specagemap}
\end{figure*}

\begin{table*}
\caption{Lobe Advance Speeds}
\label{lobespeed}
\begin{tabular}{lcccccccc}
\hline
\hline
Source&$\alpha_{inj}$&Model&Lobe&Max Age (MYrs)&$\pm$&Distance (Kpc)&Speed $\left(10^{-2}(v/c)\right)$&$\pm$\\
\hline
3C436&0.86&KP&North&5.90&0.45&208&16.8&1.3\\
&&JP&North&6.30&0.48&208&15.7&1.2\\
&&Tribble&North&6.91&0.52&208&14.4&1.1\\
&&KP&South&5.90&0.45&180&13.6&1.1\\
&&JP&South&6.30&0.48&180&14.6&1.0\\
&&Tribble&South&6.91&0.52&180&12.4&0.9\\
&0.60&KP&North&17.4&1.3&208&5.70&0.43\\
&&JP&North&16.4&1.2&208&6.05&0.46\\
&&Tribble&North&18.2&1.4&208&5.45&0.41\\
&&KP&South&17.4&1.3&180&4.94&0.37\\
&&JP&South&16.4&1.2&180&5.24&0.40\\
&&Tribble&South&18.2&1.4&180&4.72&0.36\\

3C300&0.82&KP&North&4.50&0.48&293&31.1&2.8\\
&&JP&North&5.20&0.41&293&26.9&2.5\\
&&Tribble&North&5.60&0.51&293&25.0&2.3\\
&&KP&South&5.60&0.51&140&12.0&1.0\\
&&JP&South&5.90&0.54&140&11.4&1.1\\
&&Tribble&South&6.50&0.59&140&10.3&0.9\\
&0.60&KP&North&10.5&1.0&293&13.3&1.2\\
&&JP&North&11.0&1.0&293&12.7&1.2\\
&&Tribble&North&12.8&1.2&293&11.4&1.0\\
&&KP&South&13.8&1.3&140&4.86&0.44\\
&&JP&South&13.0&1.2&140&5.15&0.47\\
&&Tribble&South&14.4&1.3&140&4.65&0.43\\
\hline
\end{tabular}

\vskip 5pt
\begin{minipage}{17.5cm}
`Source' lists the target name of the listed results. `$\alpha_{inj}$' refers to the injection index used for fitting of the `Model' column. `Max Age' is the maximum ages of the corresponding `Lobe' in Myrs. `Distance' gives the  separation between the oldest age population in the lobe and the hot spot in Kpc.`Speed' lists the derived speed as a fraction of the speed of light as detailed in Section \ref{lobespeeds}.
\end{minipage}

\end{table*}

\subsection{Spectral Age and Model Comparison}
\label{specage}

Figures \ref{3C436specagemap} and \ref{3C300specagemap} show the spectral age of the two sources and their corresponding $\chi^{2}$ values as a function of position. We see that although spectral age does change with distance along the lobe, this variation is not restricted to one axis and many unexpected features are present. Given the complexity and dynamics of these sources this is perhaps not surprising, but this is the first time these variations have been observed in great enough detail to be fully analysed. The spectral morphology between models only differs by $\simnot\,15$ per cent with respect to age with the main spectral features remaining independent of model type. This is also the case for the more commonly used injection index of 0.60. We find that spectral age increases by a factor of around 3, consistent with expectations, but also observe an increase of around 50 per cent in the $\chi^{2}$ in regions of both diffuse and small scale emission (Table \ref{restab}).

\begin{figure*}
\centering
\includegraphics[angle=0,width=8.75cm]{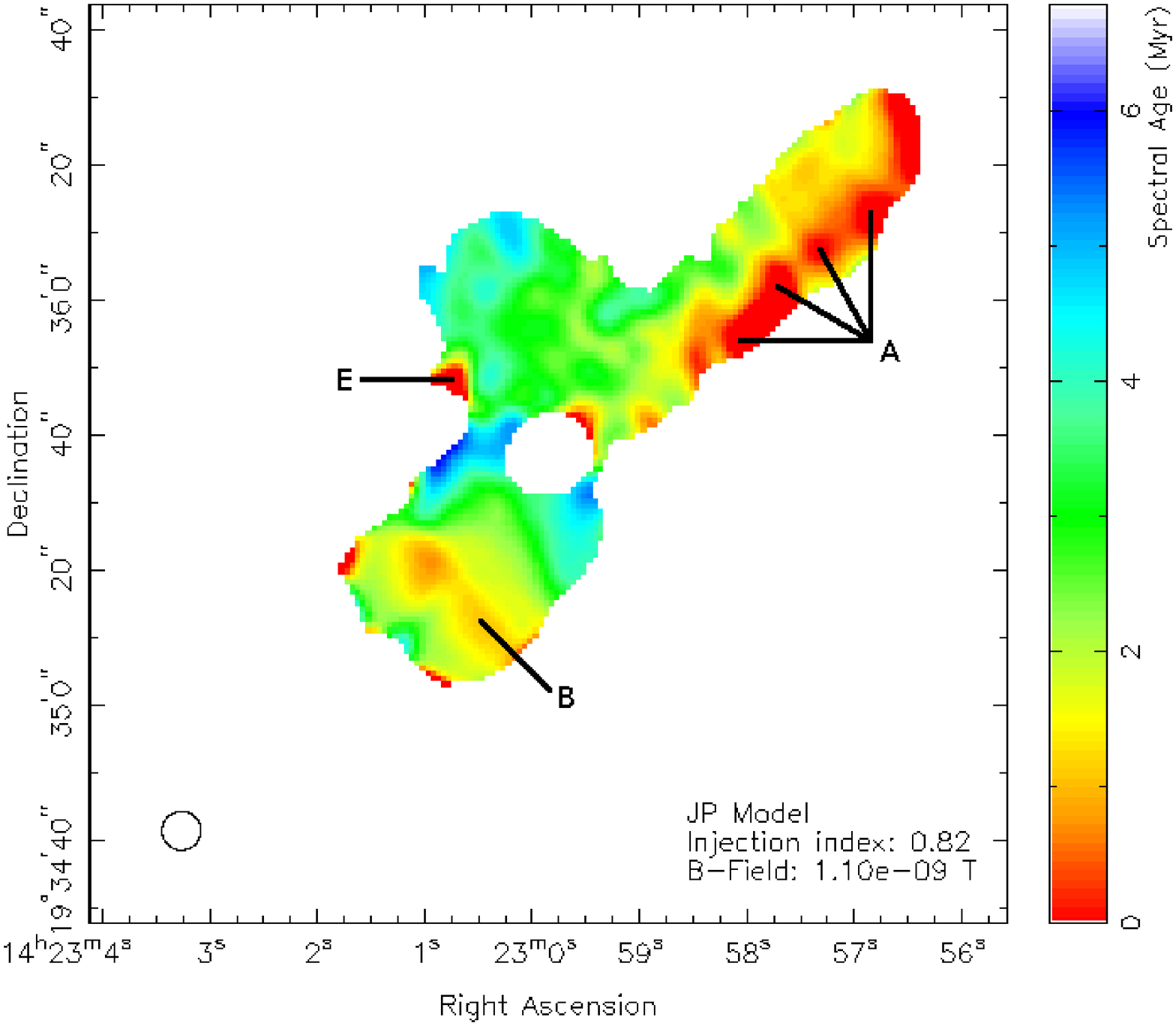}
\includegraphics[angle=0,width=8.75cm]{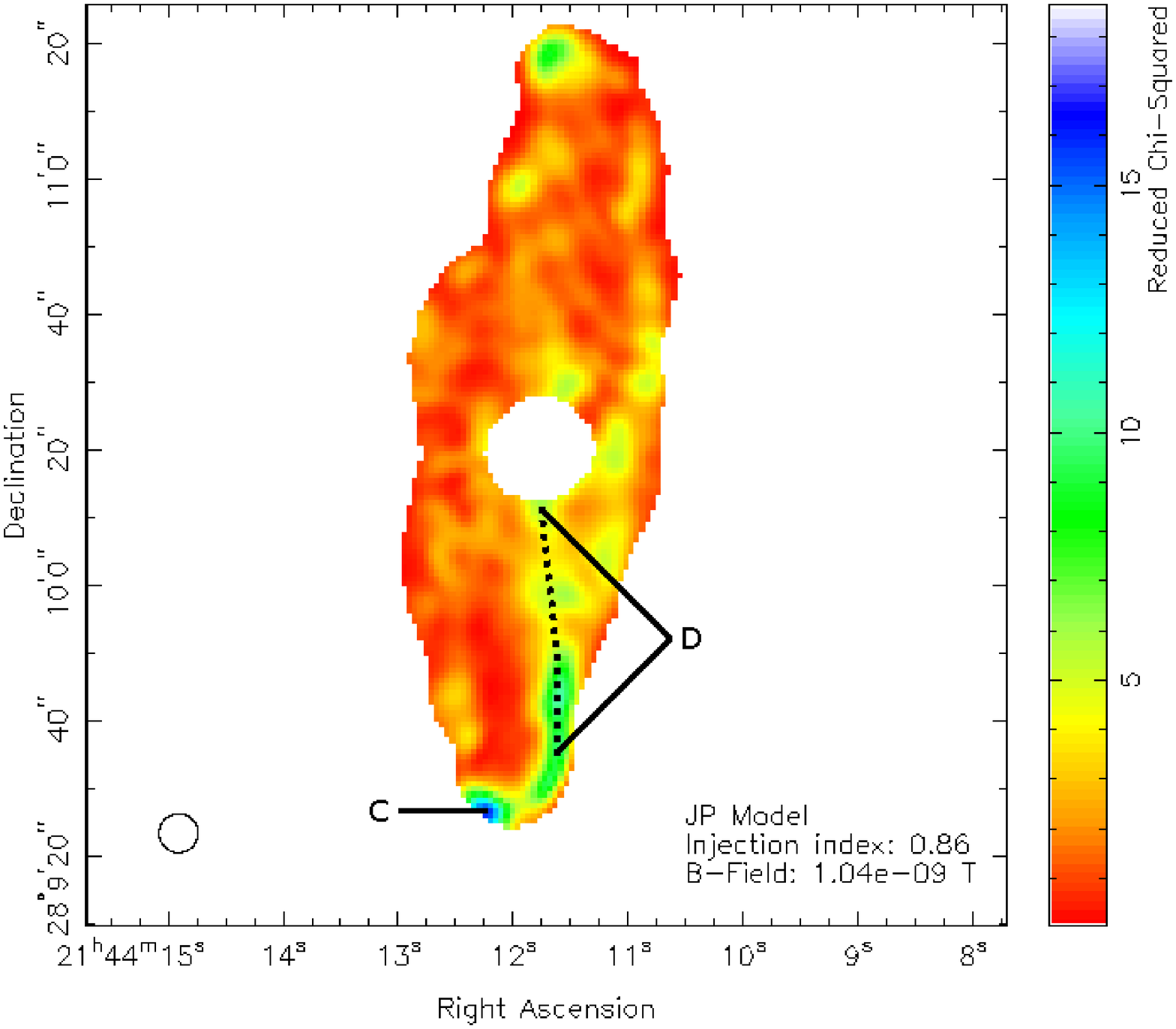}\\
\caption{Spectral age map of 3C300 (left) and reduced $\chi^{2}$ map of 3C436 (right) for the JP model fitting with a 0.86 injection index. Locations of significant interest are marked as discussed in Section \ref{specage}.}
\label{pointloc}
\end{figure*}

We see from Figure \ref{3C436specagemap} that 3C436 has a spectral morphology close to that of the classical case, with the lowest age regions located close to the hot spots and a general trend of increasing age as we move towards the galaxy core. However, we note that there is also a non-negligible variation in age across the width of both the northern and southern lobes. For 3C300 this asymmetry is even more pronounced. Not only do we observe age variations across the width of the southern lobe, we also see discrete low age regions spread along the length of the northern lobe (A of Figure \ref{pointloc}). Comparing their location to features observed on the high resolution maps of Figure \ref{highresmaps}, we find they coincide with the likely location of the northern jet. In the southern lobe of 3C300, we also see an extended region with a spectral ages of $\simnot\,1$ Myr (B of Figure \ref{pointloc}) well away from the hot spot. This region is likely to be the result of either the relic of a old hot spot, or a fast outflow as noted by \citet{hardcastle97}. What is clear is that these cross-lobe age variations and the scales on which they exist have implications for previous studies of spectral ageing where cross-lobe variations are assumed to be small. We discuss these points further in Sections \ref{imagefidelity} and \ref{spectralages}.

\begin{figure*}
\centering
\includegraphics[angle=0,width=8.75cm]{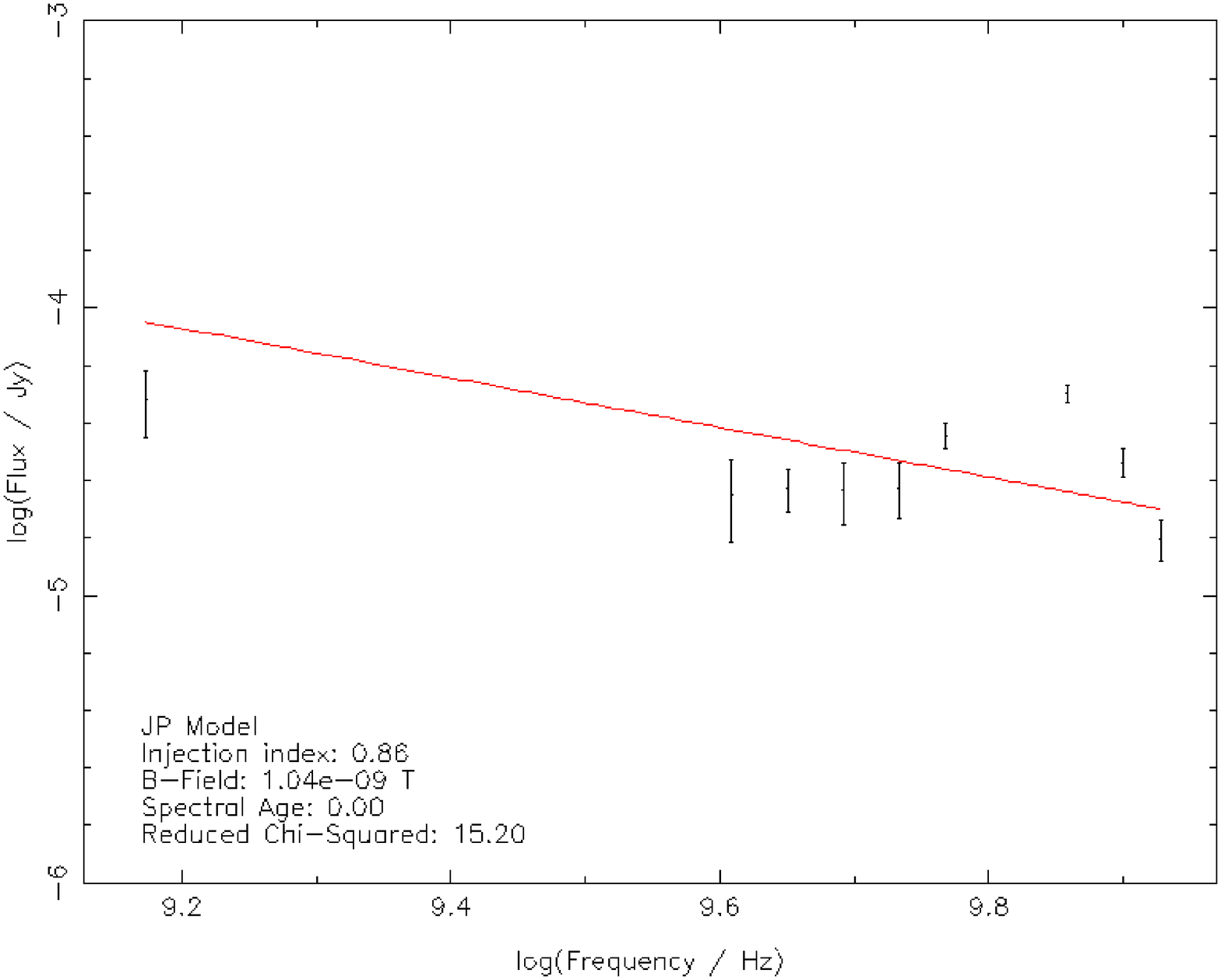}
\includegraphics[angle=0,width=8.75cm]{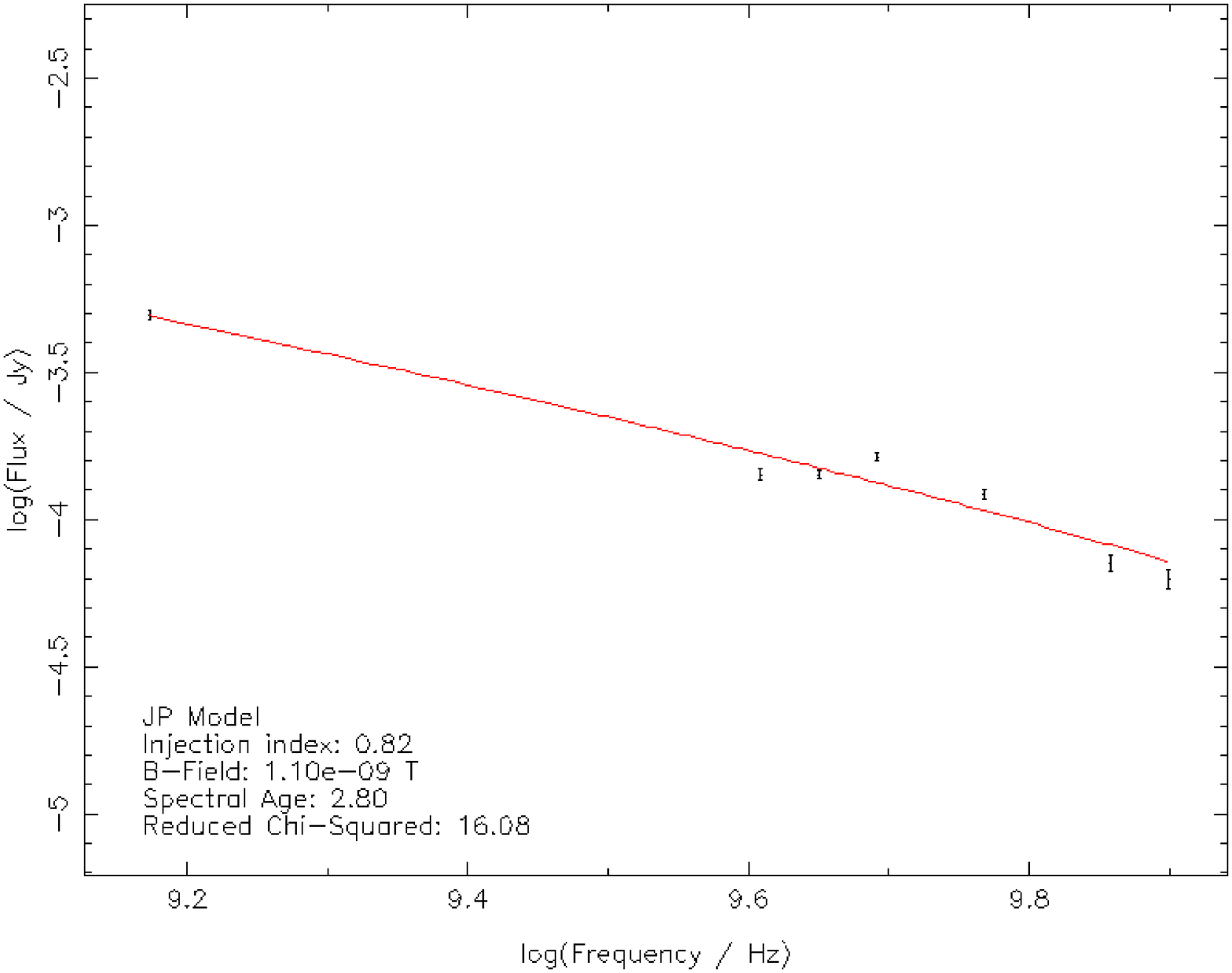}\\
\includegraphics[angle=0,width=8.75cm]{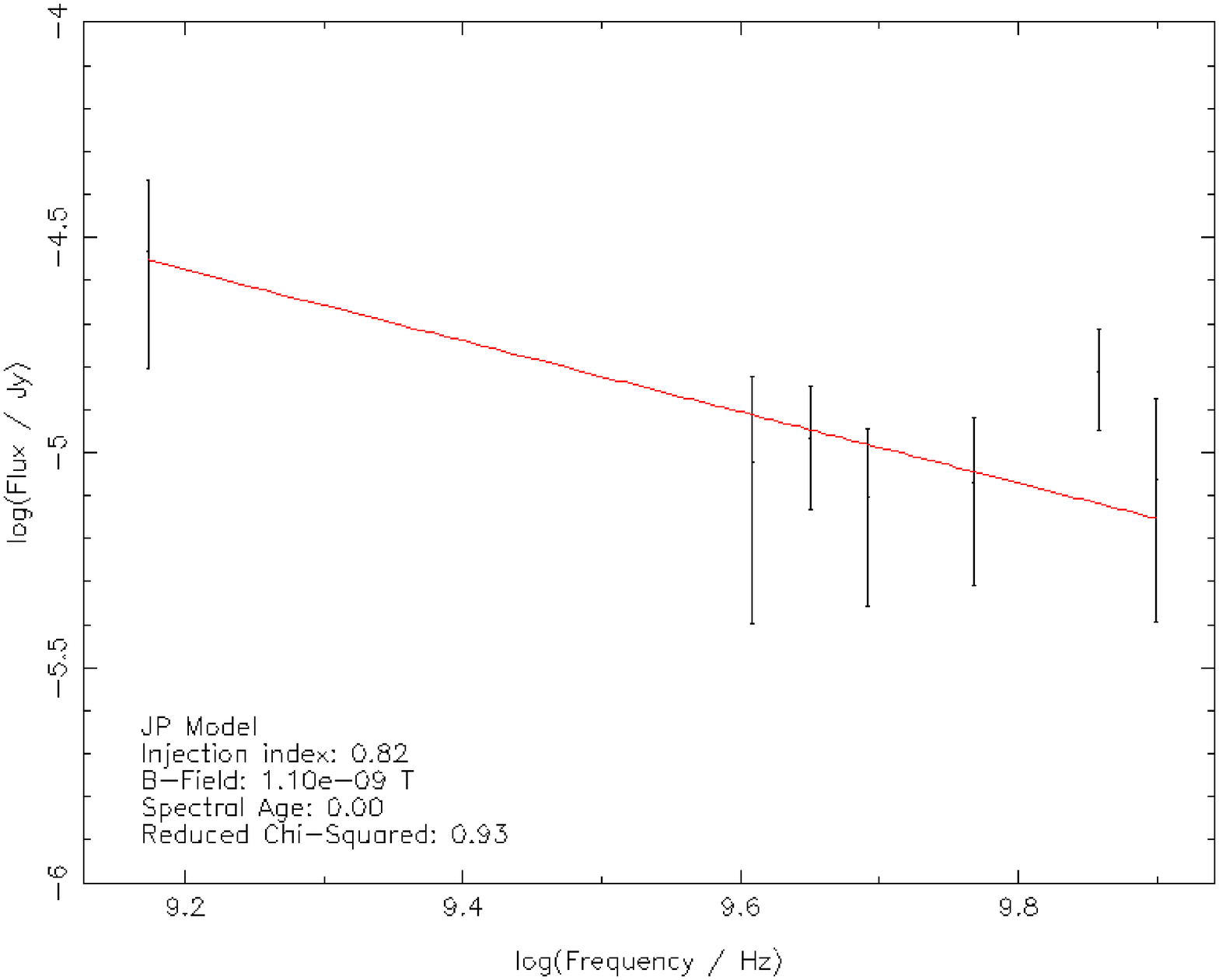}\\
\caption{Plots of flux against frequency for the high $\chi^{2}$ regions of 3C436 (top left) and 3C300 (top right) and the unexpected low ages region of 3C300 (bottom middle) as discussed in Section \ref{specage}. Red lines indicate the best-fitting JP model with parameters as noted on each plot.}
\label{highchi}
\end{figure*}

The goodness-of-fit of the models to these sources also provide some interesting results. The $\chi^{2}$ values for 3C436 of Figure \ref{3C436specagemap} and Table \ref{restab} show that for $\alpha_{inj} = 0.86$, we cannot reject any of our three models for the majority of the central lobe. However, there are also a significant number of regions for which the models do not provide a good fit. The worst fitting regions where $\chi^{2}_{red} \approx 16$ lie at the very edge of the southern lobe (C of Figure \ref{pointloc}). We note from the high-resolution map of Figure \ref{highresmaps} that within these regions an extended area of high intensity and low spectral age is present well away from the location of the classical hot spot. As these regions of high $\chi^{2}$ are so close to the edge of the source, to the hot spot and to this extended region of high intensity, we suggest that the poor model fits are likely due to edge effects and dynamic range issues resulting in large flux density variations. We see from the plot of one of these regions in Figure \ref{highchi} (top left) that the variation is noise like, with such large variations unlikely to be physical in nature. Examples of well-fitted regions for both sources are shown in Figure \ref{goodfit} for comparison. Interestingly, Figure \ref{3C436specagemap} also shows that an elongated region (D of Figure \ref{pointloc}) of the southern lobe, which we would expect to be well fitted, instead has a high $\chi^{2}$ value. Comparing these regions to the high resolution maps of Figure \ref{highresmaps} we see a clear correlation between the jet and the poor model fits. This is a somewhat unexpected result given the resolution at which these observations were made as the jet is barely visible in the individual total intensity maps. The spectral distinction between jet and lobe may provide a means of inferring the location of bright jets in future broadband investigations where high resolution maps are not available.

The $\chi^{2}$ values for 3C300 of Figure \ref{3C300specagemap} and Table \ref{restab} again provided good model fits for the central regions of the lobes for the best-fitting $\alpha_{inj} = 0.82$, but we also see many regions where the model fits can be rejected of at least the $95$ per cent significance level which must be accounted for. The area around the southern hot spot and along the southern edge of the source where $\chi^{2}_{red} \approx 15$  are the regions of greatest intensity (Figure \ref{combinedmaps}) and it is therefore again likely that these regions are affected by dynamic range issues (e.g. Figure \ref{highchi}, top right). We also observe that the more diffuse emission of the southern lobe provides a worse fit than is expected in comparison to the diffuse emission of the northern lobe. Being well separated from the hot spot it is unlikely to be subject to any significant dynamic range effects but the overall source morphology and its location within the lobe does allow for possible orientation effects which are discussed further in Section \ref{morphology}. Model fits to the diffuse emission which are observed to the side of the source are in good agreement with the models; however, from Figure \ref{3C436specagemap} we find an unexpected low age region is present (E of Figure \ref{pointloc}). Comparing these regions to the high resolution maps of Figure \ref{highresmaps}, we see no corresponding small scale structure which can account for these low ages but we note that these are the lowest flux regions of the source. From Figure \ref{highchi} (bottom) which shows the plot of one of these regions, we see noise-like variations with large error bars which will tend to push the best-fit to that of a power law. We are therefore likely pushing the limit of our image fidelity within this area and so believe that these low age regions are likely not physical in nature. We discuss image fidelity further in Section \ref{imagefidelity}.

The model fitting using the commonly assumed value of $\alpha_{inj} = 0.60$ fare much worse than those of the best-fitting injection indices. We see from Table \ref{restab} that for 3C436 all models can be rejected through median binning at greater than the 99 per cent confidence level. For 3C300 the model fits cannot be rejected at the 68 per cent confidence level, but we note that this is heavily weighted by the larger northern lobe where the models are well fitted. In the southern lobe we find that all models can be rejected at the $99$ per cent confidence level in the vast majority of regions. We note that even in the well fitted northern lobe the $\chi^{2}$ values are still significantly worse than those where $\alpha_{inj} = 0.82$. It is also interesting to note that from Figure \ref{injectmin} as we move to lower injection index values, the difference in the sum of $\chi^{2}$ between models increases, making the determination of the best-fitting model much more distinct. This again has implications for both past and future studies of spectral ageing discussed further in Section \ref{injectionindex}.

One of the key factors in creating reliable results when investigating spectral curvature is accurate flux calibration. Although the data were calibrated in the standard manner, recent results from the long term study by \citet{perley13} of the flux density scale for calibration sources commonly used by the JVLA (including those used in this paper) show deviations from the standard values employed during our initial calibration. As these variations are frequency-dependent, they could potentially have a significant impact on both the spectral ages and injection indices found within this paper. The source fluxes were therefore scaled to these corrected values (Table 13 of \citealt{perley13}) within {\sc brats} and both model fitting and injection index minimisation re-run. We find that both the spectral ages and statistical values remain similar to those derived using the standard calibration values; however, the injection index rises to 0.89 for 3C436 and 0.86 for 3C300. The impact of this re-calibration on our results is therefore minimal, but it should be carefully considered in future studies.

Although the majority of model fits cannot be rejected at any reasonable confidence level, we see from Table \ref{restab} that there is a consistency as to which of those tested provides the best description of the source. We find that for both sources and injection indices, the JP model provides a significantly worse fit to the two sources compared to those of the KP and Tribble. The KP model always provides the best fit regardless of injection index, with the Tribble model giving very similar statistical values both in terms of mean $\chi^{2}$ and median binning by confidence level. Given that the JP model has historically been considered the more physically realistic description of spectral ageing this is perhaps surprising; we discuss this finding further in Section \ref{modelcomp}.

\subsection{Imaging Methods and Fidelity}
\label{imagefidelity}

In order to determine the validity of the results presented in Section \ref{specage} and the implications which they may hold for past and future works, we must first be confident what spectral structure can be believed. It is also important that we investigate which of the current methods of image deconvolution provides the most reliable results when undertaking this form of spectral study.

\begin{figure*}
\centering
\includegraphics[angle=0,width=8.80cm]{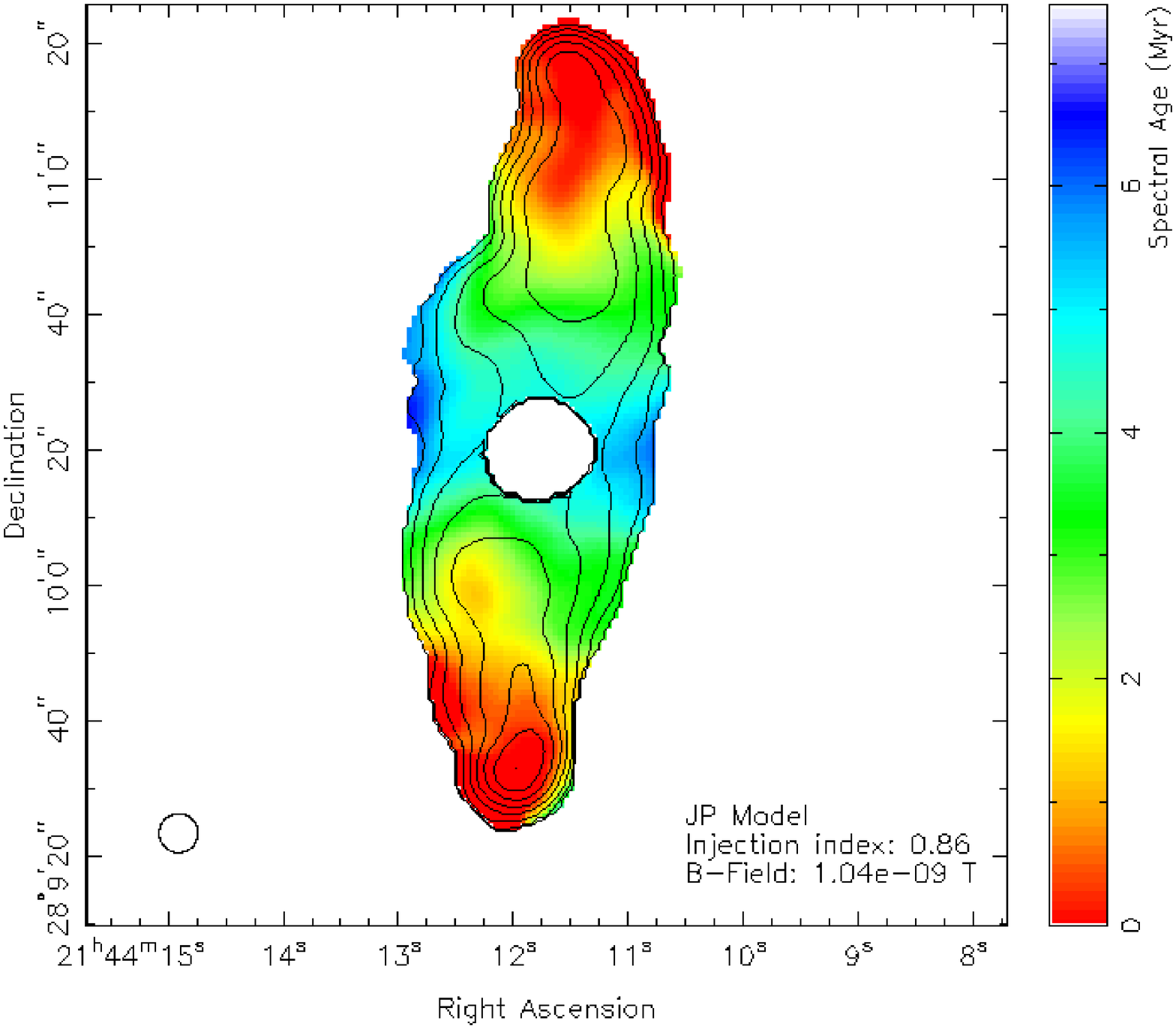}
\includegraphics[angle=0,width=8.80cm]{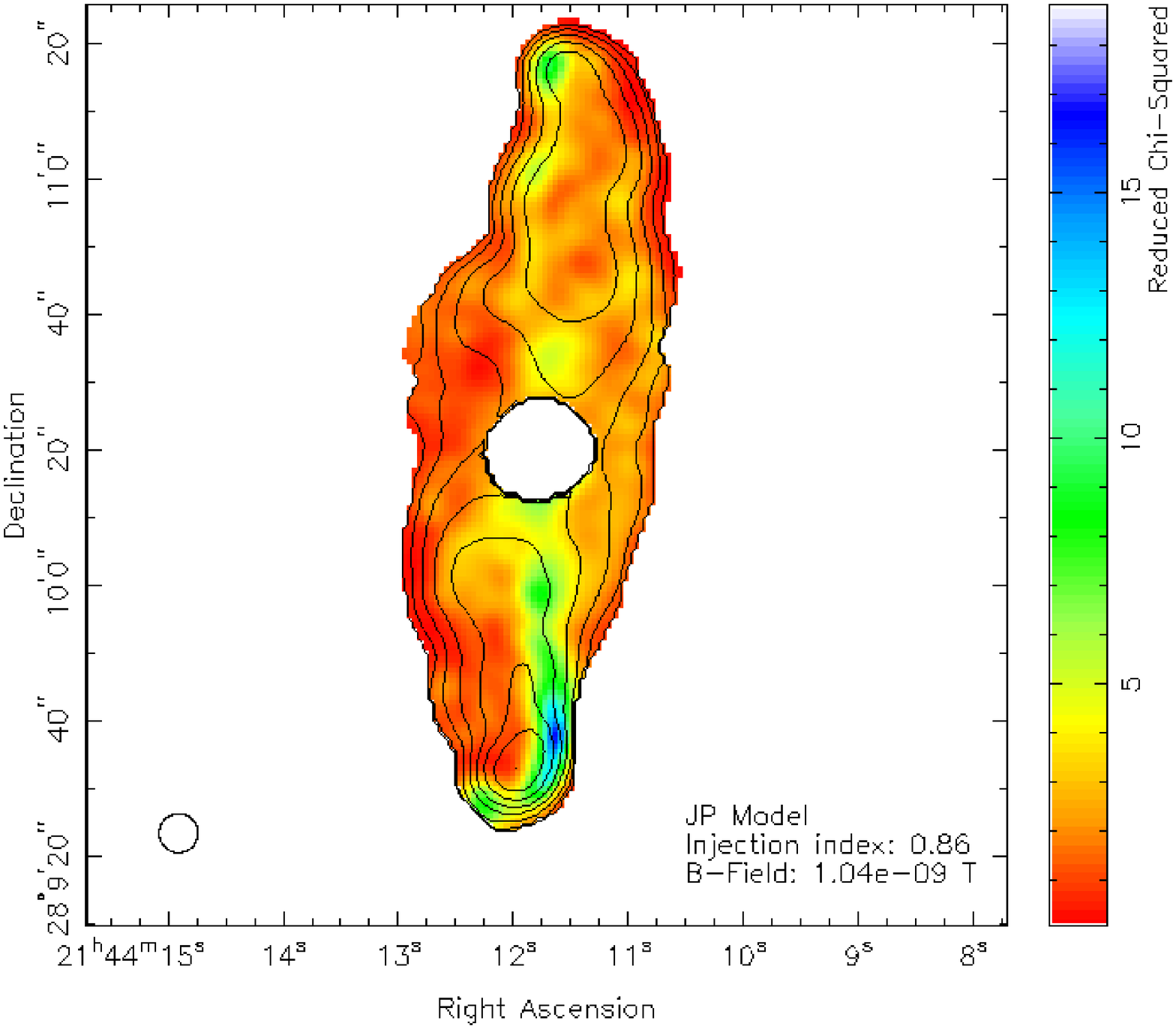}\\
\includegraphics[angle=0,width=8.80cm]{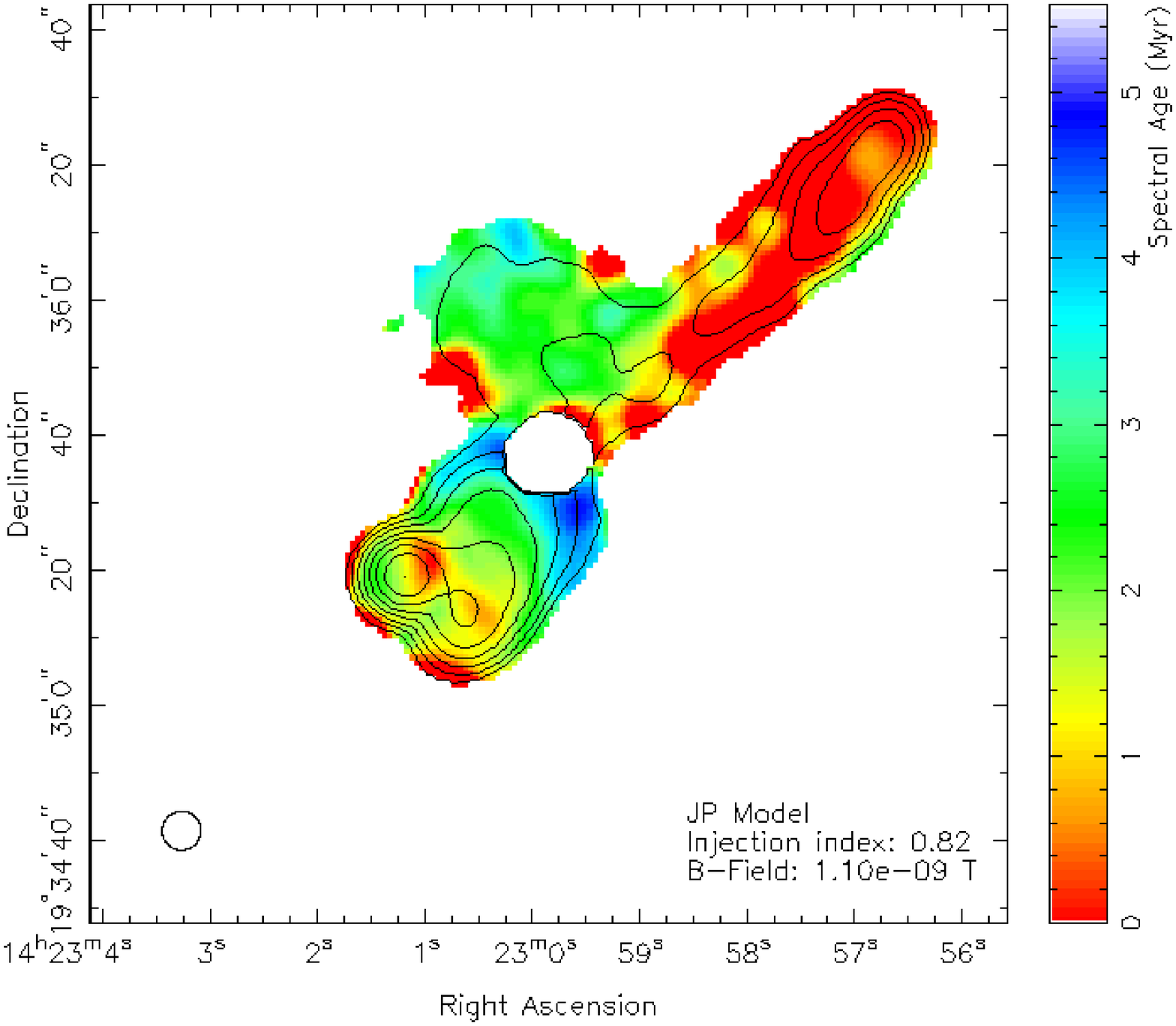}
\includegraphics[angle=0,width=8.80cm]{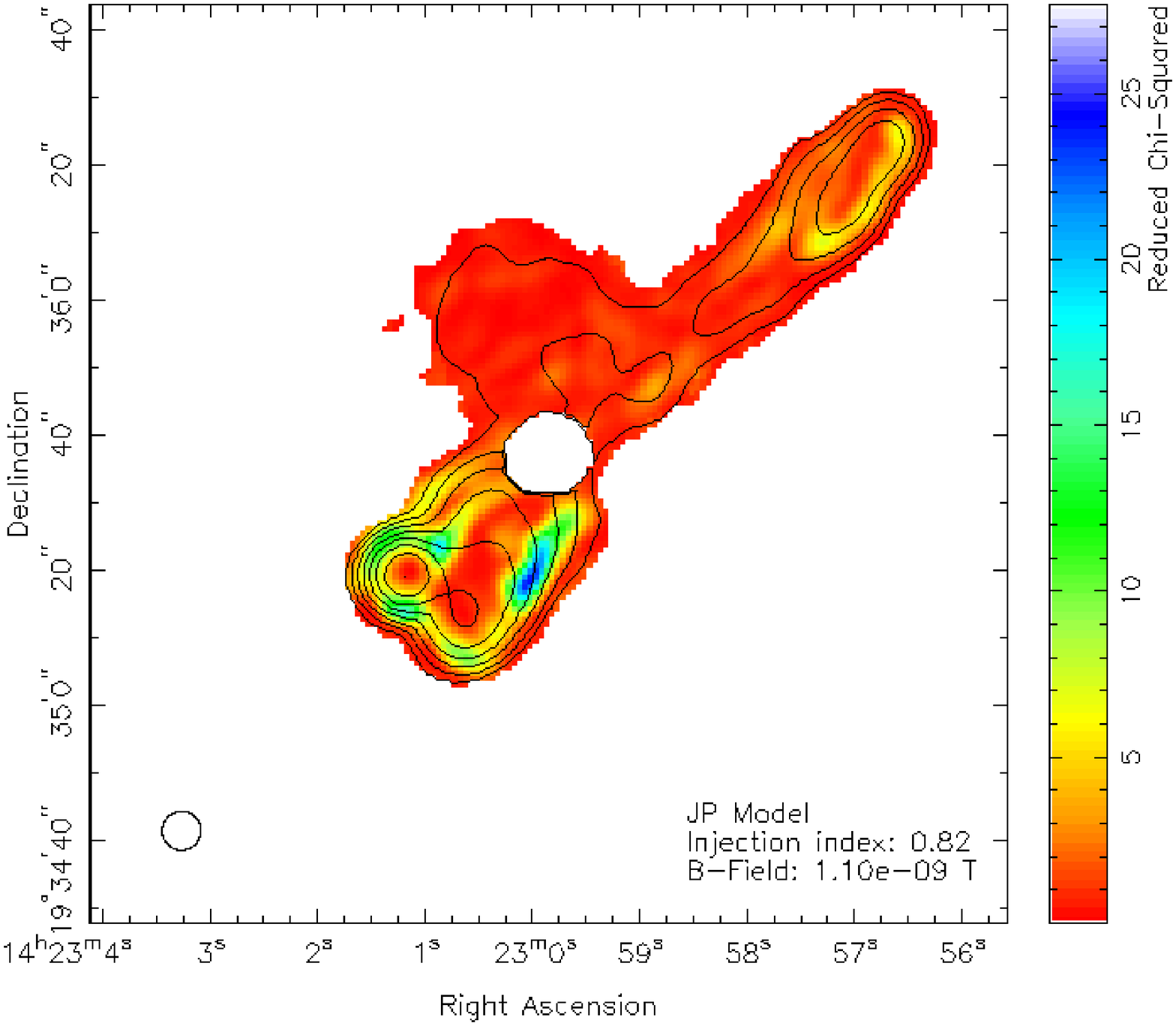}\\
\caption{Spectral ageing maps (left) and corresponding $\chi^{2}$ maps (right) of 3C436 (top) and 3C300 (bottom) with 7.2 GHz flux contours using the hybrid maximum entropy method. The JP model of spectral ageing is shown using the best fitting injection indices of 0.86 and 0.82 for 3C436 and 3C300 respectively.}
\label{memmaps}
\end{figure*}

Figure \ref{memmaps} shows the spectral age maps produced from the images deconvolved using the hybrid maximum entropy method. As the standard method provides only negligible differences in age and morphology, only the spectral age maps from the hybrid MEM deconvolution are shown for conciseness. From these images we see that the spectral morphology is similar to that produced from the standard CLEAN algorithm. As cross-lobe variations, along with the key features of Figure \ref{pointloc} are present in both the MEM and hybrid deconvolution for both sources, it is unlikely that these features are simply a result of deconvolution effects. This reinforces the assertion that our image fidelity is high and that the observed cross-lobe spectral variations are likely to be real. We also note that the MEM appears to recover more of the diffuse emission at low flux densities, vital for probing the oldest regions of plasma. However, the MEM is known to often have difficulties recovering reliable flux values at all scales (see \citealt{narayan86} for a review) and this is evident when considering the fluxes within our sample where a high amount of scatter is observed. The hybrid method improves the flux levels at smaller scales, but the uncertainty in flux imposed is likely to outweigh the benefit of an increase in the imaging of additional diffuse emission for our sources. We therefore suggest that although the MEM and hybrid methods provide a good test of image fidelity and in determining true cross-lobe variations, they are unlikely to provide more reliable ages and statistics than those obtained through standard CLEANing.

\begin{figure*}
\centering
\includegraphics[angle=0,width=8.80cm]{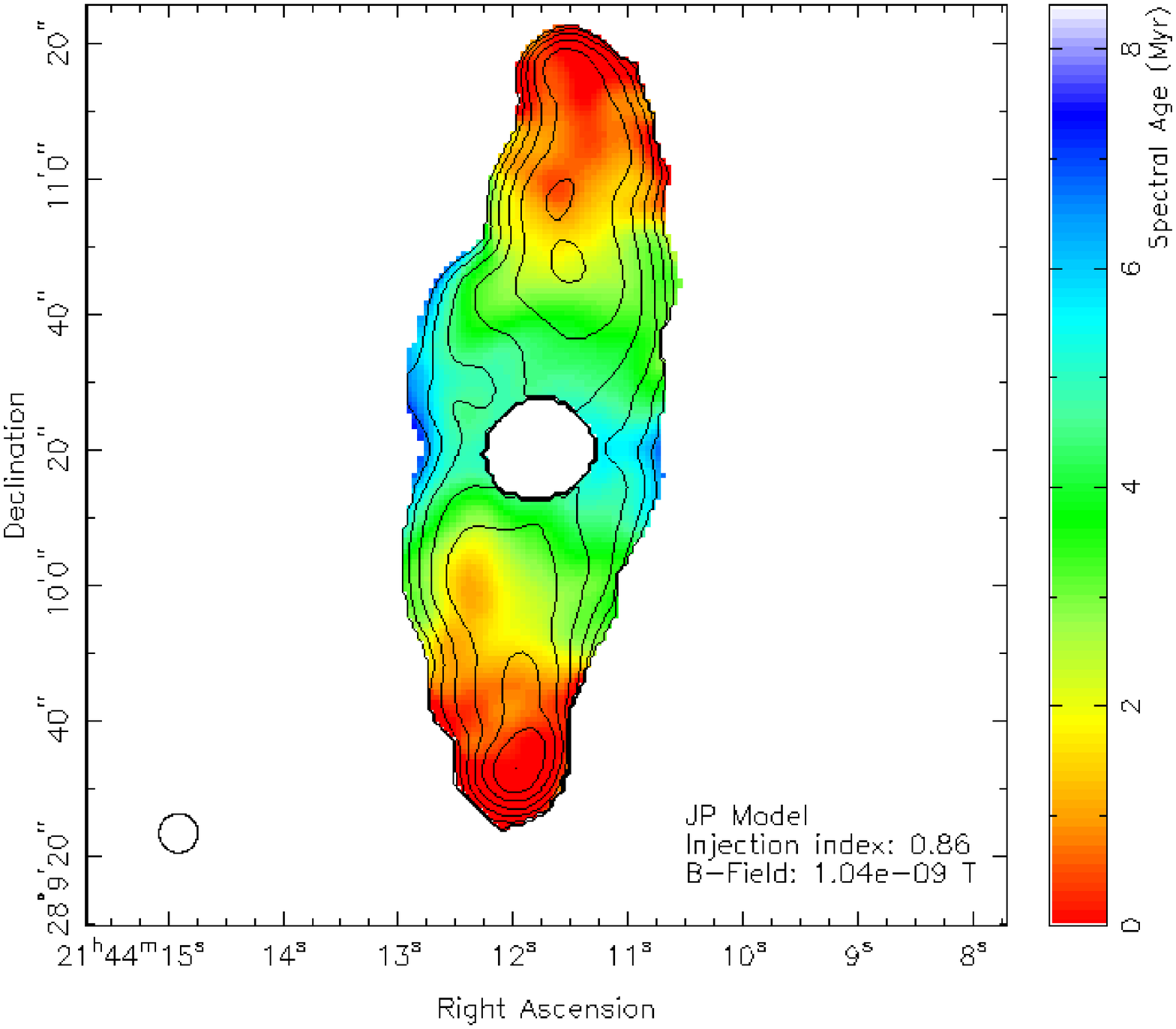}
\includegraphics[angle=0,width=8.80cm]{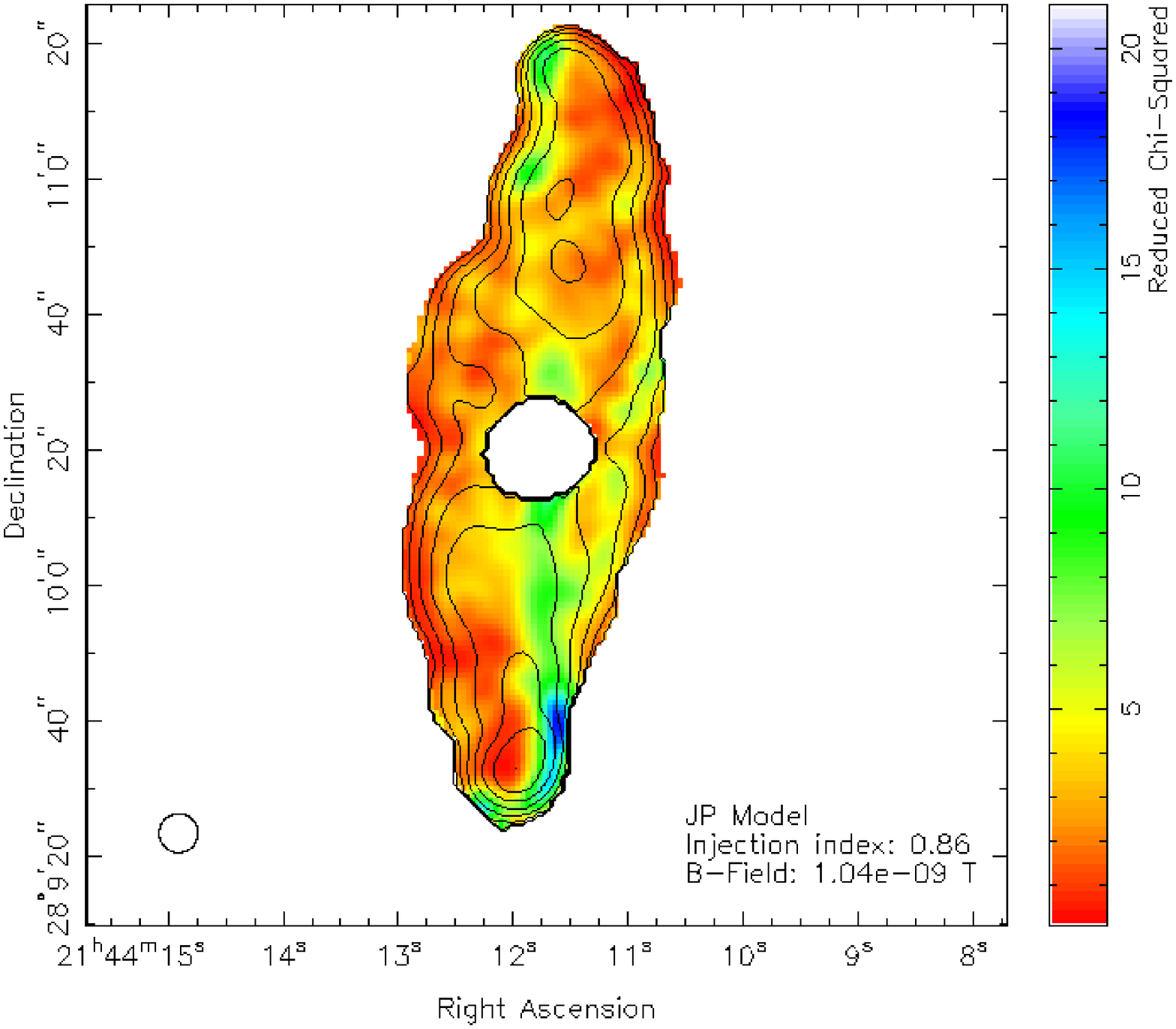}\\
\includegraphics[angle=0,width=8.80cm]{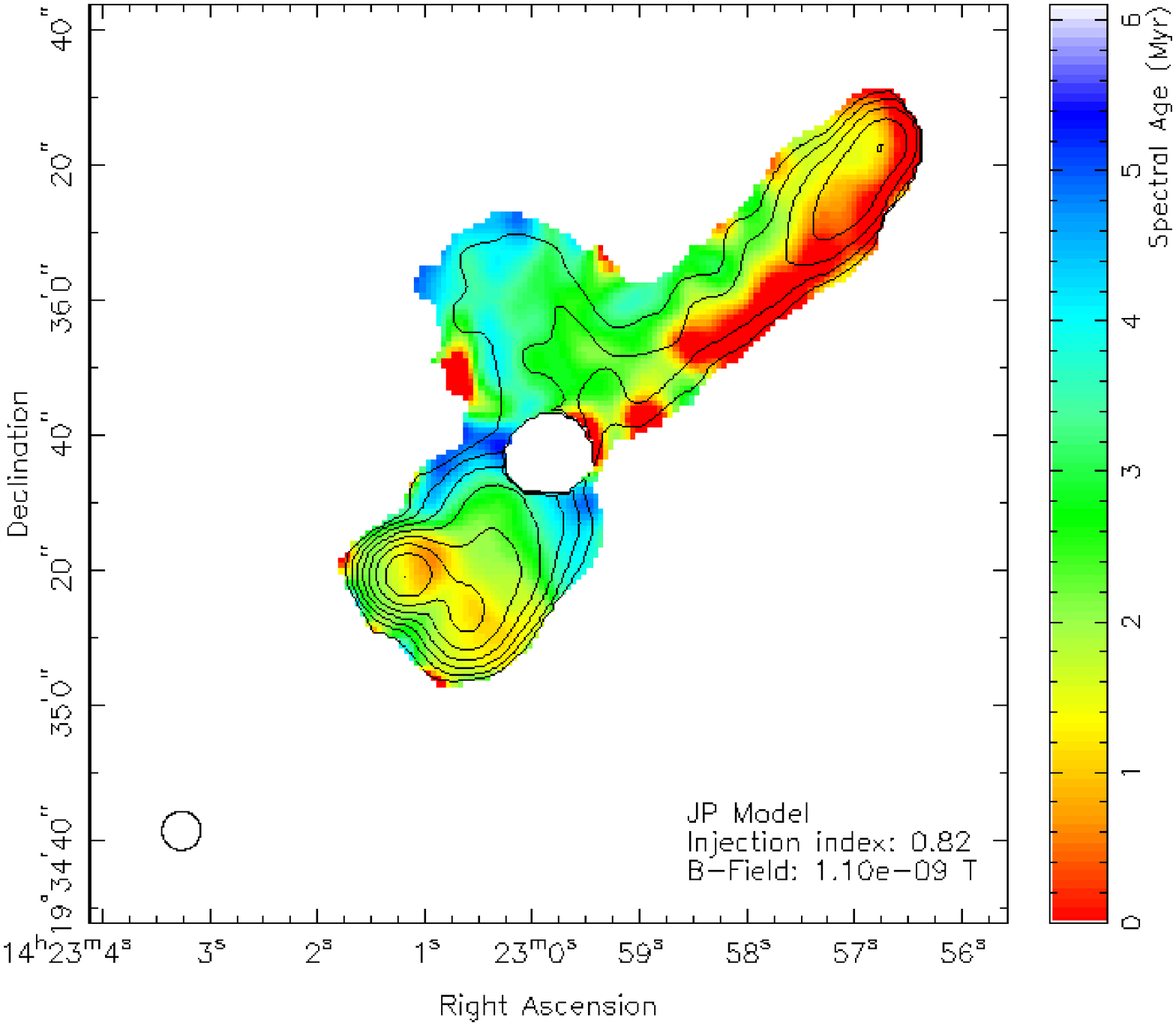}
\includegraphics[angle=0,width=8.80cm]{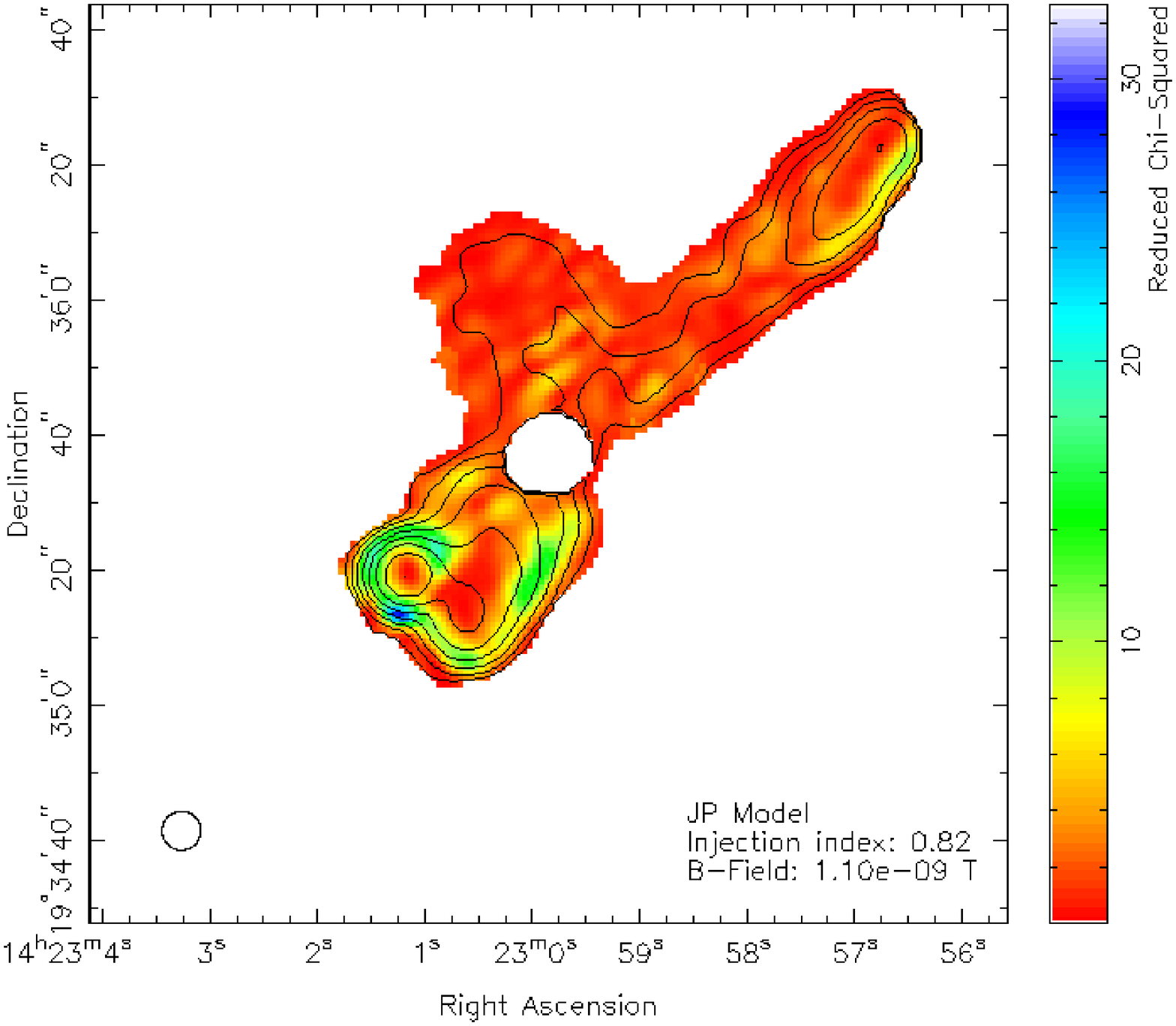}\\
\caption{Spectral ageing maps (left) and corresponding $\chi^{2}$ maps (right) of 3C300 with 7.2 GHz flux contours using multi-scale imaging. The JP model of spectral ageing is shown using the best fitting injection indices of 0.86 and 0.82 for 3C436 and 3C300 respectively.}
\label{multiscale}
\end{figure*}

\begin{figure*}
\centering
\includegraphics[angle=0,width=8.75cm]{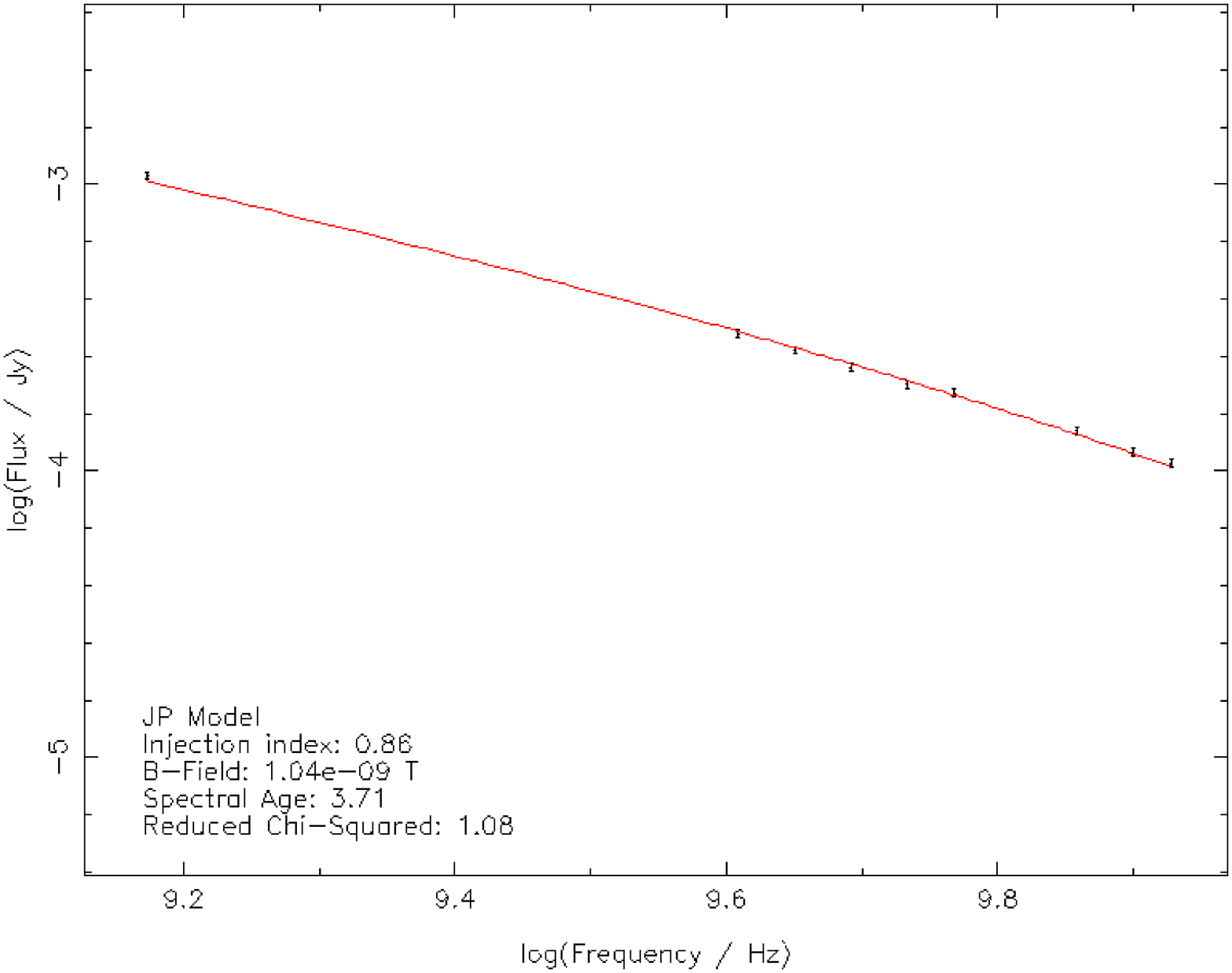}
\includegraphics[angle=0,width=8.75cm]{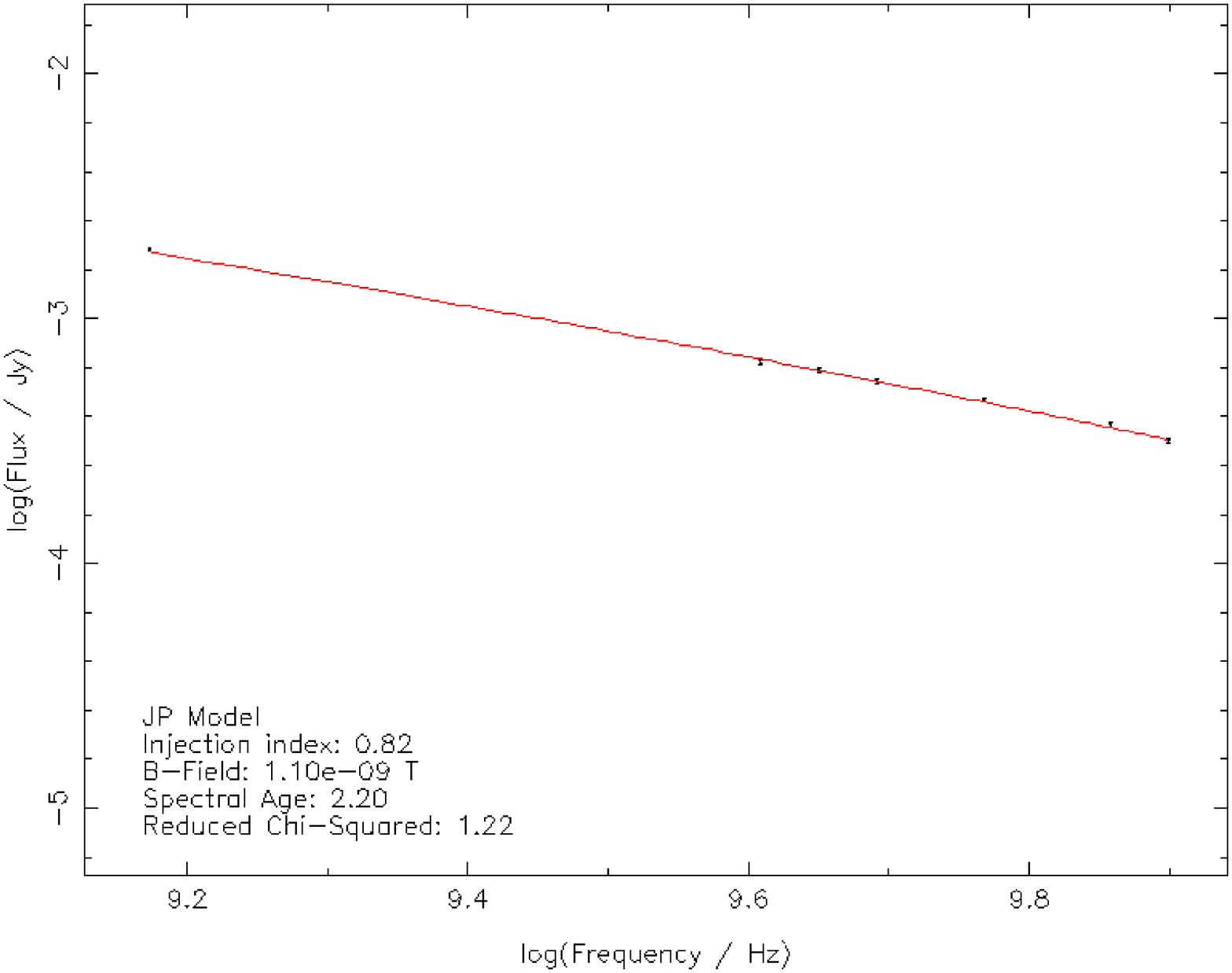}\\
\caption{Plots of flux against frequency for two well fitted regions of 3C436 (left) and 3C300 (right). Red lines indicate the best-fitting JP model with parameters as noted on each plot.}
\label{goodfit}
\end{figure*}

It is also important that we determine on what scales these variations exist if they are to be applied to future studies. The spectral age maps shown in Figure \ref{multiscale}, which use images created using multi-scale CLEANing, show that structure may be present on scales smaller than those observed when using the standard, single-scale CLEANing. Although the large scale spectral morphology remains similar, greater structure is evident on smaller scales where spectral ages are observed to vary on spatial scales of approximately the beam size. These age variations are small ($<\,1$ Myr) and so are unlikely to have a large impact on the results within this paper. Whether these variations are real or a result of deconvolution is not yet clear, but future studies should consider the possibility that variations exist within the data on scales smaller than those observed when single-scale CLEANing is used. We will investigate the optimal parameters for similar sources in the context of multi-scale CLEANing and the physical reality of these variations in greater detail using fully broad-bandwidth data in a subsequent paper currently in preparation \citep{harwood13}.

\subsection{Lobe Speeds}
\label{lobespeeds}

Assuming an initial acceleration of an electron population in the region of the hot spot which then ages freely over time, we are able to determine a characteristic speed of a given lobe, $v_{lobe}$, using the spectral age of oldest material and the current distance of the hot spot from the core. Defining a characteristic lobe speed in this way has previously been used by two of the largest spectral ageing samples by \citet{alexander87} and \citet{liu92} allowing a direct comparison to be made to lobe speeds of similar FR-II sources.

From Table \ref{lobespeed} we see that for 3C436, the best-fitting injection index provides a lobe speed of between 0.144 (Tribble) and 0.168 $c$ (KP) which are in good agreement with those found by \citet{myers85} and \citet{liu92} for similar FR-II sources. If we consider the case where $\alpha_{inj} = 0.6$, we find that due to the increased curvature (hence age) required in fitting of the models, the lobe speeds drop considerably to between 0.0545 (Tribble) and 0.0605 $c$ (JP), in line with the works of \citet{alexander87} who find significantly lower lobe speeds over their sample. 

For 3C300, we find the speed of the northern lobe to be much higher than would be expected with values between 0.250 (Tribble) and 0.311 $c$ (KP) in the best-fitting case. Although we find that where $\alpha_{inj} = 0.6$ these values drop to between 0.109 (Tribble) and 0.133 (JP), these speeds are still notably higher than those presented by \citet{alexander87} where they find for the northern lobe a speed of 0.026 $\pm$ 0.045 $c$. Assuming initial acceleration of the oldest regions of plasma occurred when the hot spot was located near the core, we are also able to derive a lower limit of the drift speed for the emission found in the side lobe of 3C300. Using the Tribble model (which provides the oldest ages) and where $\alpha_{inj} = 0.82$, we find a lower limit for the average speed of 0.113 $c$. Where $\alpha_{inj} = 0.60$ an average speed of 0.0493 $c$ is required to reach its current location. Whilst quite high, these speeds are not physically implausible. For the southern lobe of 3C300 we find for the best-fitting injection index the lobe advance speeds lie between 0.103 (Tribble) and 0.120 $c$ (KP) and where $\alpha_{inj} = 0.6$, speeds range between of 0.0465 (Tribble) and 0.0515 $c$ (JP). The large variation in advance speed between the northern and southern lobes is likely to be due to possible effects of both orientation and jet environment which are discussed further in Sections \ref{morphology} and \ref{spectralages} respectively.

\subsection{Source Reconstruction}
\label{reconresults}

\begin{figure*}
\centering
\includegraphics[angle=0,width=8cm]{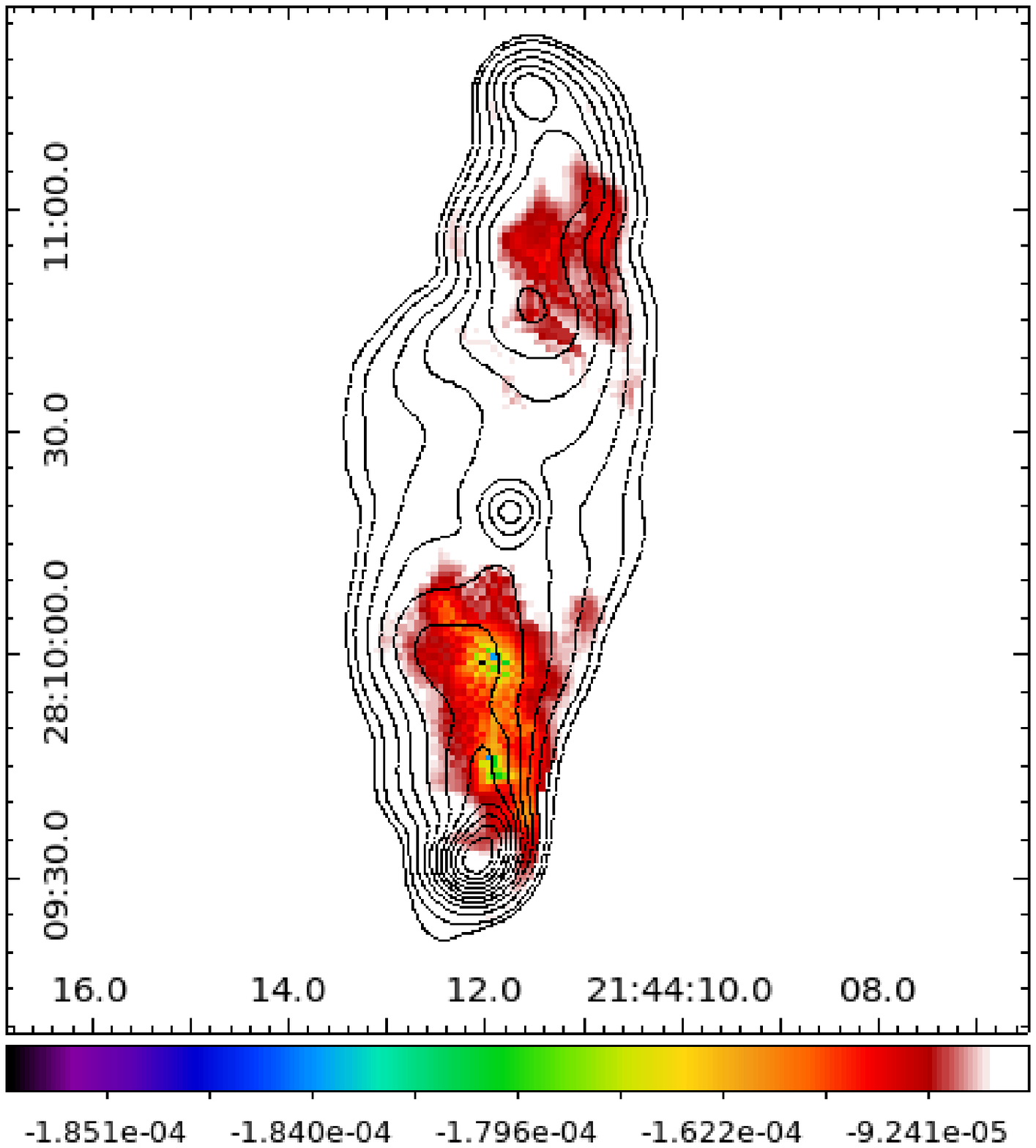}
\includegraphics[angle=0,width=8cm]{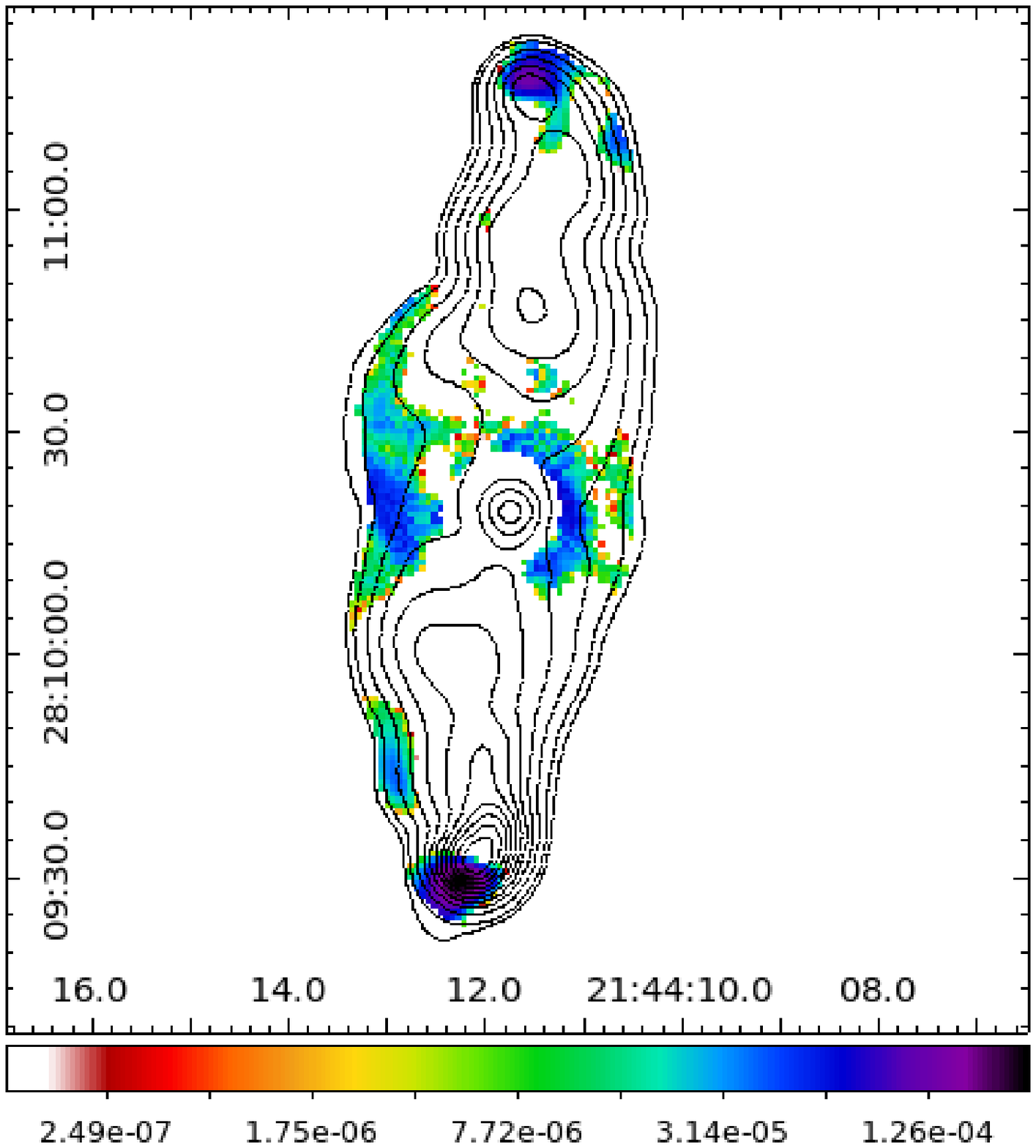}\\
\includegraphics[angle=0,width=8cm]{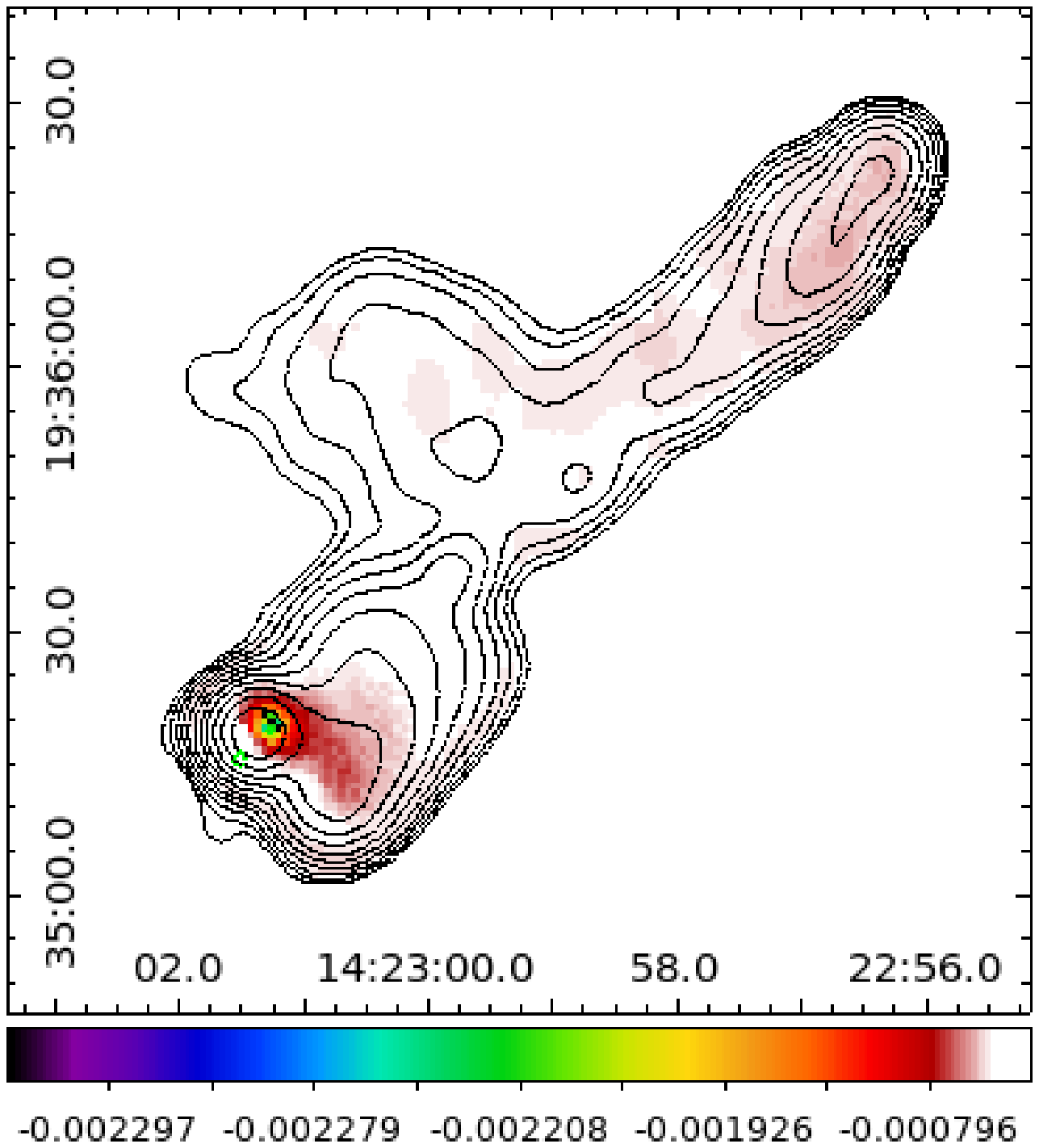}
\includegraphics[angle=0,width=8cm]{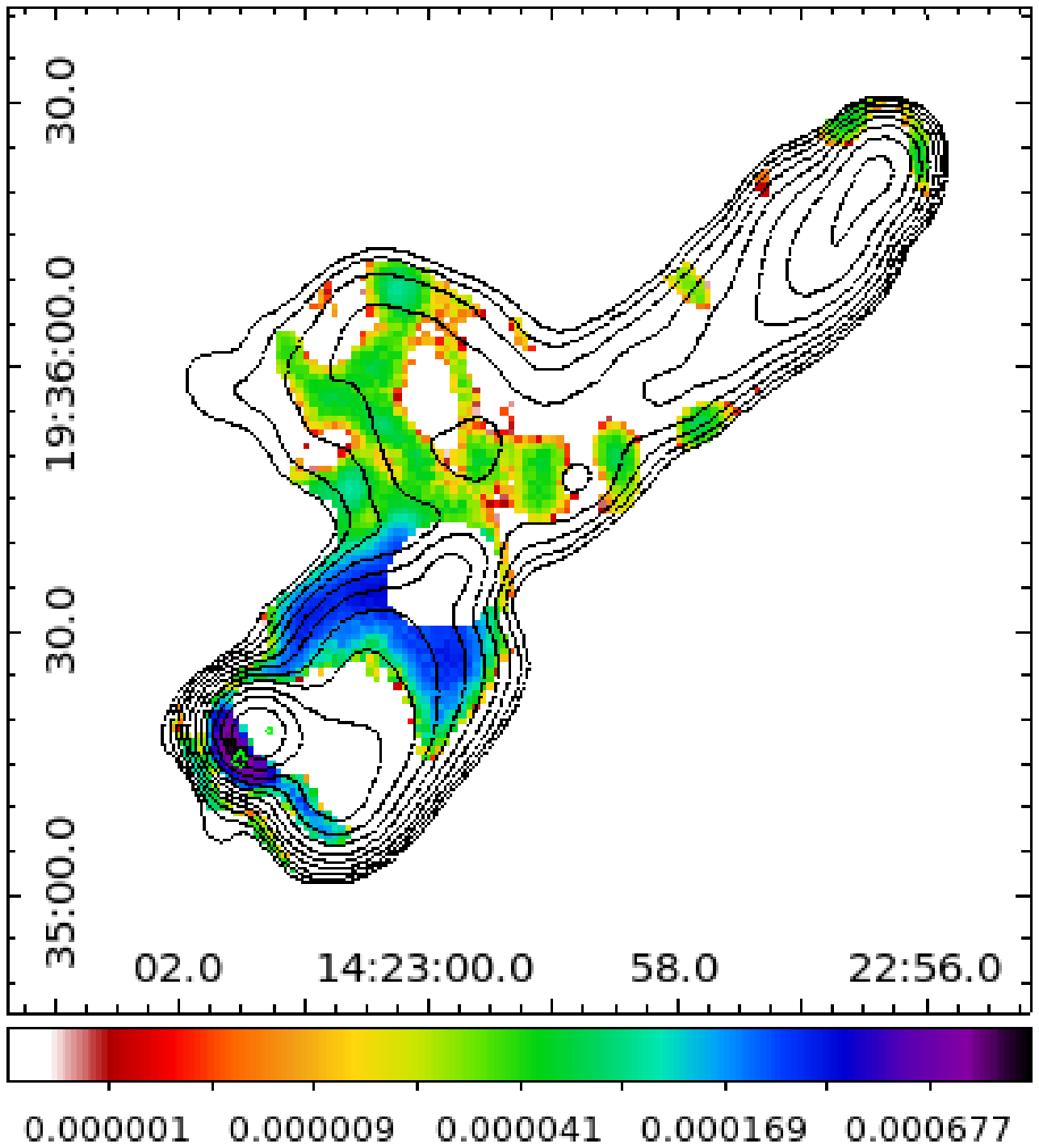}\\
\caption{Maps of residual flux from the subtraction of the reconstructed model source from the observed source as described in Section \ref{reconresults}. Model subtracted maps of 3C436 (top) and 3C300 (bottom) showing regions of model flux overestimation (left) and underestimation (right) are shown. Scale is in units of Jy beam$^{-1}$. Here the KP model has been used in reconstruction of the source using the best fitting spectral ages and normalizations. Maps are limited to the regions used in the model fitting of Sections \ref{modelfitting} and \ref{specage}. Overlaid are the 6 GHz contours of the combined frequency maps of Figure \ref{combinedmaps}. }
\label{subtractedmaps}
\end{figure*}

In principle, if the spectral age and normalization of a region within a source are known, then the flux density of that region can be determined for any given model at any given frequency. Within this paper, the regions of the source are taken on a pixel by pixel basis, hence it was possible to reconstruct the source as a whole producing images comparable to the combined radio maps of Figure \ref{combinedmaps}. The flux density of each pixel at 6 GHz was therefore calculated using the best-fitting model, age and normalization values as determined in Section \ref{specage} for both 3C436 and 3C300. In reality, differences will exist between the image as reconstructed from the model and that which is observed in regions where it does not provided an exact description of the source. The reconstructed image was therefore subtracted from the combined frequency images using {\sc brats} to provide a goodness-of-fit map as a function of position and to highlight any large scale correlations which exist. Two images were then produced for each source showing regions where flux density has been over- and underestimated, the results of which are shown in Figure \ref{subtractedmaps}.

We find, as expected, that there are slight variations between the combined frequency maps and those determined through reconstruction of the source for the best fitting KP model. From Figure \ref{combinedmaps} we note that the off-source RMS of the combined frequency maps is $27$ and $33$ $\mu$Jy beam$^{-1}$ for 3C436 and 3C300 respectively. Comparing these values to the subtracted images, we find that the model reconstruction agrees well across the majority of both sources but note some distinct regions where flux density variation is well above the RMS noise.

For 3C436, we notice that the reconstructed image considerably underestimates the flux density at the tips of both the northern and southern lobes. These correspond to the brightest regions of the source, but note from the high resolution images of Figure \ref{highresmaps} that for the southern lobe, this region may not be the location of the primary hot spot. We also see a similar discrepancy in the southern hot spot of 3C300, although in this case the model overestimates the source flux density. As discussed in Section \ref{specage}, this is unlikely to be due to the high injection indices as the model fits improve within these regions when compared to fits at lower injection indices. It is therefore possible that these bright regions we are again subject to dynamic range effects. An underestimate in flux density of the reconstructed source is also seen in regions close to the core and in the side lobe of 3C300. These flux density differences are much smaller than those found at the tips of the lobes and are found in the oldest regions of plasma. We see from Figures \ref{3C436specagemap} and \ref{3C300specagemap} that both of these areas of the source provides a good fit to the model. Bearing in mind that here we are not correcting for the on-source noise described in Section \ref{adaptiveregions}, it is likely that these underestimates are due to the large fractional error present in these fainter regions. In the case of 3C300 we also see an underestimate of flux density along the bright southern edge of the source and in regions of the southern lobe close to the core. Comparing the location of these areas to those of Figure \ref{3C300specagemap} this is not surprising as they correlate closely to the poor fitting of the model within these regions. This is likely due to possible dynamic range and orientation issues discussed further in Sections \ref{morphology} and \ref{spectralages}.

Probably the most interesting result found within these maps are the extended regions of overestimated model flux density in the southern lobe of 3C436 and the northern lobe of 3C300. These regions closely correlate with the suspected location of the jets in both the high resolution maps of Figure \ref{highresmaps}. If these are indeed the location of the jets, then it is not surprising that the model does not successfully reproduce the observed flux density, as we would not expect the spectral ageing models and the assumptions therein to be valid under such conditions. The implications of this finding in conjunction with those of Section \ref{specage} with respect to determining the age, and hence power, of sources are discussed in detail in Section \ref{discussion}.

\section{Discussion}
\label{discussion}

The dependency of spectral ageing on a number of assumptions (e.g. equipartition) and poorly constrained parameters (e.g. magnetic field strength) have historically called in to question both its reliability as a measure of age (hence power) as well as the validity of the spectral age models as a whole (e.g. \citealp{eilek96a, eilek96, blundell01}). The results we present within this paper provide the first opportunity to further investigate previously held assumptions and to constrain some of these parameters to unprecedented levels. In the following sections we discuss these parameters, our results and the influence they have on our current understanding of the dynamics of these sources.

\subsection{Source Morphology}
\label{morphology}

\begin{figure*}
\centering
\includegraphics[angle=0,width=8.75cm]{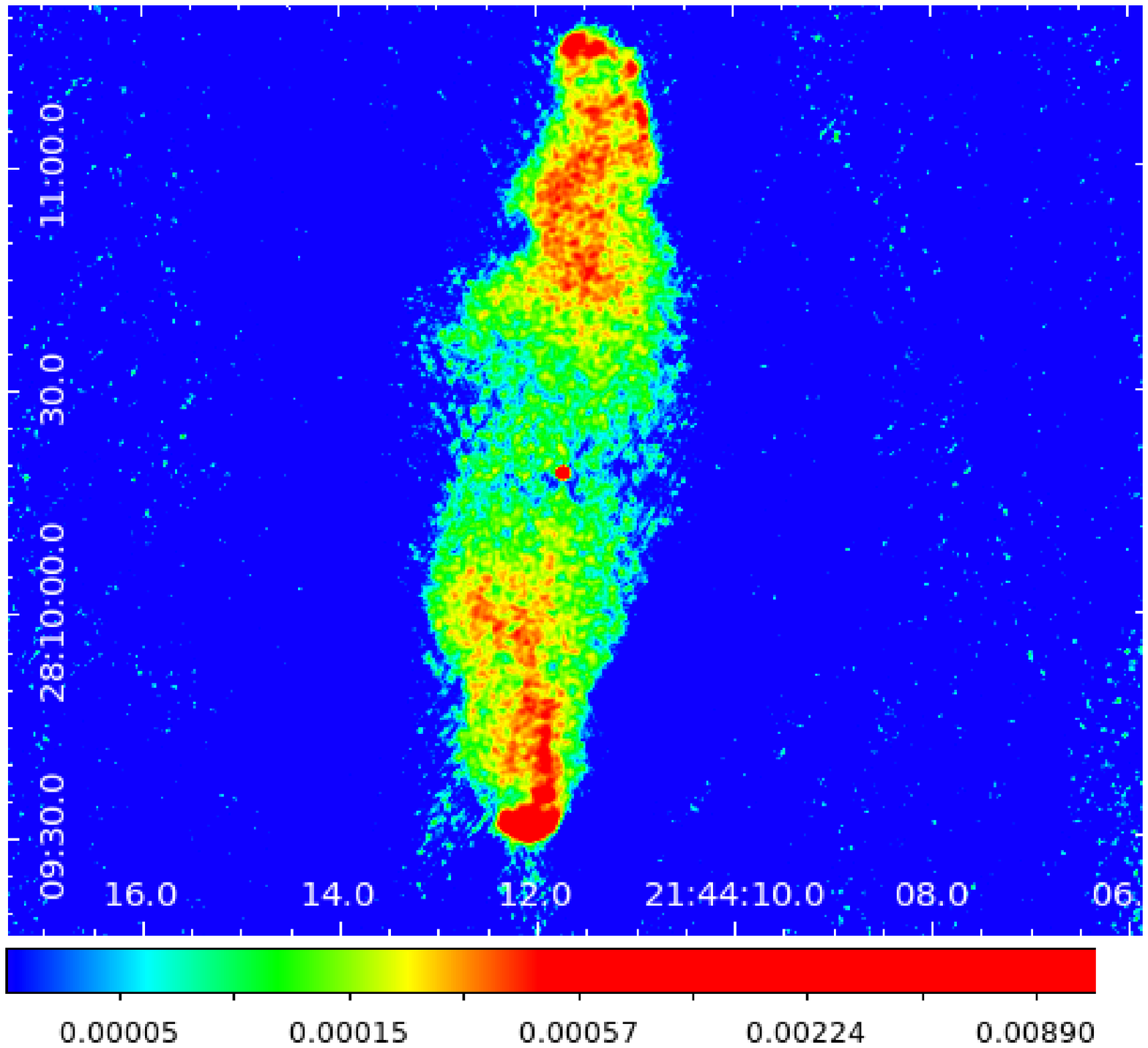}
\includegraphics[angle=0,width=8.75cm]{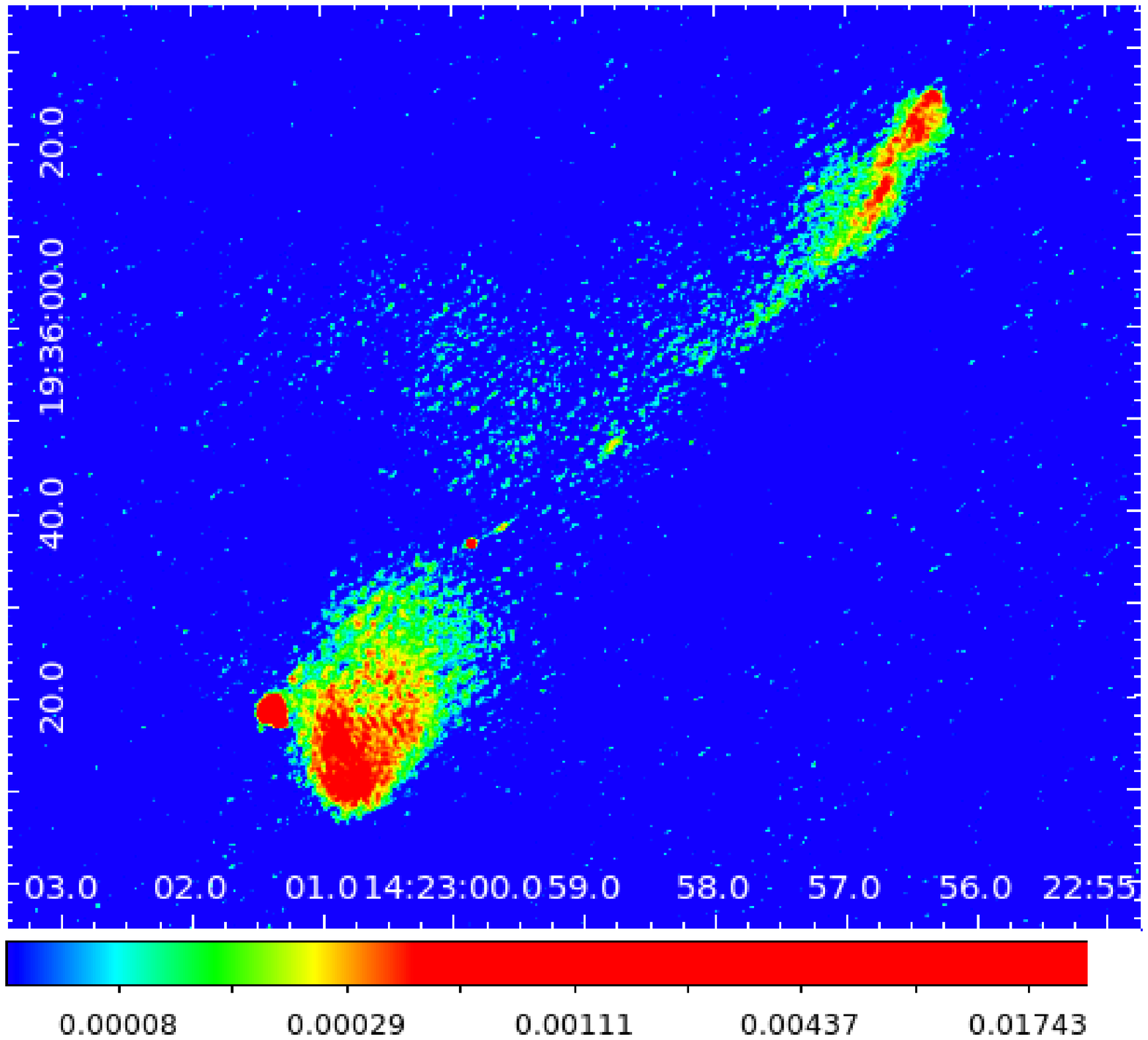}\\
\caption{High resolution total intensity maps of 3C436 (left) and 3C300 (right) by \citet{hardcastle97}. The resolution of the maps are 0.75 and 0.50 arcseconds for 3C436 and 3C300 respectively. The flux scale has been adjusted to enhance the bright small scale features and jets described in Section \ref{specage}.}
\label{highresmaps}
\end{figure*}

So far, we have assumed that the viewing angle of these galaxies is side-on. From the high resolution maps (Figure \ref{highresmaps}) of 3C436, we see that only a one-sided jet is observed from which we can infer that the source must, to at least some degree, be tilted towards the line of sight. However, sources in which the angle to the plane of the sky is large are generally expected to be non-symmetric due to time travel differences causing one lobe to be observed at an earlier stage of evolution than the other. The symmetry of 3C436 in terms of lobe size and classical double structure therefore suggests that this angle is likely to be small and so the assumption of a side-on viewing angle is probably still a good approximation. The irregular morphology of 3C300 means that this assumption is far less clear cut. We see from Figure \ref{combinedmaps} that the southern lobe of 3C300 is significantly shorter than than that of the northern lobe and contains bright emission away from the hot spot (B of Figure \ref{pointloc}). We also clearly observe a distinctive region of diffuse emission to the side of the source, which we assume to be associated with the northern lobe.

One strong possibility for this asymmetric morphology is the environment in which the galaxy resides. If the external gas density of the medium through which the jet passes differs significantly between the northern and southern lobes, then we would also expect to observe differing lobe morphologies. If a jet resides in a low density environment, the jet can easily propagate through the medium, hence we would expect an elongated structure as is observed in the northern lobe of 3C300. Conversely, if the external gas density is high we would expect a much shorter, wider lobe to be present. Unfortunately, determining whether 3C300 resides in such an asymmetric environment would require X-ray observations of the surrounding medium; however, the derived jet speeds and spectral ages, discussed further in Section \ref{spectralages}, do provide some indirect evidence that this may be the case.

Although the observations made within this paper do not allow us to directly determine the cause of this irregular morphology, the possible observable effects must be considered. As mentioned in Section \ref{specage}, the poorly fitted regions within the southern edge of the lobe of 3C300 could be a result of observing angle. The general assumption for all investigations of spectral analysis of this type is that the point to point age variations along the line of sight will be small enough to be negligible (Section \ref{adaptiveregions}). For a classical double such as 3C436 this assumption is likely to hold as we see from Figure \ref{3C436specagemap} that age variations are small for any region size (i.e. one pixel deep) slice across the width of the lobe. Age variations along the line of sight would therefore remain small for any side-on viewing angle. However, if we rotate the viewing angle of the source so that we are looking significantly `head on', such as may be the case for 3C300, we will observe the effects of non-negligible variation in age that we know to exist along the length of the source. We would therefore encounter the problems of modelling a superposition of ages as outlined in Sections \ref{adaptiveregions} but along the line of sight, rather than along the length or width of the source. In areas such as the hot spot, the observed flux density will be dominated by the young, bright emission of a given (small) region with only a minor fraction contribution being made from any aged regions present, hence a good fit will still be found. For more diffuse and low flux regions such as the southern edge of 3C300, this will not be the case and the effects of superposition can no longer be ignored. It is also possible that if the source lies significantly away from the plane of the sky, then the asymmetry between the lobes may be due to light travel time effects. In this scenario, we would observe the southern lobe at a much earlier time than its northern counterpart; hence, it may not yet have evolved in to the extended structure seen in the northern lobe.

The absence of diffuse emission to the side of the southern lobe of 3C300 may also be due to the effects of orientation. If the southern lobe does in fact have a morphology similar to the characteristic L-shape of the northern lobe, but is rotated either towards or away from the line of sight, it would not be immediately evident that such a feature exists. Given the potentially large age variations and size of this region, this emission could add considerably to the superposition of spectra observed, so invalidating the underlying assumption that along any given line of sight, age variations are small. We therefore believe that in the southern lobe of 3C300, both dynamic range effects caused by the extended region of high intensity emission, along with a possible contribution from the observing angle effects mentioned in this section are likely to be the cause of poor fits in the southern lobe, rather than any physical property of the source.

The observed low age regions (3C300) and poor $\chi^{2}$ (3C436) which correlate closely to the suspected location of the jets, is an unexpected result at such moderate resolutions and is the first observation that we are aware of such features in FR-II galaxies. As the jet spectra are thought to take the form of a power law \citep{bridle84, treichel01} similar to those found in the jets of more extensively studied FR-I jets \citep{laing06b, laing08, laing13}, it is likely that these regions are a superposition of the aged and jet spectra. In the case of 3C300 the low age regions correlate closely to bright features along the jet, hence, the jet emission is likely to dominate the spectral profile in these areas. In the case of 3C436 we do not observe the jet in either flux density or spectral age, therefore the regions must still dominated by the aged plasma at these resolutions. As the flux density of the ageing spectra decreases faster with increasing frequency than a power law of the jet, this leads to a flattening of the spectral profile at higher frequencies. It is therefore possible for an age to be well-fitted at lower frequencies, but increasingly deviate from the model at as we increase in frequency leading to poor $\chi^{2}$ values (particularly when the spectral age is also low such as is seen in 3C436). In either case, our results provide a possible method for inferring the location of jets where no high resolution images are available. The jet also gives rise to a potentially significant source of cross lobe age variations which should be carefully considered in future investigations.

\subsection{Injection index}
\label{injectionindex}

In determining the spectral age of FR-II sources through the two most commonly used single injection JP and KP models, as well as the more complex Tribble model, a key assumption is made about to the initial electron energy distribution and radiative losses. It is assumed that the initial distribution (Equation \ref{initialpowerlaw}), takes the form of a power law of index $\delta$. If we are to assume these models to be correct, then by extension the observable initial spectral energy distribution of these models given by $\alpha_{inj} = 2 \delta + 1$, should also take the form of a power law; the so-called injection index. As noted in Section \ref{modelfitting}, previous studies have assumed the injection index to take values $\simnot\,0.5$ to $0.7$ \citep{jaffe73, carilli91, katz93, orru10} but it is clear from the results of Section \ref{modelprms} that these values do not provide the best-fits for any of the commonly used models tested within this paper. It is interesting to note that the study of spectral index at low frequencies by \citet{laing80} as well as subsequent early studies of spectral ageing based on these works (e.g. \citealt{alexander87a}) all find spectral indices at low frequencies of around 0.8 for FR-II sources of a similar redshift. In this section we consider possible explanations for the high values of the injection index obtained within this work.

One possibility is that the derived injection indices are being driven to high values through lack of curvature at low-frequencies. If we consider the spectral profiles of the Tribble model and the JP model on which it is based, at very low frequencies (i.e. a few tens of MHz) we see from the loss Equation \ref{emiss} that for any given age the Tribble model produces a more highly curved spectrum due to the integration of losses over a Maxwell-Boltzmann distribution. We would therefore expect a lower best-fitting injection index value to be found for the Tribble model compared to that of the JP model if the cause of the high values were due to a lack of low-frequency curvature, rather than the convergence of all models to a single injection index value within a source. In theory, this can also be reproduced by increasing the age of a given region so a greater low-frequency curvature is present; however, this would also force the model at GHz frequencies to be poorly fitted. As it is already well established that these models can provide a good description of FR-II sources, we find that this is unlikely to be the case.

The low-age regions (particularly those of zero age) also provide key insight into the injection index values found within this paper. These regions are well described by spectra close to that of a power law; hence, if the injection index was somehow being forced by the ageing models to higher values through a lack of low-frequency curvature, we would expect these regions to still provide the best-fit around the commonly assumed value of 0.6 as they are less heavily influenced by loss effects. From Figure \ref{3C436specagemap} and Table \ref{restab}, we see that this is clearly not the case. Across all sources and models, the $\chi^{2}$ values within these low age regions are better fitted by the higher injection indices. We therefore find it unlikely that a lack of low-frequency curvature in the models is the underlying cause of this unexpected result. We note that this does not preclude a need for increased curvature at low frequencies in providing the best description of spectral ageing as a whole, but it does not appear to be influencing our determination of the initial electron energy distribution.

Probably the most likely cause of such a bias to high injection indices comes from the data itself. The large frequency gap between the archival VLA observations and the higher frequency JVLA observations mean that the 1.4 GHz data hold a higher weighting in the determination of an injection index. We find that when only the JVLA data is considered the best-fitting injection index falls to the more commonly assumed values; however we cannot reject these variations based on selectively excluding data points solely to fit with convention. Historically, spectral ageing has been determined through colour-colour plots (e.g. \citealp{katz93, hardcastle01}), or, through spectral break analysis (e.g. \citealp{burch79, alexander87a, liu92, konar06}) which suffer from the same frequency space limitations. As these studies are also limited to large spatial regions, injection index values can only be determined reliably in regions of particle acceleration i.e. the hot spots, where ageing effects should be negligible. As we find no valid reason to exclude these low-frequency data points and, within this paper, we are able to determine the injection index based on many small regions, over multiple sources, fitted to multiple models, all of which converge to a common injection index, we cannot confidently reject these findings. This leaves three plausible solutions to the problem; 1) the injection indices do conform to the assumed values of $\simnot\,0.6$ but are forced to higher values by erroneous 1.4 GHz data; 2) the higher injection indices found within this paper are correct and previous studies are biased towards lower values; or, 3) the models themselves are poorly fitted at low frequencies, resulting in an observed injection index which is higher than the physical reality.

What is clear is that these findings call into question what has previously been held to be a reliable assumption. As we can find no cause of a bias towards higher than expected injection indices, we must consider the possibility that there may be either some physical reality behind these values or that previous studies where a lower injection index has been assumed may also be subject to this same effects leading to biased statistics in model fitting. We see from Figure \ref{injectmin} that with an injection index of 0.6, there is a large variation in the goodness-of-fit between models but when considering these fits for the best-fitting injection index, they become far less pronounced. Therefore regardless of the underlying reason for the high injection indices, a given model may be favoured over another due to an incorrect assumption, rather than its ability to describe a given source. This is a particular risk when combined with the cross-lobe variations in age discussed in subsequent sections.

Most importantly, this uncertainty in injection index has major consequences in determination of the underlying jet power and dynamics. As the injection index influences the spectral ages both directly through a reduced need for curvature at the oldest ages, and indirectly through the determination of the magnetic field strength, there is also a considerable variation in any derived values. Although we cannot draw any definitive conclusions with respect to the FR-II population as a whole based on such a small sample, finding the cause of this discrepancy is key if we are to determine the validity of spectral ageing models. Further investigations are currently underway to expand this investigation to a larger selection of FR-II class galaxies, which we also hope to extend to higher frequencies. Determination of the initial electron energy distribution may also be possible using LOFAR observations, which will let us investigate the low-frequency curvature and infer the initial electron energy distribution to a much greater confidence.

Some insight into the origin of these high injection indices may be provided by comparisons of the properties of the jets of FR-I and FR-II sources. In the case of FR-Is, it is found that the spectral indices of inner jets are $\simnot\,0.6$ (e.g. \citealp{young05}). The jets of FR-II sources tend to be much less well studied at multiple frequencies than their FR-I counterparts, but previous investigations (e.g. \citealp{clarke92} for 3C219, \citealp{schilizzi01} for 3C236, \citealp{kataoka08} for 3C353) have shown that spectral indices range from  $\simnot\,0.5$ to $\simnot\,1.0$ with increasing distance along the jet. An investigation by \citet{laing13} has also recently shown that the spectral index of (re)accelerated particles for the inner jets of FR-I sources decreases with decreasing bulk flow speed within their sample. Given that the jets of FR-II galaxies are generally thought to require a higher internal Mach number than FR-Is in order to form hotspots, it is possible that these higher speeds may also result in intrinsically higher injection indices. Further investigation requires detailed multi-frequency study of FR-II jets and hotspots which is beyond the scope of the present paper.

\subsection{Spectral Ages and Lobe Speeds}
\label{spectralages}

\begin{figure*}
\centering
\includegraphics[angle=0,width=8cm]{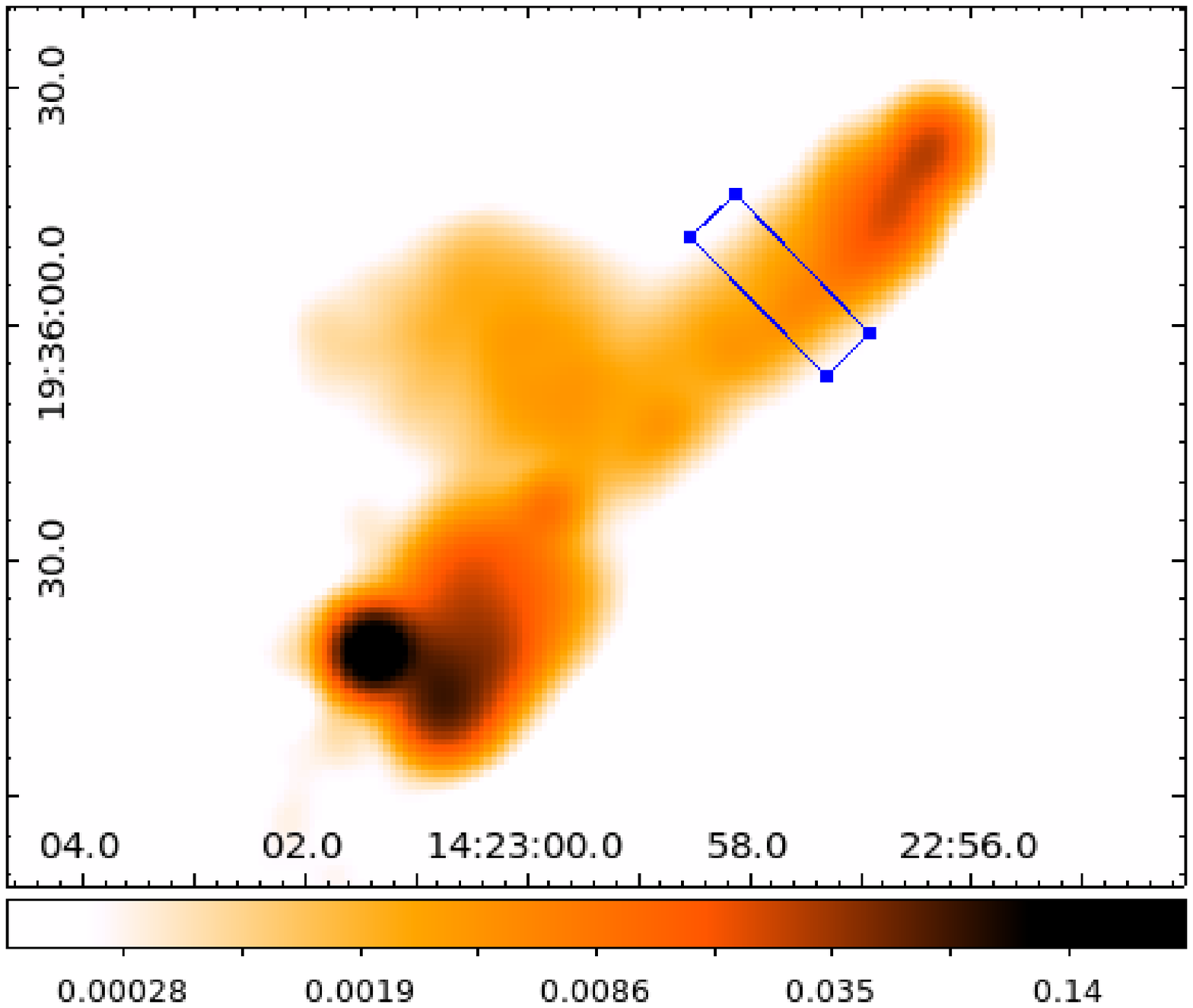}
\includegraphics[angle=0,width=9.5cm]{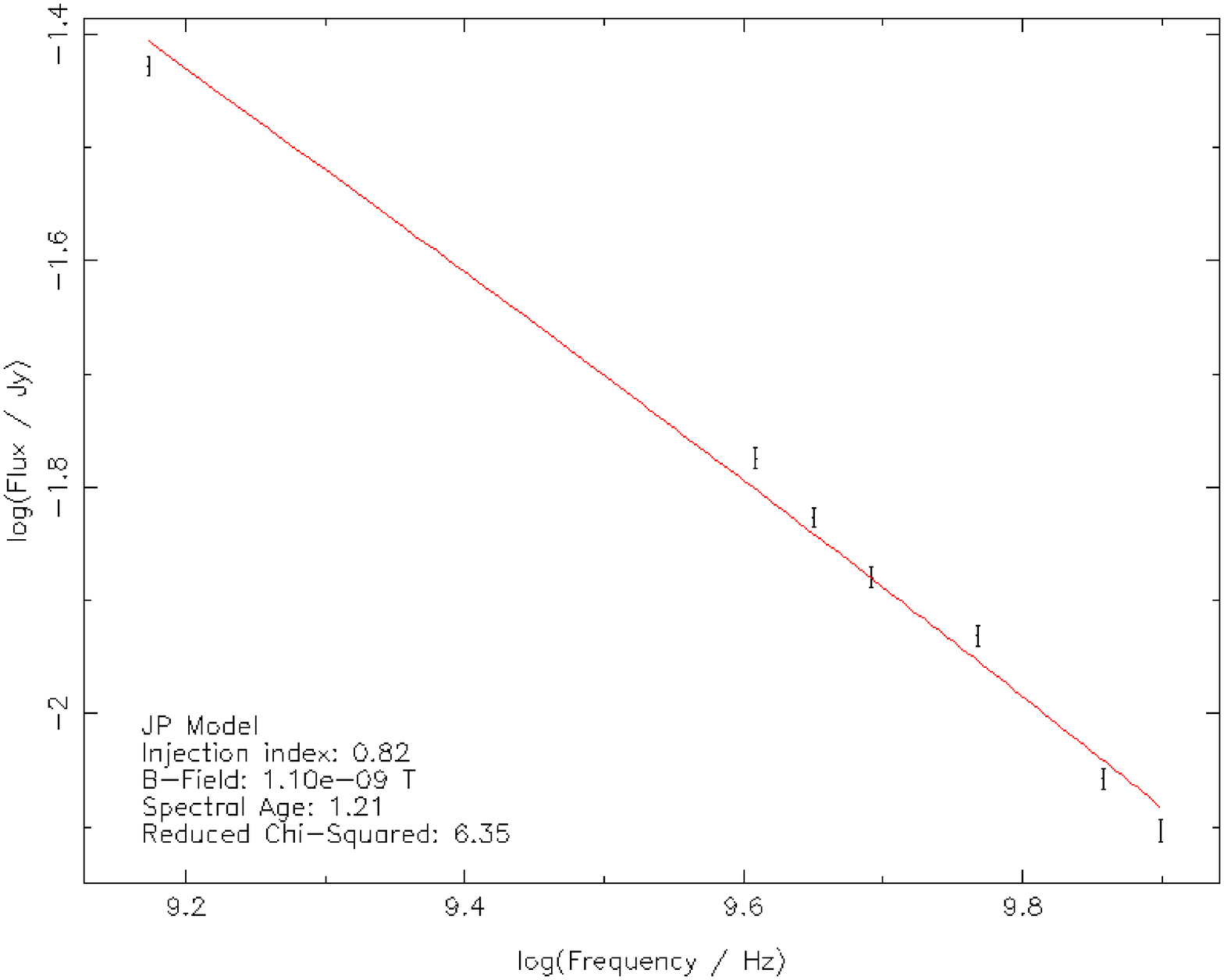}\\
\caption{Example of how spatially large regions can bias the true spectral age. Left: Combined frequency map of 3C300 at 5 GHz with a single, spatially large region overlaid. Right: A plot of the resulting JP model fit and statistics.}
\label{singlereg}
\end{figure*}

It has been known since the mid 90's \citep{eilek96a}, that spectral ages of both FR-I and FR-II radio sources underestimate the ages as determined from a dynamical view point. Later works such as those by \citet{blundell00} have suggest that spectral and dynamical ages may be reconciled, but only for sources with ages $\ll 10$ Myrs. Many alternative models such as \emph{in situ} acceleration \citep{eilek96a, carilli91} and magnetic field variations \citep{katz93, eilek96, eilek97} have therefore been proposed as possible explanations for the observed spectra of these sources. However, as has been discussed by many authors (e.g. \citealp{rudnick94, blundell00, hardcastle05, goodger08}) these methods are still unable to fully account for what is observed. Although previous determination of spectral ages on which these differences are inferred have a good physical basis, they are derived from a somewhat naive view of the source. As the capabilities of the older generation of radio telescopes only allow large regions of the lobe to be considered, often spanning across its entire width (e.g. \citealp{alexander87, machalski09}), a uniform age and normalization across the width of the lobe (or at minimum within the specified region) has to be assumed to avoid the problems associated with the superposition of synchrotron spectra raised in Section \ref{adaptiveregions}. It is clear from Figures \ref{3C436specagemap} and \ref{3C300specagemap} that this is not the case and there will be a non-negligible superposition of electron energy distributions from electron populations of varying ages. This is especially prevalent in the northern lobe of 3C300 where from the flux density maps of Figure \ref{combinedmaps} it is not immediately evident that multiple low age regions exist. If we consider a reasonably large cross section of the lobe as one single region (as has historically been the case), we observed the integrated spectra for a large number of ages ranging between 1 and 3 Myrs. This superposition creates a deviation from the true model spectrum (Section \ref{adaptiveregions}), leading to an unreliable measure of the spectral age of the source. Figure \ref{singlereg} shows the effects of these large regions when applied to our data. We clearly see that when measured in this way, the observed spectral age does not provide an accurate representation of the true spectral age. We also see that the goodness-of-fit is far worse for the single region when compared to the multiple small regions used within this paper, as one would expect if this were a result of the issues associated with superposition. We see a similar, although less pronounced, situation present in the lobes of 3C436. In this case although this new form of analysis allows us to see that notable age differences are present across the width of the source, taking large regions based on the flux density maps of Figure \ref{combinedmaps} may again cause a superposition of electron populations leading to an unreliable estimate of the age. We therefore strongly suggest that future investigation in to spectral ageing should carefully consider spectral morphology of a sample if large regions are to be considered, or preferably if the data allow, use small regions such as within this paper to account for any small scale variations.

We find that the lobe speeds of 3C436 are consistent with those found by \citet{myers85, alexander87} and \citet{liu92} for similar FR-II sources which use similar characteristic lobe speeds to those derived within this paper; however, it initially appears that the northern lobe of 3C300 is advancing much faster than one might expect. As discussed in Section \ref{morphology}, it is possible that the northern lobe resides in a region of low external gas density, with the southern lobe in a much denser environment. As a shock front moving through a low density medium is able to travel faster than one which is slowed by interaction with a dense external medium, the high advance speed observed in the northern lobe is perhaps not surprising. The complimentary low advance speed of the southern lobe and high drift speed of the northern side lobe also provides good support for this dichotomy in environment. It is currently not possible to conclusively deconvolved the effects of orientation, light travel time effects and interaction with the external medium; however, given that the spectral ages of both the northern and southern lobes of 3C300 are comparable, hence orientation effects with respect to `head-on' viewing angle are likely to be small, we suggest that differing external gas densities is the most likely cause for both the morphological asymmetry and advance speeds of 3C300.

 The advance speeds of FR-II sources can also be considered through arguments of lobe asymmetry. \citet{scheuer95} proposes that assuming the brightest jet is observed in the approaching lobe, then for a source which is intrinsically symmetric the ratio of apparent lobe lengths, $Q$, is related to the viewing angle $\theta$ by \begin{equation}\label{scheuereq} Q = \frac{(1 + v_{a} \cos{\theta})}{(1 - v_{a} \cos{\theta})}\end{equation} where $v_{a} \equiv v/c$ ($\beta$ of \citealt{scheuer95}) is the lobe advance speed. From their sample of 43 sources, they find a typical advance speed of only a few percent the speed of light. This is much slower than either of our sources, but we note from  Figure 5 of \citet{scheuer95} that these values range from between $-0.2$ and $0.4$c, well within the range of our derived values. This form of analysis is only suitable for finding the `typical' advance speed over a large sample and so cannot be directly applied to individual target sources. This is particularly evident from those considered within this paper, where the possible non-intrinsic asymmetry of 3C300 makes it immediately unsuitable for this type of analysis, and the inverted lobe asymmetry of 3C436 (i.e. the jet is located in the short of the two lobes) means an advance speed cannot be derived reliably in this way. However, their analysis does raise an important issue in determination of advance speeds derived through spectral ageing analysis. \citet{scheuer95} suggest that the speeds determined from spectral age could be dominated by the back flow of ageing plasma, rather than the advance of the lobe. If this is the case, the inferred speeds of a source would be much greater than the physical reality.

The possible causes of low ages and fast advance speeds discussed above do not, however, fully account for the difference of up to 10 times the dynamical age found by \citet{eilek96a}. This issue is further complicated by the fact that the high injection indices found within this paper only serves to increase this discrepancy. The widely used self-similar analytic model of \citet{kaiser97} which assumes an atmosphere where the density, $\rho$, decreases as a power law such that \begin{equation}\label{kaisereq} \rho_{x}= \rho_{0} \left( \frac{d}{a} \right )^{-\beta}\end{equation} where $d$ is the distance from the AGN, $a$ is the scale length and $\beta$ is the density power law index, can provide a basis for determining the dynamical age of our sources as a test of this discrepancy. Through numerical modelling \citet{hardcastle13} (herein H\&K) determine the range over which the \citet{kaiser97} lobe length / time relations hold. They find that the lobes of sources similar to those studied within this paper grow as \begin{equation}L \simnot\,2.0 t^{9/(5-3\beta)}\end{equation} where $L$ is the lobe length in Kpc and $t$ is the ages of the source in Myrs (Figure 3 of H\&K and discussion therein). The possible asymmetric environment of 3C300 means that this relation may be unreliable, but for 3C436 we are able to find the dynamical age. Using the model of H\&K where $\beta=0.75$ and considering the northern lobe, we find the dynamical age to be $\simnot\,62$ Myrs. This gives to an dynamical advance speed of only $\simnot\,0.016c$; around one tenth the advance speed derived from the oldest spectral ages of the Tribble model with the best fitting injection index. It should be noted that the average value for $\beta$ in clusters is $\simnot\,0.6$ \citep{mohr99, sand04, croston08, newman13} and is unlikely to be any greater than the assumed value, hence we can take this to be a lower limit for the advance speed.

One possible solution to the large difference in spectral and dynamical ages lies in the assumptions we have so far made with respect to the magnetic field. Across all models, we have assumed that the field strength is in equipartition, but this may not be the case. X-ray studies by \citet{croston05} of FR-II sources find that the strength in the lobes varies between $\simnot\,0.3$ and $1.3$ B$_{eq}$, with a peak around $0.7$ B$_{eq}$, hence if the lobe field strength is lower than equipartition, much younger spectral ages will be derived. The spectral break frequency, $\nu_{T}$, is related to the age of the source, $t$, by \citep{hughes91} \begin{equation}\label{specbreak}\nu_{T} = \frac{(9/4)c_7 B}{(B^{2} + B_{CMB}^{2})^{2} t^{2}}\end{equation} where  $B$ is the field strength in the lobe, $B_{CMB} = 0.318(1+z)^{2}$ nT is the equivalent magnetic field strength for the CMB and $c_7 = 1.12 \times 10^3$ nT$^{3}$ Myr$^{2}$ GHz is a constant defined by \citet{pacholczyk70}. Rearranging Equation \ref{specbreak} in terms of $t$ and differentiating, we see that a maximum age occurs when $B = B_{CMB}/\sqrt(3)$, which in the case of 3C436 equates to a field strength of $\simnot\,0.25$ B$_{eq}$; however, even with a magnetic field strength below the lower bound of \citet{croston05}, the oldest regions of plasma still only rise to $\simnot\,15$ Myrs for the best fitting injection indices and $\simnot\,40$ Myrs where $\alpha_{inj} = 0.6$. We therefore find unlikely that magnetic field strength in isolation can account for the difference between the dynamical and spectral ages, although they potentially play a major role in reconciling these values.

An alternative solution to this disparity in age is that the lobes of these sources are instead best modelled by a continuous supply of newly accelerated particles, rather than through the single injection models used within this paper. The most widely used model of this continuous injection (CI) is that detailed by \citet{pacholczyk70}. Within this model it is assumed that a constant supply of particles are injected into the lobes which then subsequently age through standard JP losses. In order to fit this model to observations we must assume that the injected particles are confined to the fitted region; hence only the integrated flux density can be considered. Consequently, we would expect to observe that a CI model is well-fitted to all lobe spectra since the integrated flux density encompasses both aged emission (the lobes) and a source of continuously injected particles (the hotspot). We therefore do not consider fitting of the CI model within this paper but it is still possible to infer the effect of the CI model from \emph{in situ} particle acceleration, from sources such as turbulence within the lobes, on the measured spectral age. As within these models particles are constantly being accelerated throughout the lobes, the supply of high energy electrons is replenished and the spectra should be observed to be flatter than those given by the single injection models. Higher ages are therefore required to account for the observed curvature, potentially closing the gap between the spectral and dynamic ages, but the physical reality of such a model is not so clear cut. If there is indeed a significant underlying contribution from \emph{in situ} particle acceleration, then as the flux density observed from high energy electrons in aged regions of plasma will decreases faster than the power law distribution of newly accelerated particles, we would expect to observe a flattening of the spectrum at high frequencies ($\simnot\,90$ GHz). However, no such flattening is observed at these short wavelengths \citep{hardcastle08} and the data are instead well fitted by a JP model of spectral ageing. As no specific physical process provides a clear means of causing such acceleration in the lobes of such sources, and the only evidence for such a model is the discrepancy itself, we do not favour \emph{in situ} acceleration models.

Irrespective of this age disparity, one feature which is common to all lobes and across all sources is the distinctive transition from low to high ages which are a clear prediction of the single injection model. It therefore appears that spectral ageing does have a physical basis. Whether the dynamical ages are currently being over estimated or the lobes are somehow being kept spectrally young remains an outstanding question; however, it is apparent that determination of intrinsic properties such as jet power will require careful consideration to account for the points raised within this paper if we are to ultimately determine the dynamics of these sources.

\subsection{Model Comparison}
\label{modelcomp}

The KP and JP models of spectral ageing have been a longstanding basis for the spectral analysis of the lobes of FR-II radio galaxies. It is therefore not surprising that for the best-fitting injection indices, neither model can be fully rejected based purely on the statistical tests performed within this paper. However, as we note in Section \ref{specage} there is a clear disparity in which model provides the best description of the source. Across all models and injection indices, the KP model provides a significantly better fit to the observations than the JP model. Previous studies have also shown, most notably \citet{carilli91}, that this is not an isolated case and may be common in many other sources. Whilst this in itself is not a cause for concern, the time averaged pitch angle of the JP model is a much more physically realistic case than the fine tuning of parameters that would be required to maintain the fixed pitch angle of the KP model. Should we believe the consistently better fitting KP model, or the worse fitting but probably more realistic JP model? 

Thankfully, the low-field-strength, high-diffusion Tribble model tested within this paper may be able to provide the first steps towards answering this question. As this model is based on the original JP electron energy distribution and losses, we are able to maintain the physically realistic case that the JP model provides. The results are also comparable to that of the KP model in all statistical tests undertaken but the Tribble model is more physically realistic. Although we do not suggest that the assumption made by the Tribble model is a highly accurate description of the field throughout the source, we do propose that it is likely to be much closer to the true case than a simple uniform field. The Tribble model therefore appears to bridge the gap between the JP and KP models, providing a basis for determining a more accurate description of the FR-II population as a whole.

\section{Conclusions}
\label{conclusions}

We have presented within this paper a variety of new methods which are implemented by the {\sc brats} software package for the analysis of the new generations of broadband radio data. Using new JVLA observations along with archival data at GHz frequencies, we have applied these methods to address many of the outstanding issues of spectral ageing in the lobes of FR-II radio galaxies. We have tested two of the most commonly used single injection models of spectral ageing (KP and JP) along with the Tribble model, which attempts to account for magnetic field variations throughout the source. The key points made within this paper are as follows:

\begin{enumerate}
\item We present the first high spectral resolution analysis of spectral ageing in FR-II galaxies. Our analysis further supports that spectral ageing has at least some physical basis in reality, although we suggest that classical (JP and KP) models provide a somewhat naive view of the source.

\item Lobe advance speeds agree well with previous studies of similar FR-II sources. We suggest the elongated morphology and fast advance speed of the northern lobe of 3C300 is environmental in nature.

\item Using dynamical models, we confirm the findings of \citet{eilek96a} that a discrepancy exists between those ages determined spectrally and those determined dynamically. We place an upper limit on this discrepancy of 10 times the spectral age for our sources. We find that although lower magnetic field strengths cannot in isolation fully account for this disparity, they potentially play a major role in reconciling these values.

\item In both sources observed, the KP model provides a better fit than the (more physically realistic) JP model of spectral ageing: However, we find that the model first suggested by \citet{tribble93} and subsequently investigated by \citet{hardcastle13a}, provides an equivalent goodness-of-fit as the KP model whilst maintaining the physical reality of the JP model. We therefore suggest that the Tribble model may provide the next step towards accurately describing spectral ageing in the lobes of FR-II galaxies.

\item We strongly suggest that many of the assumptions previously held when undertaking spectral ageing investigation must be viewed with caution and, going forward, methods should be adapted to take in to consideration the finding within this paper. In particular:

\begin{itemize}
\item The best-fitting injection indices of the JP, KP and Tribble models all converge to 0.86 for 3C436 and 0.82 for 3C300. This is much higher than the commonly assumed values of between 0.5 to 0.7. We discuss possible causes for this variation but suggest further investigations are made at low frequencies in an attempt to determine a reliable range of injection index values.

\item We find that even in the close to ideal case of 3C436, non-negligible age variations are present across the width of the lobe. We show how determining a lobe's spectra based on large regions spread along the length of the lobe (as has historically been the case) can bias model fitting towards low ages and poor fits.

\item A non-negligible contribution from the jet of FR-II sources is observed even at moderate resolutions. Whilst this may provide a method of jet location where high resolution maps are not available, this can again result in significant age variations over the width of the source and poor model fits.

\end{itemize}

\end{enumerate}

Future investigations over a wider sample of FR-II sources using truly broadband data in which many more frequency points are available will attempt to constrain many of the findings within this paper, particularly discerning between spectral ageing models. Extension of this work to low frequencies as part of the LOFAR nearby AGN legacy project will address the issue of low-frequency curvature and determination of reliable injection indices.

\section{Acknowledgements}
\label{acknowledgements}

JJH wishes to thank the STFC for a studentship and the University of Hertfordshire for their generous financial support. JJH also wishes to thank Volker Heesen for his constructive feedback and comments during development of the BRATS software. We wish to thank staff of the NRAO Jansky Very Large Array of which this work makes heavy use. The National Radio Astronomy Observatory is a facility of the National Science Foundation operated under cooperative agreement by Associated Universities, Inc. We also wish to thank the anonymous referee for their constructive comments and suggestions for improving this paper.

\bibliographystyle{mn2e}

\bibliography{./references}

\end{document}